%% file: main.tex
\documentclass[sigconf,9pt]{acmart}

\AtBeginDocument{%
  }

\usepackage{enumitem}
\usepackage{subcaption} 
\usepackage{graphicx}
\usepackage{makecell}
\usepackage{xspace}
\usepackage{tikz}
\usepackage{pgfplots}
\pgfplotsset{compat=1.18}
\usetikzlibrary{patterns}
\pgfplotsset{compat=1.18}
\usepackage{pgfplotstable}
\usepackage{grffile} 
\usepackage{caption}

\begin{document}

\newcommand{\name}{\textit{WAVES}\xspace} 

\title[Wearable Active Vibroacoustic Elastodynamic Sensing]{\name: \underline{W}earable \underline{A}ctive \underline{V}ibroacoustic \underline{E}lasticity
\underline{S}ensing for Low-Cost and Non-Invasive Muscle Stiffness Monitoring}


\author{Shubham Rohal}
\affiliation{%
  \institution{University of California, Merced}
  \city{Merced}
  \state{CA}
  \country{USA}
}
\email{srohal@ucmerced.edu}

\author{Dong Yoon Lee}
\affiliation{%
  \institution{University of California, Merced}
  \city{Merced}
  \state{CA}
  \country{USA}
}
\email{dlee267@ucmerced.edu}

\author{Shijia Pan}
\affiliation{%
  \institution{University of California, Merced}
  \city{Merced}
  \state{CA}
  \country{USA}
}
\email{span24@ucmerced.edu}

\begin{abstract}
Muscle stiffness is a physiologically meaningful biomarker associated with a wide range of musculoskeletal and neurological conditions. 
However, existing muscle stiffness assessment methods, such as shear wave elastography (SWE), manual palpation, and handheld mechanical probes, remain impractical for continuous, everyday monitoring due to bulky hardware, operator dependence, and limited self-use. 
Meanwhile, wearable muscle-sensing modalities such as Electromyography (EMG) and Mechanomyography (MMG) primarily capture muscle activation rather than passive mechanical properties of tissue at rest, leaving a critical gap in portable monitoring of resting muscle stiffness.

In this work, we present \name, a wearable system for continuous monitoring of muscle stiffness via active vibroacoustic sensing. 
\name injects controlled vibroacoustic excitation into soft tissue and estimates muscle stiffness by measuring shear-wave propagation velocity through Time-Difference-of-Arrival (TDoA) analysis across spatially distributed wearable sensors. 
To achieve accurate TDoA estimation with low-cost wearable hardware, we develop a physiology- and quality-guided vibroacoustic sensing framework that effectively screens and refines TDoA for accurate stiffness estimation. 
We evaluate \name through characterization experiments using ballistic jelly and end-to-end on-body evaluation with human participants.
Results demonstrate that \name can robustly estimate muscle stiffness across multiple users and different superficial muscles while enabling lightweight, wearable deployment for everyday use.
\name is able to achieve a muscle stiffness estimation error of $13.1\%$ or less across multiple users and multiple muscle groups.
\end{abstract}


\maketitle

\input{Introduction}
\input{Background}

\input{Overview}

\input{Evaluation}
\input{Related_Work}

\input{Discussion}

\input{Conclusion}
\bibliographystyle{ACM-Reference-Format}
\bibliography{ref}
\end{document}

%% file: Introduction.tex
\section{Introduction}
Muscle health monitoring is important in people's lives -- from athletes managing training load and injury recovery, to supporting workers performing repetitive sedentary/physical tasks, helping older adults maintain mobility and functional independence throughout aging, and tracking rehabilitation progress between clinical visits \cite{Francis2017Dec, Strasser2020Oct, Chastin2012Jan}. 
It is difficult for people to self-monitor muscle health because changes in muscle condition are often subtle and go unnoticed until pain, fatigue, or injury has already occurred, resulting in significantly harder recovery \cite{Vieira2025Oct}.

One of the most critical biomarkers for muscle health is the resting muscle stiffness, which reflects a wide range of physiological conditions and musculoskeletal changes \cite{proske2001muscle, gajdosik2001passive}.
For example, athletes may experience increased passive muscle stiffness during training or fatigue, which elevates the risk of non-contact strain injuries \cite{witvrouw2003muscle, Ekstrand2016Jun}.
Office workers with a sedentary lifestyle may develop neck and upper-extremity musculoskeletal disorders due to posture-induced muscle stress \cite{chaudhuri2024development}.
Despite differences in manifestation, these conditions share a common underlying physical phenomenon --  changes in the viscoelastic properties of the tissue -- reflecting mechanical alterations in the muscle and fascia that often occur independently of voluntary contraction \cite{Orizio1999Aug}. 
Although muscle stiffness is not the sole indicator of muscle health, it provides a physiologically meaningful and broadly applicable biomarker that enables quantitative assessment of underlying mechanical tissue conditions beyond instantaneous muscle activation \cite{Alfuraih2019Dec}.

The industry gold standard for muscle stiffness measurement — shear wave elastography (SWE) \cite{taljanovic2017shear} — is primarily used in clinical settings due to its high cost and bulky form factor.
SWE uses an ultrasound probe to image subcutaneous (in between skin and muscle) layers and assess their stiffness; however, beyond its physical and financial barriers, the method also requires a trained sonographer to operate, making it unsuitable for continuous monitoring.
Similarly, manual palpation is a clinical diagnostic method in which physicians palpate patients' muscles to qualitatively assess muscle stiffness \cite{Nolet2021Dec, Davidson2019Oct, Kothari2014Feb}, leading to subjective and inconsistent measurements.
Portable devices, such as Myoton \cite{BibEntry2023Jan}, estimate muscle stiffness by applying a brief mechanical impulse to the skin and measuring the resulting damped oscillatory response of the underlying tissue \cite{taljanovic2017shear, ferraioli2022perform}. 
Crucially, operating such a device effectively requires the second person to manually align the measurement probe perpendicular to the muscle, limiting its self-operating capabilities.
Wearable Doppler Elastography (WDE) \cite{kenny2021novel, liu2024wearable} directly measures tissue motion and mechanical wave propagation using miniaturized ultrasound sensing. 
However, these systems still rely on ultrasound imaging hardware, stable acoustic coupling, and relatively complex sensing pipelines, which limit their practicality for low-cost, lightweight everyday use.

\begin{figure}[!t]
\centering
  \includegraphics[width=\linewidth]{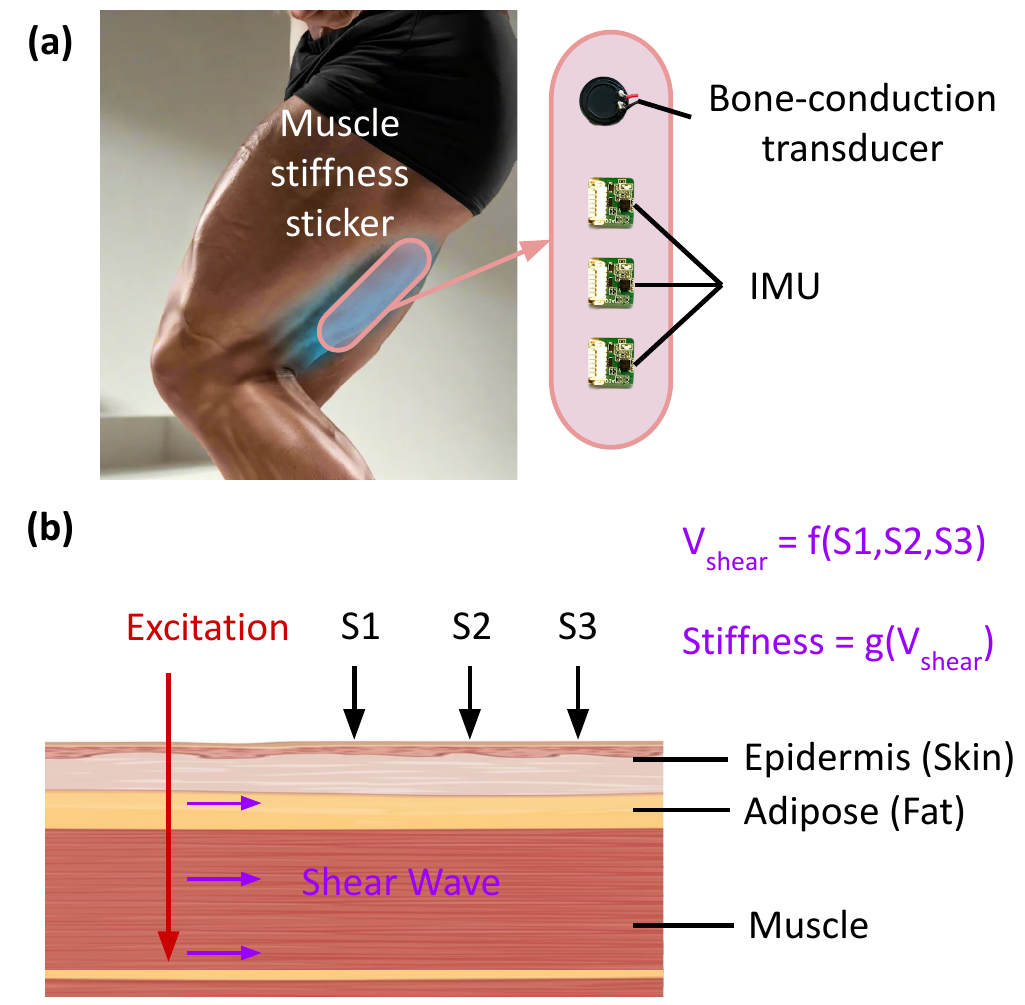}
    \caption{\name sensing principle illustration. }
  \label{fig:teaser}
\end{figure}

A broader class of portable muscle-sensing technologies includes wearable approaches such as Electromyography (EMG) \cite{de2006electromyography, mills2005basics}, and Mechanomyography (MMG) \cite{mokaya2016burnout, yasukawa2015introduction}.
While these modalities are portable and widely adopted for monitoring muscle activity, they are not designed to measure muscle stiffness at rest. 
EMG measures the electrical depolarization of muscle fibers, whereas MMG captures the mechanical vibration and deformation generated during voluntary contraction. 
Although muscle contraction can transiently increase muscle stiffness, contraction-induced stiffness differs fundamentally from the passive mechanical and viscoelastic properties of muscle and fascia tissues at rest. 
Consequently, these modalities cannot reliably capture persistent changes in resting stiffness, such as fascial densification or chronic muscle tightness, which may occur without significant electrical or mechanical activation signals.
This leaves a critical gap in the continuous, portable, and everyday monitoring of resting muscle stiffness.

We present \name, a wearable sticker-based system that measures and tracks muscle stiffness via active vibroacoustic sensing. 
Skin-mounted stickers are already widely used in everyday life for a variety of health and wellness applications -- from adhesive bandages for wound protection, to kinesiology (KT) tape for musculoskeletal support, and skin-mounted biosensors such as continuous glucose monitors (CGMs) and electrocardiogram (ECG) patches for continuous physiological monitoring.
\name adopts the sticker form factor to enable high-quality vibroacoustic elasticity sensing on the body parts, as illustrated in Figure \ref{fig:teaser} (a).
By injecting an acoustic excitation signal into the muscle and measuring its propagation via spatially distributed sensors.
The muscles' stiffness directly affects shear-wave propagation speeds \cite{bernabei2020shear}.
\name calculates the shear wave velocity as the Time Difference of Arrival (TDoA) between sensor pairs and estimates the muscle stiffness accordingly, as depicted in Figure \ref{fig:teaser} (b).
In this way, muscle stiffness can be estimated without relying on purely data-driven models that are often sensitive to inter-person physiological variability.

However, this is non-trivial due to several challenges.
(\textbf{C1}) Accurately estimating the Time-Difference-of-Arrival (TDoA) between closely spaced wearable sensors using low-cost vibroacoustic hardware is difficult due to limited spatial separation, low signal-to-noise ratio (SNR), and hardware-induced timing uncertainty.
(\textbf{C2}) Vibroacoustic propagation in soft tissue is highly heterogeneous and physiologically dependent. Variations in tissue composition, anatomical structure, posture, and muscle state introduce complex attenuation, dispersion, and propagation dynamics, making stiffness-related propagation features difficult to robustly characterize across users and real-world conditions.
(\textbf{C3}) Wearable deployment introduces unintended mechanical coupling and spurious propagation paths through the skin interface, adhesive structure, and body movement, contaminating tissue propagation measurements and reducing the robustness of stiffness estimation in everyday use.
To tackle these challenges, we co-design physiology-guided hardware with sensing-quality-aware TDoA estimation to enable reliable measurement of shear-wave propagation under wearable constraints. Specifically, we design a deployment-quality enhancement and validation pipeline to account for constraints in low-cost wearables. 
We compensate for noisy measurements by screening and refining TDoAs that deviate from physiological and geometric constraints.
By adopting this physics-driven approach, we reduce the impact of inter-subject variability and improve robustness across users.

We conduct both characterization (on ballistic jelly) and end-to-end on-body experiments to evaluate the system's performance across different people and multiple muscle groups. 
\name achieves an average stiffness estimation error of less than 13\% across multiple participants and muscle groups. This level of accuracy falls within the 20\% error margin commonly considered acceptable for non-clinical and everyday physiological monitoring applications \cite{warneke2025discussion}.

In summary, the contributions of this work include:
\begin{itemize}
    \item We present \name, a wearable sticker-based active vibroacoustic sensing system for continuous, unobtrusive monitoring of muscle stiffness in everyday settings.
    
    \item We introduce a wearable vibroacoustic active sensing framework that models soft-tissue propagation and estimates muscle stiffness through physics-informed shear-wave velocity analysis without relying on purely data-driven models.
    
    \item We co-design a physiology-guided wearable hardware and a sensing-quality-aware estimation pipeline that mitigates spurious mechanical coupling, motion interference, and low-SNR propagation challenges for robust on-body muscle stiffness estimation.
    
    \item We evaluate \name through experiments on ballistic jelly and human subjects, demonstrating accurate and robust muscle stiffness estimation across diverse sensing conditions.
\end{itemize}

The rest of the paper is organized as follows. First, we introduce background knowledge on the causes of muscle stiffness and on acoustic wave propagation and reflection in muscles in Section \ref{sec:background}. Then, we present the \name system design and implementation in Section \ref{sec:sys}. Next, we report the experimental results in Section \ref{sec:eval} to evaluate the design's effectiveness. Finally, we discuss the future directions in Section \ref{sec:discuss} and conclude in Section \ref{sec:conclusion}.

%% file: Background.tex
\begin{figure*}[!t]
    \centering
    \includegraphics[width=0.92\linewidth]{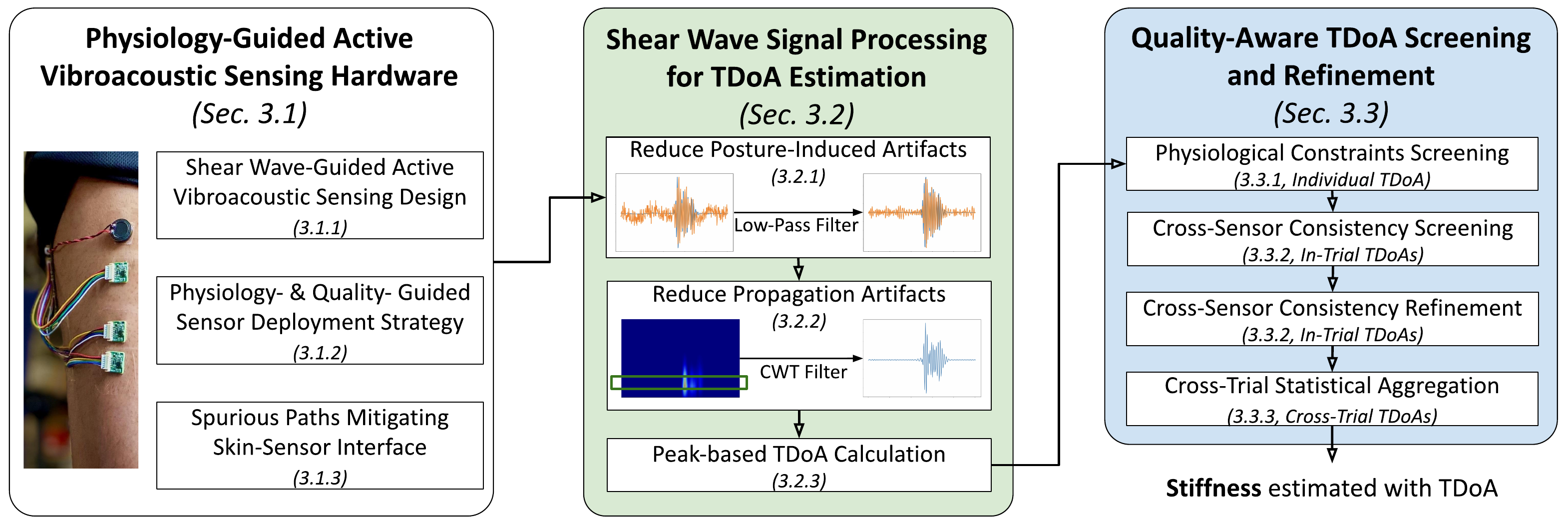}
    \vspace{-2ex}
    \caption{The system overview of \name, consisting of three main modules. First, we adopt physiology-guided wearable hardware to enable high-quality data acquisition. The acquired signals are then used to estimate shear-wave velocity through Time Difference of Arrival (TDoA) analysis. Finally, physiology- and consistency-guided screening and refinement are applied to improve TDoA accuracy for robust stiffness estimation. }
    \vspace{-1ex}
    \label{fig:overview}
\end{figure*}

\section{Background} \label{sec:background}
This section introduces the physiological significance of muscle stiffness, the need for continuous monitoring, and the acoustic principle for shear-wave-based stiffness measurement, which motivate and inspire the design of \name.

\paragraph{Muscle Stiffness: Causes and Needs}
A resting muscle in a healthy state should be soft and flexible, similar to a rubber band. 
This flexibility allows for healthy blood flow to absorb shock.
However, under various physiological and pathological conditions, muscles may lose their ability to fully relax and instead remain partially contracted, entering a state referred to as prolonged stiffness \cite{nakamura2011acute}.
Unlike the stiffness due to muscle contraction, this stiffness is involuntary and electrically silent. As a result, electromyography (EMG) becomes ineffective at capturing such changes.
Here, we list several common causes of elevated resting muscle stiffness. 

\textbf{Prolonged sitting.} 
During prolonged sitting, postural muscles such as the trapezius remain in a low-level, sustained contraction \cite{westgaard1999motor}. 
This continuous activation leads to metabolic accumulation and fascial dehydration/cross-linking \cite{larsson1988muscle}. 
Consequently, the local muscle–fascia complex exhibits elevated passive tension and increased viscoelastic stiffness \cite{purslow2002structure}. 
Therefore, continuous monitoring of these muscles' stiffness can reveal sustained low-level mechanical loading before pain symptoms appear, enabling early detection and intervention.

\textbf{Over-training.} When muscles are repeatedly loaded beyond recovery capacity, it causes micro-damage, and the damaged fibers trigger inflammatory responses and edema \cite{proske2001muscle}. 
This will elevate intramuscular pressure and tissue density, while nociceptor-mediated protective reflexes further increase baseline muscle tone \cite{stauber2004factors}. 
Together, these processes generate passive tension and increase the muscle's effective viscoelastic stiffness even at rest \cite{lacourpaille2014time}.
Therefore, continuous tracking of muscle stiffness can inform persistent mechanical changes after exercise, help identify inadequate recovery, and reduce the risk of strain injury.


\paragraph{Acoustic-based Muscle Stiffness Measurements}
Traditional Shear Wave Elastography (SWE) leverages shear wave propagation in tissue to quantify stiffness by measuring wave velocity.
An ultrasound transducer generates a localized ultrasound push pulse (acoustic radiation force) to induce shear waves -- propagate at a lower velocity -- in tissues and to track their propagation using fast plane-wave ultrasound imaging \cite{taljanovic2017shear}. 
Ultrasound imaging will generate tissue displacement maps, enabling precise tracking of shear-wave-induced displacement (micrometer scale) and the calculation of shear speed.
The shear modulus of the tissue is calculated as 
$G = \rho v^2$,
where $\rho$ is the tissue density and $v$ is the shear speed \cite{taljanovic2017shear}.
The velocity measurement of the reflected signal can be translated to the Young's modulus ($E$) or stiffness of the medium:
\begin{equation}\label{eq:stiffness}
    E= 2(1+\nu). \rho .v^2
\end{equation}
where $\nu$ is Poisson's ratio and $\rho$ is the density of the medium.
For incompressible media, the Poisson's ratio is $\nu = 0.5$, i.e., $E\approx 3. \rho .v^2$ \cite{nitta2021review}.  
In this work, we leverage the established relationship between shear-wave velocity and tissue stiffness to enable wearable monitoring of muscle stiffness. 

%% file: Overview.tex
\begin{figure*}[!t]
    \centering

    \begin{subfigure}[b]{0.30\linewidth}
        \centering
        \begin{minipage}[t][0.22\textheight][t]{\linewidth}
            \centering
            \input{Figures/pancake_1}
        \end{minipage}
        \caption{High quality shear wave ($r_H=5.86$)}
    \end{subfigure}
    \hfill
    \begin{subfigure}[b]{0.30\linewidth}
        \centering
        \begin{minipage}[t][0.22\textheight][t]{\linewidth}
            \centering
            \input{Figures/pancake_2}
        \end{minipage}
        \caption{Medium quality shear wave ($r_H=2.02$)}
    \end{subfigure}
    \hfill
    \begin{subfigure}[b]{0.30\linewidth}
        \centering
        \begin{minipage}[t][0.22\textheight][t]{\linewidth}
            \centering
            \input{Figures/pancake_3}
        \end{minipage}
        \caption{Low quality shear wave ($r_H=1.08$)}
    \end{subfigure}

    \caption{Quantifying deployment quality with Hodogram analysis of wavelet-filtered vibroacoustic signals at different accelerometers shown in Figure~\ref{fig:teaser}. The metric $r_H$, calculated as the ratio between the ranges of the $z$ and $x$ axes, quantifies shear-wave quality relative to other vibration noise. Low $r_H$ values indicate deployments with poor signal quality that require physical adjustment.}
    \label{fig:signal-quality-placement}
\end{figure*}
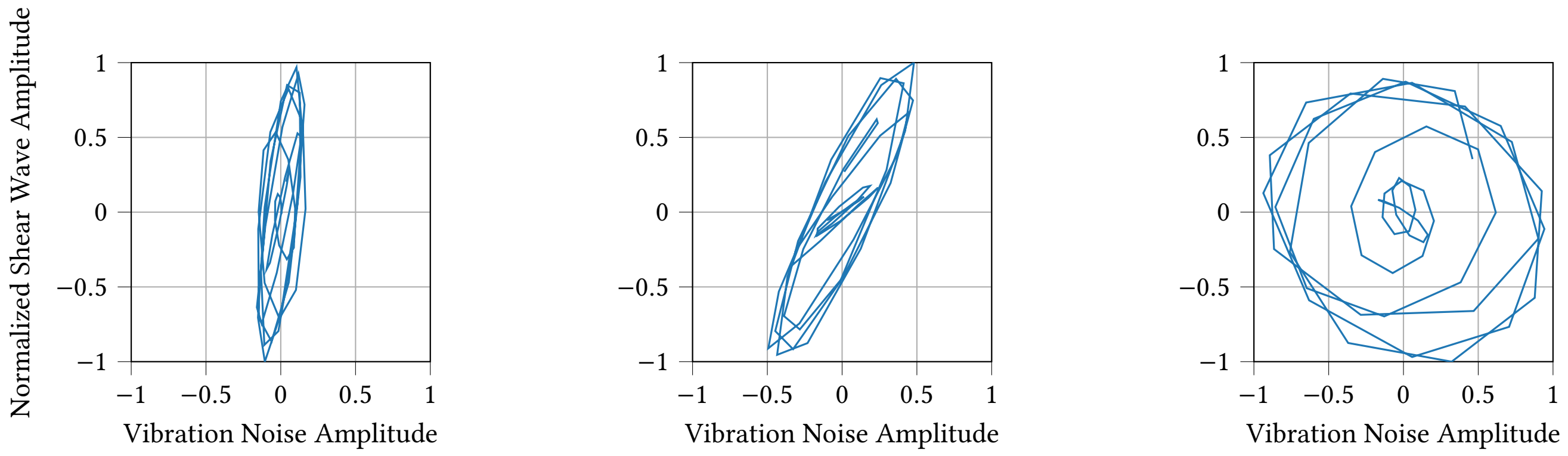

\section{\name System Design}\label{sec:sys}
The design of \name is inspired by the shear wave elastrography (SWE) -- the gold standard for muscle stiffness measurement -- where muscle stiffness directly affects the propagation dynamics of mechanically induced shear waves in soft tissues \cite{bernabei2020shear}.
Stiff tissues generally exhibit higher shear-wave propagation velocities due to their increased resistance to deformation \cite{bernabei2020shear}.
Based on this principle, \name actively injects controlled vibroacoustic excitation into the muscle and estimates the muscle stiffness by analyzing the propagation characteristics of the resulting mechanical waves, i.e., their velocity.
In contrast to bulky and expensive ultrasound imaging systems, \name achieves a lightweight wearable form factor through physiology-guided active vibroacoustic sensing with compact wearable hardware (Section \ref{sec:sensing}).
However, measuring shear wave velocity in muscle with low-cost actuators and sensors is challenging for a few reasons -- heterogeneous tissue propagation, attenuation, and dispersion of vibroacoustic signals, limited spatial separation between wearable sensors, and motion-induced interference during on-body deployment.
\name leverages physics-informed filtering and propagation-aware signal analysis to robustly estimate shear-wave velocity (Section~\ref{sec:velocity}), and applies quality-aware screening and refinement strategies to mitigate sensing uncertainty and physiological variability (Section~\ref{sec:signal-quality}).
Figure \ref{fig:overview} depicts the overview with key modules in \name.

\subsection{Physiology-Guided Active Vibroacoustic Sensing Hardware Design} \label{sec:sensing}
\begin{figure}[!t]
    \centering
    \includegraphics[width=0.9\linewidth]{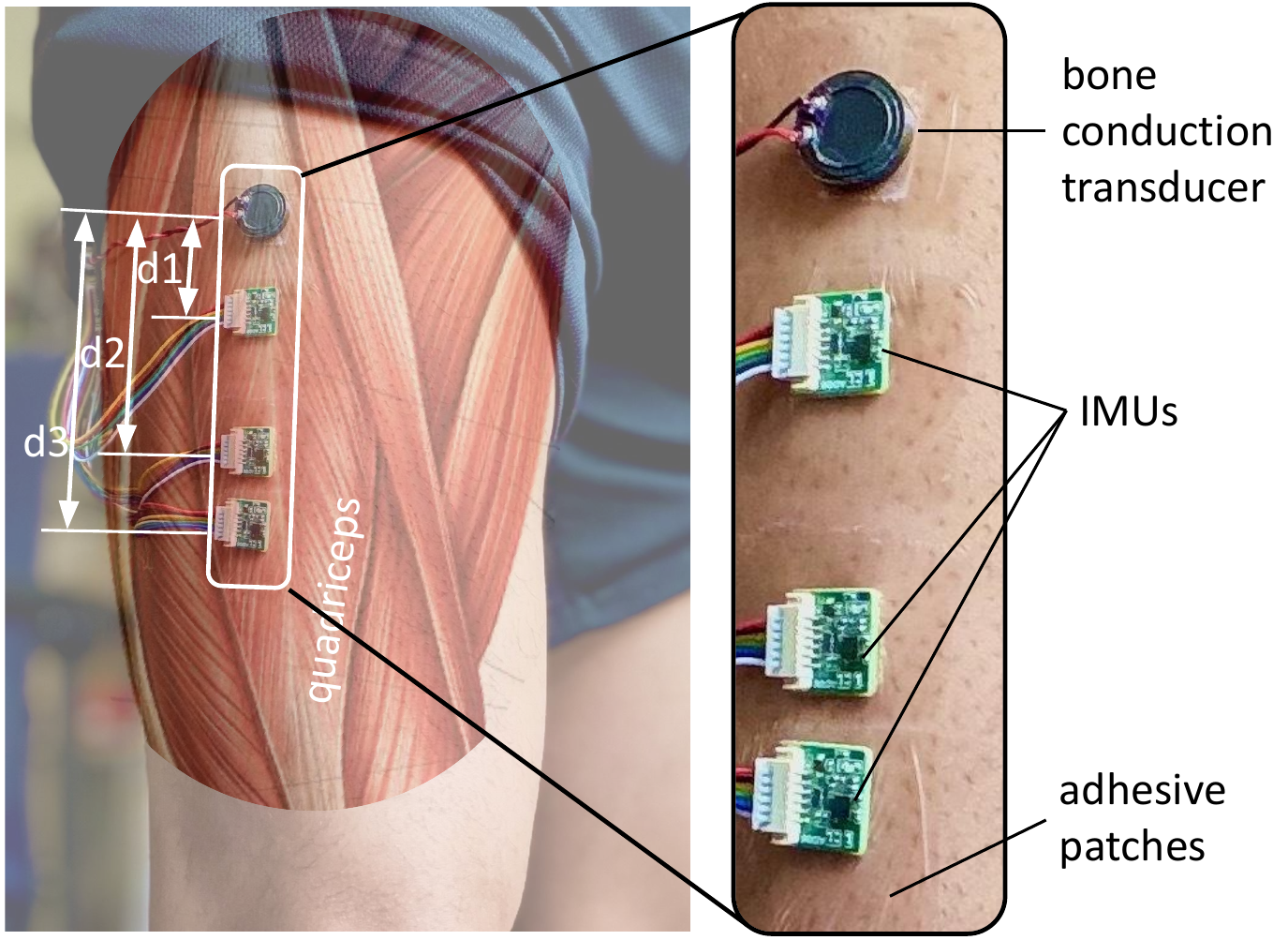}
    \caption{\name form factor and mounting mechanism. }
    \label{fig:mounting}
    \vspace{-3ex}
\end{figure}

\name adopts a sticker-based active vibroacoustic sensing design that consists of one bone-conduction transducer and multiple inertial measurement units (IMUs) to measure shear wave propagation speed on the target muscle. 
The transducer and the IMUs are placed in a linear array on the muscle with carefully derived distances between them, as shown in Figure \ref{fig:mounting}.

\subsubsection{Shear Wave-Guided Active Vibroacoustic Sensing Design} 
Since the sensing principle of \name is based on the propagation properties of mechanically induced shear waves, we first need to understand the shear wave generation and propagation in viscoelastic soft tissues.

\paragraph{Generation Model.}
Generating shear waves in soft tissue requires localized out-of-plane mechanical excitation \cite{Arda2013Jan}.
Biological soft tissues are often modeled as viscoelastic, nearly incompressible materials.
When localized out-of-plane mechanical excitation is applied to the surface, the resulting displacement field is governed by the Navier-Cauchy equation \cite{Liu2014Jan}:
\begin{equation}
    \rho \frac{\partial^2u}{\partial t ^2} = (\lambda+\mu)\nabla(\nabla.u)+\mu \nabla^2u
\end{equation}
where $u$ denotes the displacement vector field, $\rho$ is the density of material, $\lambda$ and $\mu$ are the Lame parameters describing compressional and shear behavior, respectively. 

\paragraph{Propagation Model.} 
Because biological soft tissues are nearly incompressible, $\lambda \gg \mu$, causing volumetric deformation to be strongly constrained while transverse shear propagation becomes the dominant measurable mechanical response \cite{Arda2013Jan}. 
Therefore, shear-wave propagation provides a more informative and measurable indicator for stiffness estimation in soft tissue \cite{Blank2022Mar}.
The displacement field generated by the localized out-of-plane mechanical excitation is often modeled using Helmholtz decomposition, which can be separated into compressional (P-wave) and shear (S-wave) components:
\begin{equation}
    u = \nabla \phi + \nabla \times \psi
\end{equation}
where $\nabla \phi$ represents longitudinal compressional waves and $\nabla \times \psi$ represents transverse shear waves.
This indicates that the P-wave particle motion is parallel to the propagation direction, whereas the S-wave particle motion is perpendicular to propagation \cite{Dankbaar1985Oct}.
The transverse displacement associated with S-wave propagation produces measurable surface-normal vibration that can be captured by skin-mounted IMU sensors.

\paragraph{Active Sensing Design}
The penetration depth and propagation behavior of the generated shear waves are strongly frequency dependent \cite{Wang2020Mar}.
Higher-frequency excitation produces shorter wavelengths and improved temporal localization but experiences stronger attenuation and shallower penetration depth.
In contrast, lower-frequency excitation propagates deeper into tissue with reduced attenuation, enabling mechanically induced waves to interact with deeper muscle structures \cite{Wang2020Mar}.
Therefore, the actuation frequency in \name must balance penetration depth, propagation stability, and temporal resolution for reliable wearable stiffness sensing.

We select actuation transducers that can effectively provide controllable mechanical excitation normal to the skin surface, making commonly available rotational vibromotors, such as eccentric rotating mass (ERM) actuators \cite{Xie2026Jul}, unsuitable for reliable shear-wave generation. 
Bone-conduction transducers, on the other hand, are designed to generate localized out-of-plane vibroacoustic excitation through direct mechanical coupling with solid media.
They use an internal vibrating mass driven by an electromagnetic actuator to generate controllable normal displacement at the contact surface \cite{Lee2020May}.
This enables efficient mechanical coupling of vibration energy into soft tissue and produces repeatable shear-wave propagation suitable for stiffness sensing.

To generate temporally localized and distinguishable propagation events, \name uses short-duration actuation pulses instead of continuous excitation.
Short pulses reduce temporal overlap between successive propagation responses and improve the separability of wavefront arrivals across sensing units, which is critical for reliable TDoA estimation.
However, unlike ultrasound-based SWE systems that generate near-impulsive excitation through acoustic radiation force, the pulse duration achievable by \name is constrained by the dynamics of the bone-conduction transducer and the mechanical loading imposed by the soft tissue interface.
Specifically, the actuator requires sufficient excitation duration to overcome inertial response and establish stable mechanical coupling before generating a measurable shear-wave event.
Therefore, the pulse duration in \name is determined by jointly considering the hardware response characteristics and the expected propagation delay range between sensing units.
We select the minimum duration sufficient to generate stable, repeatable excitation at the target frequency.

\begin{figure}
    \centering
    \begin{subfigure}[t]{0.43\linewidth}
        \centering
        \input{Figures/normalized_event}
        \vspace{-2ex}
        \caption{Raw signal pairs.}
        \label{fig:sensor1_eg}
    \end{subfigure}
    \hspace{2.5ex}
    \begin{subfigure}[t]{0.43\linewidth}
        \centering
        \input{Figures/normalized_event_filtered}
        \vspace{-2ex}
        \caption{Filtered signal.}
        \label{fig:sensor2_eg}
    \end{subfigure}
    \vspace{2ex}  
    \begin{subfigure}[t]{0.43\linewidth}
        \centering
        \input{Figures/cwt_peaks}
        \vspace{-2ex}
        \caption{CWT filtered signal.}
        \label{fig:cwt_eg}
    \end{subfigure}
    \hspace{2.5ex}
    \begin{subfigure}[t]{0.43\linewidth}
        \centering
        \input{Figures/cwt_peaks}
        \vspace{-2ex}
        \caption{Peak detection.}
        \label{fig:peak_detection}
    \end{subfigure}
    \vspace{-2ex}
    \caption{Signal examples. (a) and (b) depict the shear wave actuation signals captured by two sensors at distances of 3.5cm and 7cm from the actuator. (c) is the continuous wavelet transform (CWT) for the signal depicted in (a). (d) shows the peak detection on the signals in (c).}
    \label{fig:sig_process_eg}
    \vspace{-2ex}
\end{figure}
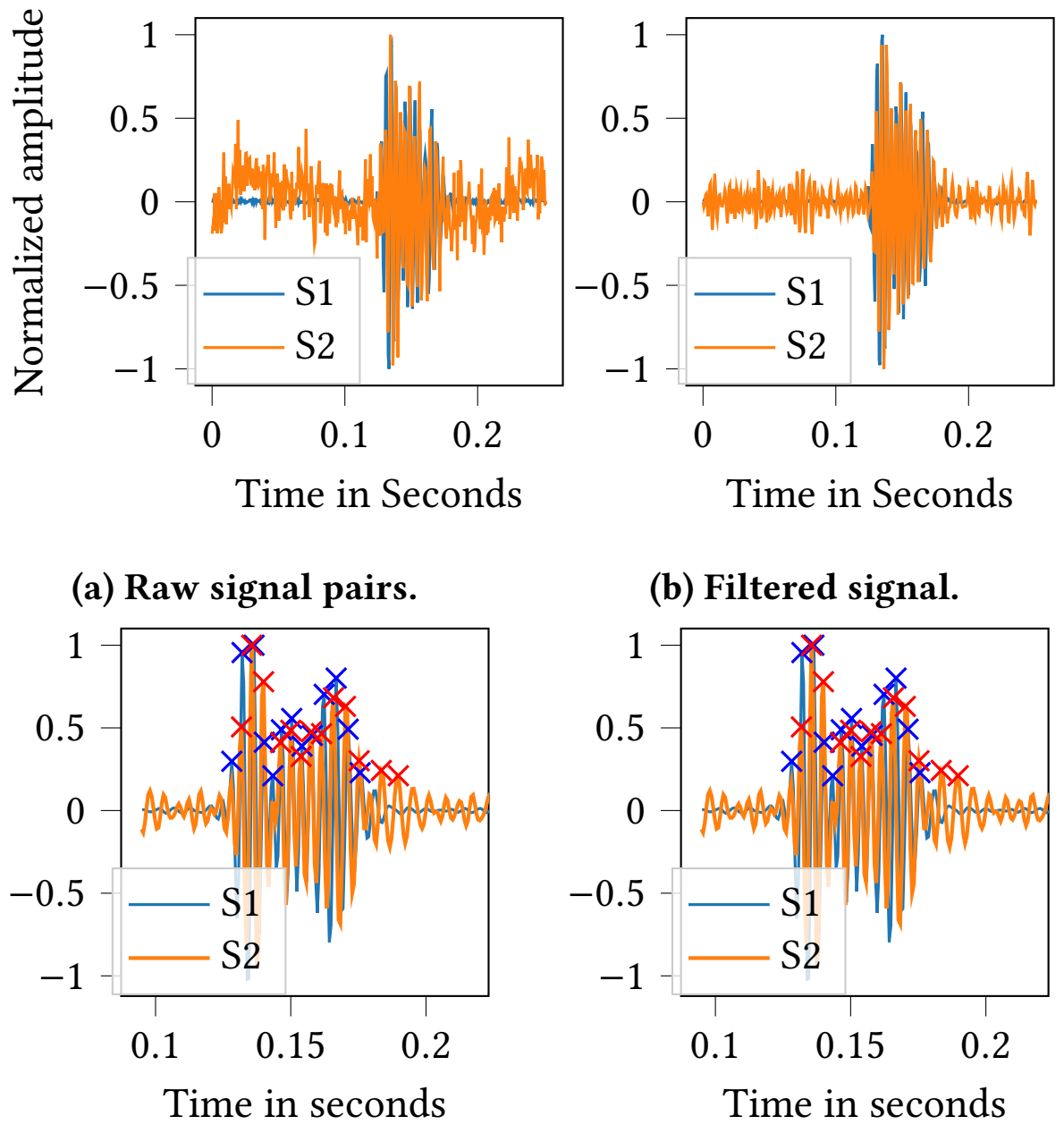

\subsubsection{Physiology- and Quality- Guided Sensor Deployment Strategy}\label{sec:sys_quality}
The hardware design must jointly account for the expected shear-wave velocity range in soft tissue, anatomically feasible sensor placement on target muscles, and the sampling-rate requirements to resolve small TDoA differences between closely spaced sensors.

The propagation velocity of shear waves in biological soft tissue is substantially lower than compressional waves and typically falls within a limited range -- 2 to 8 m/s -- determined by tissue stiffness and physiological conditions \cite{Blank2022Mar}.
Because \name estimates muscle stiffness through propagation delay measurements, the spatial separation between sensing units directly affects the measurable TDoA range.
Smaller sensor spacing would reduce propagation delay differences between sensors and increase the required sampling rate for accurate measurement.
Excessive spacing increases propagation attenuation and exposure to heterogeneous tissue dynamics, resulting in higher noise impact on the sensor signal.
Therefore, the hardware configuration -- sensor layout and sampling rate -- needs to balance TDoA resolution and sensing reliability under wearable deployment constraints.

On the other hand, sensor placement is also constrained by anatomical and physiological factors.
Because shear-wave propagation in muscle is anisotropic, propagation behavior depends strongly on the direction relative to muscle fibers. 
Muscle fibers form aligned mechanical structures, so along-fiber propagation encounters fewer structural boundaries and less scattering than cross-fiber propagation.
As a result, longitudinal propagation more consistently reflects the intrinsic mechanical behavior of the target muscle fibers.
To address these constraints, \name adopts a compact skin-mounted sensing geometry that places sensing units along physiologically meaningful propagation paths while minimizing rigid structural coupling between components.

To systematically quantify sensing quality and assess the effectiveness of the shear-wave propagation captured at a distance $d$, \name leverages hodogram analysis on the wavelet-filtered vibroacoustic signals.
A hodogram visualizes the multi-axis particle motion trajectory of the sensed vibration signal over time, enabling characterization of propagation coherence, directional consistency, and structural interference patterns \cite{Slack1993Oct}.
Because shear-wave propagation induces structured transverse motion, high-quality propagation signals typically produce concentrated and geometrically consistent hodogram trajectories, whereas noisy or distorted measurements exhibit scattered and irregular motion patterns.

Figure~\ref{fig:signal-quality-placement} illustrates three representative hodograms generated from signals captured at different sensing distances.
At the shortest propagation distance, as shown in ~\ref{fig:signal-quality-placement} (a), the received signals exhibit coherent directional motion with concentrated hodogram trajectories, indicating strong mechanical coupling and stable shear-wave propagation.
As the propagation distance increases, attenuation, dispersion, and heterogeneous tissue interactions progressively distort the propagation waveform, resulting in broader and less structured hodogram patterns.
At excessive sensing distances, the measured signals become dominated by noise and spurious propagation artifacts, producing highly scattered hodogram distributions with weak directional consistency.

We use Hodogram-based signal-quality assessment to identify reliable sensing configurations and to reject low-confidence propagation measurements.
The mechanically induced shear wave generates a dominant transverse displacement normal to the skin surface.
As a result, valid propagation events should exhibit motion geometry dominated by the z-axis, which can be quantified via Hodogram axis range ratio $r_H=\frac{z_{max}-z_{min}}{x_{max}-x_{min}}$.
Measurements with a z-to-x ratio below 1.25 are rejected because they typically indicate weak shear-wave coupling, excessive structural vibration, or spurious propagation artifacts.

\subsubsection{Spurious Propagation Path Mitigating Skin-Sensor Interface}
A key challenge in wearable vibroacoustic sensing is that the wearable device itself may introduce unintended mechanical propagation paths in addition to the target tissue propagation path.
Because mechanically induced waves can propagate through rigid structural components, loose skin coupling, and wearable substrates, the sensing signals may, in this case, contain structural vibrations unrelated to intrinsic tissue mechanics.
Such spurious propagation paths distort propagation-delay estimation and reduce the reliability of stiffness sensing.
On the other hand, many existing wearable sensing systems adopt garment-based form factors that prioritize flexibility and freedom of movement.
However, these loosely coupled interfaces introduce unstable skin contact and additional structural vibration paths that are undesirable in propagation-based muscle sensing, where stable, localized mechanical coupling is critical for reliable wave transmission and measurement.

To address this challenge, \name adopts a compact, adhesive, skin-mounted interface that minimizes the transmission of structural vibrations outside the target tissue path while maintaining stable mechanical coupling between the sensing components and the skin surface. Figure \ref{fig:mounting} depicts the skin-mounted interface on a person's thigh, where the IMU array is aligned with the muscle fiber direction and attached to the skin via the adhesive patches \cite{Liao2026Mar}. 
The adhesive patch material ensures stable mechanical coupling between the skin and the sensing components for efficient shear-wave generation and transmission, while mechanically isolating adjacent components to reduce vibration leakage and cross-component interference.

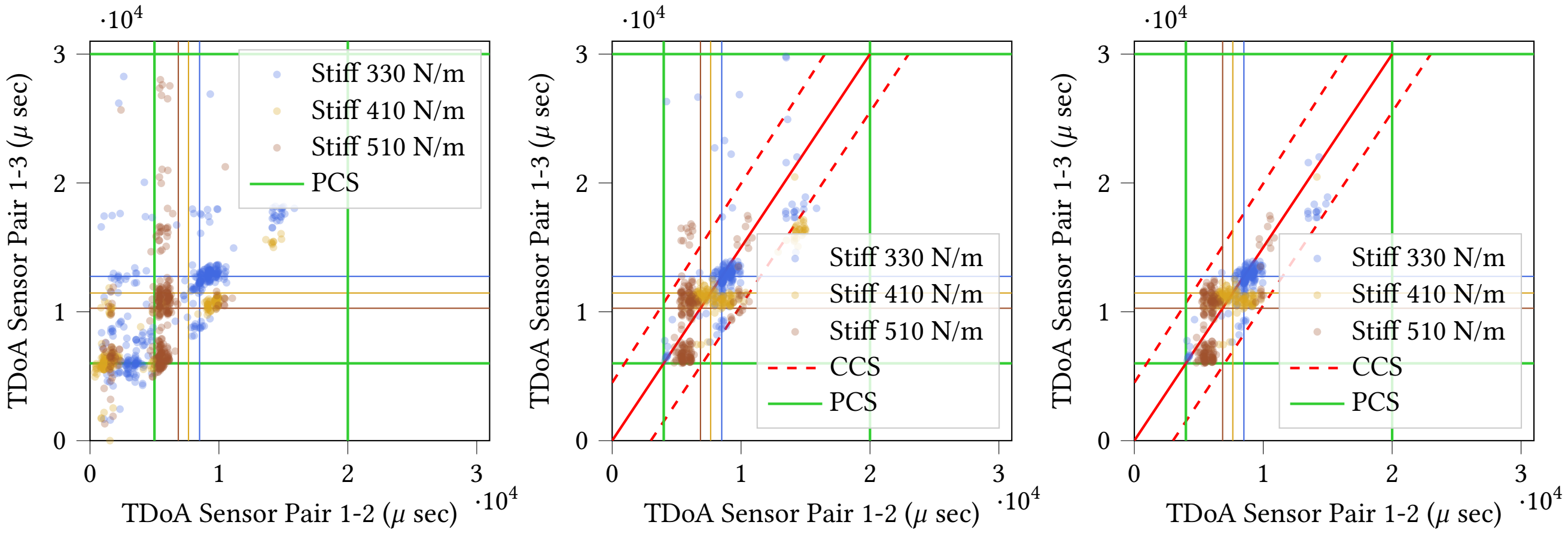
\begin{figure*}[t]
    \centering

    \begin{subfigure}[t]{0.32\linewidth}
        \centering
        \begin{minipage}[t][0.18\textheight][t]{\linewidth}
            \centering
            \input{Figures/raw_prediction}
        \end{minipage}
        \caption{Physiological Constraint Screening (PCS).}
        \label{fig:sensor1_eg}
    \end{subfigure}
    \begin{subfigure}[t]{0.32\linewidth}
        \centering
        \begin{minipage}[t][0.18\textheight][t]{\linewidth}
            \centering
            \input{Figures/corridor_prediction}
        \end{minipage}
        \caption{Cross-Sensor Consistency Screening (CCS).}
        \label{fig:sensor2_eg}
    \end{subfigure}
    \begin{subfigure}[t]{0.32\linewidth}
        \centering
        \begin{minipage}[t][0.18\textheight][t]{\linewidth}
            \centering
            \input{Figures/filtered_prediction}
        \end{minipage}
        \caption{Screened trials after PCS and CCS.}
        \label{fig:cwt_eg}
    \end{subfigure}
\vspace{-2ex}
    \caption{Quality-aware TDoA estimation screening.}
    \vspace{-2ex}
    \label{fig:estimation_quality}
\end{figure*}

\subsection{Shear-Wave Signal Processing for Time Difference of Arrival (TDoA) Estimation}\label{sec:velocity}
The raw signals captured by the IMU array are then processed to estimate the shear-wave velocity. 
However, directly estimating shear-wave velocity from wearable IMU measurements is challenging due to low SNR, propagation attenuation, waveform dispersion, posture-induced interference, and structural vibration artifacts introduced during on-body deployment.

To address these challenges, \name adopts a physics-informed signal processing pipeline that explicitly incorporates the propagation characteristics of shear waves in viscoelastic soft tissue.
Instead of relying on purely data-driven feature extraction, the system leverages propagation-aware filtering, wavelet-domain temporal analysis, and physiological signal quality constraints to robustly estimate Time Difference of Arrival (TDoA) between sensing units.
The resulting TDoA measurements are then used to derive shear-wave propagation velocity for downstream stiffness estimation.

\subsubsection{Reduce Posture-Induced Artifacts}
Although \name is designed to estimate resting muscle stiffness under passive muscle conditions, posture changes may still introduce motion artifacts and alter the propagation characteristics of the captured vibroacoustic signals.
In order to reduce such posture-induced motion artifacts, \name applies zero-phase filtering \cite{Wang2015Apr}. 
In this way, the temporal alignment of the propagation wavefront is preserved without phase distortion that leads to TDoA estimation errors.
Figure \ref{fig:sig_process_eg} depicts the example signals, where the raw signal and filtered signal are presented in (a) and (b), respectively.  

\subsubsection{Reduce Propagation Artifacts}
Although the actuation signal is designed to be narrowband, imperfections in actuator response and mechanical coupling introduce frequency leakage, causing different spectral components to propagate with different attenuation and phase characteristics in soft tissue.
To effectively isolate the target propagation band, \name applies Continuous Wavelet Transform (CWT)-based filtering with the corresponding scale.
CWT jointly preserves time and frequency information and has been proven effective for impulsive and dispersive vibroacoustic signal TDoA estimation \cite{mirshekari2018occupant}. 
\name adopts the first-order Gaussian Wavelet $\psi(t) = -t e^{-t^2}$, because it is commonly used for transient onset detection \cite{Plett2007May}.
The wavelet scale corresponding to the excitation frequency is extracted for further analysis.
Figure \ref{fig:sig_process_eg} demonstrates the CWT filtered signal.

\subsubsection{Peak-based TDoA Calculation}
One of the well-known challenges for TDoA estimation is the noisy onset detection. 
To acquire a smooth and stable characterization of the propagation model, we first apply the Hilbert Transform \cite{ulrich2006envelope} to extract the envelope of the extracted scale, denoted as $en(t)$.
An adaptive peak threshold based on the statistical distribution of the envelope signal -- mean $\mu_{en}$ and standard deviation $\delta_{en}$  -- is then derived as $th_{peak} = \mu_{en} + \delta_{en}$.
The first detected peak satisfying $p > th_{peak}$ is treated as the propagation onset.
This adaptive thresholding strategy improves robustness to signal-amplitude variation due to propagation and inter-subject physiological variability.

\subsection{Quality-Aware TDoA Screening and Refinement}\label{sec:signal-quality}
Although propagation-aware signal processing can estimate TDoA with mitigated propagation and physiological variability, there may still be uncertainty factors that affect the measurement.
To further refine the estimation for higher accuracy, we leverage the physical relationships among multiple TDoA measurements from different sensor pairs to quantify estimation quality.
The intuition is as illustrated in Figure \ref{fig:estimation_quality}, where the x-axis is the TDoA measured between sensor pair 0-1 and the y-axis is the TDoA measured between sensor pair 0-2. 
We can derive the relationship between $TDoA_{1-2}$ and $TDoA_{1-3}$ as below
\begin{equation}\label{eq:TDoA_ratio}
    \frac{d2-d1}{d3-d1} = \frac{TDoA_{1-2}}{TDoA_{1-3}}
\end{equation}
where $d1$, $d2$, $d3$ are distances between the actuator and sensors as depicted in Figure \ref{fig:mounting}.
Therefore, the ideal estimation values should fall on the line that satisfies this relationship, which is marked by the red line in Figure \ref{fig:estimation_quality} (a).
The blue, yellow, and brown solid lines in Figure \ref{fig:estimation_quality} (a) marks the TDoA calculated from the ground truth value of the muscle stiffness at 330 N/m, 410 N/m, and 510 N/m, respectively, where $v = \frac{\Delta d}{TDoA} = f_{s-v}^{-1}(v)$. 
The intersection of these lines and the TDoA ratio line in red indicates the ideal measurements under corresponding conditions.
However, due to various physical factors, such as sensor noise, tissue heterogeneity, propagation attenuation, and residual structural vibration artifacts, the TDoA estimation may be erroneous. 
\name conducts the estimation assessment based on the following criteria to determine whether the estimation is reliable or not.

\subsubsection{Physiological Constraint Screening (PCS): Individual TDoA Estimation Quality Assessment and Refinement}
First, the estimated propagation velocity must fall within the expected shear-wave velocity range in biological muscle tissues.
When the TDoA measurement-based velocity estimation falls outside of the physiologically plausible propagation velocity range, an erroneous peak detection is likely to occur.
Since active sensing via bone-conduction transducers uses short-duration actuation pulses (on the 1/10-second timescale), while resting muscle stiffness changes occur at a much coarser timescale (hours to weeks), the system can discard these erroneous measurements and acquire new ones.
Therefore, for individual TDoA measurement, we conduct the physiological-based filtering as shown in Figure \ref{fig:estimation_quality} (b).
The green lines are the TDoA ranges calculated using the known parameter $\Delta d$ and the range of shear-wave propagation speeds in soft tissues.
Individual estimates that fail this criterion are rejected because they typically indicate poor coupling, excessive structural vibration, or spurious propagation paths.

\subsubsection{Cross-Sensor Consistency Screening (CCS) and Refinement (CCR)}
Although the remaining measurements satisfy individual quality constraints, estimation uncertainty may still arise from waveform distortion and onset-detection errors.
Fortunately, \name employs multiple sensors, yielding multiple pairs of TDoA measurements.
Because all sensor pairs observe the same propagation event, their TDoA measurements must satisfy the geometric relationship described in Equation~\ref{eq:TDoA_ratio}.
As aforementioned, the red line in Figure~\ref{fig:estimation_quality} represents the ideal relationship between TDoA measurements from different sensor pairs, where the ratio of the measured TDoAs matches the ratio predicted by the physical relationship of the sensor placements.
To account for noise and propagation artifacts, we consider measurements whose TDoA ratios fall within a bounded region around the ideal ratio as reliable candidates for stiffness estimation.
The bounded region is marked by the red dashed lines in Figure~\ref{fig:estimation_quality}(b), and Figure~\ref{fig:estimation_quality}(c) shows the measurements that satisfy the proposed propagation-consistency criterion.
\name further projects the measurements onto the ideal ratio relationship, refining the estimates by enforcing physical consistency among TDoA measurements.

\subsubsection{Statistical Aggregation across Repeated Measurements}\label{sec:stat}
The propagation consistency analysis helps identify reliable measurements. 
However, there may still be measurement variability due to various factors, such as transient changes in coupling, viscoelastic damping, and measurement noise.
Because the time scale for muscle stiffness changes much more coarsely than that of individual sensing events, multiple measurements can be viewed as repeated observations of the same physiological state.
Therefore, \name aggregates repeated measurements to further reduce the uncertainty of the estimation.
\name employs a Gaussian Mixture Model (GMM) for clustering to aggregate multiple TDoA estimates by identifying the dominant cluster and its centroid as the aggregated TDoA value. 
This aggregated TDoA value is then converted into shear-wave velocity using the known sensor placement geometry.
This repeated-measurement aggregation effectively reduces estimation variance, thereby improving robustness to transient physiological and sensing artifacts.

%% file: Figures/pancake_1.tex
\begin{tikzpicture}[baseline=(current bounding box.south)]

\definecolor{darkgray176}{RGB}{176,176,176}
\definecolor{steelblue31119180}{RGB}{31,119,180}

\begin{axis}[
height = 4.9cm,
width = 4.9cm,
tick align=outside,
tick pos=left,
x grid style={darkgray176},
xlabel={Vibration Noise Amplitude},
xmajorgrids,
xmin=-1, xmax=1,
xtick style={color=black},
y grid style={darkgray176},
ylabel={Normalized Shear Wave Amplitude},
ymajorgrids,
ymin=-1, ymax=1,
ytick style={color=black}
]
\addplot [semithick, steelblue31119180]
table {%
0.00348986589644794 -0.0129516267804472
0.00506429874379487 0.00760813945667852
0.00408818413305382 0.0492820195423774
-0.00371530427256498 0.0990992884357996
-0.0199806607115646 0.120954904290336
-0.037144565957827 0.0724511007782693
-0.0385439401706752 -0.0584730247767871
-0.0106984707579607 -0.22234808654905
0.0406932287977884 -0.315305833693961
0.0876699175478804 -0.237471596560236
0.0962849203501921 0.0244039518880932
0.050585798518653 0.350849779546338
-0.0337405051583456 0.534942002948288
-0.115150434305076 0.413951675435958
-0.147909979139165 0.00373330111657765
-0.108007022828556 -0.473505627277056
-0.0090975490073762 -0.711120304462589
0.101465462810337 -0.520528323992394
0.165863040412873 0.0222175781375434
0.147062362404149 0.595601548147236
0.051748887151662 0.825060456534687
-0.0697732734754337 0.5353893608187
-0.150561710098984 -0.111921677541732
-0.146317168116746 -0.708533935985833
-0.0614884503833933 -0.865055856407186
0.0537583264946498 -0.469425355707998
0.131991562187995 0.235084720897048
0.126856748646112 0.798708518283025
0.0420969413673063 0.849755835744252
-0.0679451632018395 0.342259670236475
-0.132171987999059 -0.399624913553863
-0.109728486910621 -0.887910530893933
-0.0157152293938689 -0.795720256697901
0.0904294160862879 -0.175386631583008
0.142045873496988 0.575299032813176
0.105159898081553 0.968791188715984
0.000426131963647201 0.74204155744699
-0.107801642250932 0.0327037134209552
-0.152363031687578 -0.704992996782125
-0.106421835833108 -1
0.00218544815359245 -0.674337720843243
0.110978421835506 0.0520802101837045
0.15893628703716 0.719093424468033
0.118595703194543 0.924451082682236
0.0112550873296795 0.569749026673809
-0.103154609705771 -0.0925470374156894
-0.159767968587436 -0.636778347124524
-0.12735483442752 -0.74654585332466
-0.0268280555225029 -0.405919352476958
0.0820864983292525 0.12418987244088
0.137334410009648 0.502839650390538
0.111925906843556 0.526903647642002
0.0288540306818914 0.234036302279944
-0.0566259545401895 -0.152154793471958
-0.0948556759138594 -0.382387481043196
-0.0722096170327269 -0.339067812460142
-0.0137818461990921 -0.0916937944612324
0.0391033163941222 0.173807289719963
0.05689029972687 0.288756932322347
0.0366680540069409 0.203670928895317
};
\end{axis}

\end{tikzpicture}

%% file: Figures/pancake_2.tex
\begin{tikzpicture}[baseline=(current bounding box.south)]

\definecolor{darkgray176}{RGB}{176,176,176}
\definecolor{steelblue31119180}{RGB}{31,119,180}

\begin{axis}[
height = 4.9cm,
width = 4.9cm,
tick align=outside,
tick pos=left,
x grid style={darkgray176},
xlabel={Vibration Noise Amplitude},
xmajorgrids,
xmin=-1, xmax=1,
xtick style={color=black},
y grid style={darkgray176},
ylabel={\phantom{Normalized Shear Wave Amplitude}},
ymajorgrids,
ymin=-1., ymax=1.,
ytick style={color=black}
]
\addplot [semithick, steelblue31119180]
table {%
0.0142830170100363 0.270051945547489
0.238686486920946 0.595640668444086
0.231214822373306 0.620824585725537
0.00953946584224424 0.28960200808877
-0.25957904925796 -0.246918046893263
-0.388543140198362 -0.695114729426698
-0.282956013897368 -0.784356932075512
0.00910692987368968 -0.431611876839402
0.324136785011759 0.194307260603169
0.474238464275476 0.747513477999219
0.360594689689147 0.891770819872658
0.0419842828992043 0.51037023813609
-0.294225258824028 -0.192908787388579
-0.447131466387279 -0.795338122063798
-0.327985548862826 -0.916281503220209
-0.0145530096560529 -0.466744698937988
0.296992400071458 0.282588347378427
0.41149044665689 0.862101984871728
0.255697446112028 0.896591070174955
-0.0730103002132283 0.351379166032657
-0.365708175682233 -0.432893060555583
-0.434523234486933 -0.954227079013187
-0.231206837725489 -0.87589058387065
0.12491944404198 -0.244410830208626
0.421622967045039 0.544144446628948
0.480426019453237 1
0.261520000411678 0.849473764439501
-0.111517349635267 0.202080036709021
-0.423098417070821 -0.531152428699289
-0.49335215934959 -0.910597985488167
-0.285065376238024 -0.744082781211359
0.0756713682183284 -0.184194486019761
0.376307909760326 0.399004377513258
0.445246892899776 0.665951182811714
0.254437230463412 0.511590642493381
-0.0655087883706915 0.10380123818728
-0.310243353173702 -0.257300707670729
-0.334769968909783 -0.355450698648234
-0.146640882235862 -0.193312605428615
0.105297002047491 0.0459356350565359
0.240325106811429 0.16379651728615
0.176705587046826 0.0967319548157925
-0.0129244584825073 -0.0606851906819226
-0.165966356041801 -0.156601284705192
-0.163075983810447 -0.109596533656035
-0.0194819679612487 0.0371610838848458
0.139103678259464 0.164194111815492
0.186313360223191 0.17537047570186
0.0921284009733344 0.0678501245536972
-0.0663580756952761 -0.0756082033336985
-0.173731500268252 -0.156172181474418
-0.16148832546509 -0.130678132947683
-0.0487323924717066 -0.0326630646104073
0.0815430585138988 0.0642167175484777
0.145555405948081 0.103101565733799
0.113601861823606 0.0763127196071549
0.0206589658812769 0.0167692068251661
-0.0672667140421621 -0.0340298111540471
-0.0998496510342468 -0.0527391934690778
-0.068951785705102 -0.040467312355284
};
\end{axis}

\end{tikzpicture}

%% file: Figures/pancake_3.tex
\begin{tikzpicture}[baseline=(current bounding box.south)]

\definecolor{darkgray176}{RGB}{176,176,176}
\definecolor{steelblue31119180}{RGB}{31,119,180}

\begin{axis}[
height = 4.9cm,
width = 4.9cm,
tick align=outside,
tick pos=left,
x grid style={darkgray176},
xlabel={Vibration Noise Amplitude},
xmajorgrids,
xmin=-1, xmax=1,
xtick style={color=black},
y grid style={darkgray176},
ylabel={\phantom{Normalized Shear Wave Amplitude}},
ymajorgrids,
ymin=-1., ymax=1.,
ytick style={color=black}
]
\addplot [semithick, steelblue31119180]
table {%
0.461799350099339 0.354908179189779
0.343365709357216 0.809844516230596
-0.137413184073425 0.89151229411531
-0.633523972931853 0.462810395325067
-0.759407094610609 -0.267254253797976
-0.369709785055173 -0.874953649912064
0.322329249358705 -1
0.877693955214631 -0.572946194586392
0.924986480531073 0.14100708536367
0.411624992642066 0.706508767611454
-0.35389609333638 0.793388245619733
-0.893880971828262 0.380529428544204
-0.865464960110605 -0.246934265218027
-0.285662139893578 -0.686777854718666
0.469379470194582 -0.661081616897471
0.902164888660959 -0.17715830741011
0.724539572355111 0.469789318717307
0.0575512577653946 0.86498224454867
-0.650793396993572 0.733150416154687
-0.937105693508112 0.127221562504948
-0.63110421353391 -0.589609855997705
0.058081557312066 -0.968758894462365
0.706271573238748 -0.767060278593614
0.94246169240258 -0.112395466135239
0.650550665926957 0.576490670209202
0.0143992712510449 0.872382470898917
-0.599435754436363 0.624070428325433
-0.856747761477176 0.0343631599931039
-0.64634379485748 -0.508858373914575
-0.128687228138452 -0.697200499587481
0.382485935583361 -0.469338557949046
0.617708267590946 -0.00268386579542424
0.498673805863175 0.420013989124801
0.153369660978329 0.573105525048126
-0.190223338787569 0.402080325952839
-0.349678252060102 0.0394483233420987
-0.280375333050663 -0.28835293052222
-0.07201621838416 -0.407312327397294
0.127442698245783 -0.29341638168456
0.203730886180196 -0.0584039064292169
0.134272246892225 0.143849253443709
-0.0117954347333221 0.208925396693947
-0.127048503671476 0.12458838898066
-0.139895797389047 -0.0337633254296502
-0.059949269398503 -0.146929837174656
0.0385455745005005 -0.127365809948182
0.0800360505260076 0.0142322068982498
0.0437808768887917 0.172142365014764
-0.0293457124116712 0.228424712729026
-0.0742376751135848 0.144721174862985
-0.0483792808639649 -0.0175814804871676
0.0398183254515129 -0.155684273638975
0.133440307257716 -0.200728602663502
0.162806621877163 -0.15163026280625
0.0973550974963049 -0.0568954181997771
-0.0271239526357636 0.0282536831234782
-0.135046735201061 0.0745221432722507
-0.169705394287279 0.0825296960740851
-0.129639642256491 0.066541051490823
-0.0550513705287642 0.0400918695199206
};
\end{axis}

\end{tikzpicture}

%% file: Figures/normalized_event.tex
\begin{tikzpicture}

\definecolor{darkgrey176}{RGB}{176,176,176}
\definecolor{darkorange25512714}{RGB}{255,127,14}
\definecolor{lightgrey204}{RGB}{204,204,204}
\definecolor{steelblue31119180}{RGB}{31,119,180}

\begin{axis}[
height=4.3cm,
width=4.3cm,
legend cell align={left},
legend style={fill opacity=0.8, draw opacity=1, text opacity=1, at={(0.45,0.35)},  draw=lightgrey204},
tick align=outside,
tick pos=left,
x grid style={darkgrey176},
xlabel={Time in Seconds},
xmin=-0.01258015, xmax=0.26418315,
xtick style={color=black},
y grid style={darkgrey176},
ylabel={Normalized amplitude},
ymin=-1.1, ymax=1.1,
ytick style={color=black}
]
\addplot [semithick, steelblue31119180]
table {%
0.000335 -0.00540051590204901
0.000968 0.0141842615840496
0.001601 0.00112774325998387
0.002234 -0.00117634820896891
0.002867 -0.00565652606526599
0.0035 -0.0105207191663885
0.004133 -0.00373644984113867
0.004766 -0.00680857179974238
0.005399 -0.0188410494709402
0.006032 0.00112774325998387
0.006665 -0.0124407953905158
0.007298 -0.00514450573883204
0.007931 0.00253579915767724
0.008563 -0.00732059212617633
0.009196 -0.00360844475953018
0.009829 -0.0065525616365254
0.010462 -0.0162809478387705
0.011095 -0.0023283939434453
0.011728 -0.0101367039215631
0.012361 -0.0033524345963132
0.012994 -0.0141048614514262
0.013627 0.00778400750362524
0.01426 -0.0129528157169498
0.014893 -0.00424847016757262
0.015526 -0.000792332964143447
0.016159 0.00701597701397432
0.016792 -0.00386445492274716
0.017425 -0.00181637361701135
0.018058 0.00304781948411119
0.01869 0.0136722412576157
0.019323 -0.00578453114687448
0.019956 0.00547991603467246
0.020589 -0.00271240918827077
0.021222 0.0140562565024411
0.021855 0.000487717851941432
0.022488 0.0143122666656581
0.023121 0.00343183472893665
0.023754 0.00701597701397432
0.024387 -0.00386445492274716
0.02502 0.00432787030019607
0.025653 -0.0065525616365254
0.026286 -0.00322442951470472
0.026919 0.000871733096766896
0.027552 0.00138375342320085
0.028185 0.00996009389096954
0.028817 0.00778400750362524
0.02945 0.0109841345438374
0.030083 -0.00642455655491691
0.030716 0.00855203799327617
0.031349 -0.0087286480238697
0.031982 0.0146962819104836
0.032615 0.015336307318526
0.033248 -0.00604054131009145
0.033881 0.00996009389096954
0.034514 0.00663196176914885
0.035147 0.0029198144025027
0.03578 0.0071439820955828
0.036413 -0.00348043967792169
0.037045 0.010344109135795
0.037678 0.00599193636110641
0.038311 -0.000536322800926471
0.038944 0.00496789570823851
0.039577 0.0117521650334884
0.04021 -0.00360844475953018
0.040843 -0.00834463277904423
0.041476 -0.00949667851352062
0.042109 -0.00616854639169994
0.042742 0.0175123937058703
0.043375 -0.00604054131009145
0.044008 0.00112774325998387
0.044641 0.0212245410725165
0.045273 -0.00424847016757262
0.045906 0.00483989062663002
0.046539 -0.00168836853540286
0.047172 0.0168723682978279
0.047805 -0.01039271408478
0.048438 0.00176776866802631
0.049071 0.00279180932089421
0.049704 -0.000920338045751935
0.050337 -2.43024744925192e-05
0.05097 0.0130322158495732
0.051603 -0.00117634820896891
0.052236 -0.0151289021042941
0.052869 0.0125201955231393
0.053501 0.00381584997376212
0.054134 -0.0151289021042941
0.054767 0.00407186013697909
0.0554 0.00317582456571968
0.056033 0.00304781948411119
0.056666 0.000999738178375384
0.057299 -0.00386445492274716
0.057932 0.000487717851941432
0.058565 0.00983208880936105
0.059198 -0.00348043967792169
0.059831 -0.00117634820896891
0.060464 -0.0045044803307896
0.061097 0.00752799734040827
0.061729 0.00752799734040827
0.062362 0.00547991603467246
0.062995 -0.00898465818708667
0.063628 -0.00360844475953018
0.064261 0.00739999225879978
0.064894 -0.0087286480238697
0.065527 0.0123921904415308
0.06616 -0.0065525616365254
0.066793 -0.00399246000435564
0.067426 -0.00220038886183681
0.068059 -0.00360844475953018
0.068692 0.00483989062663002
0.069325 0.00624794652432339
0.069957 -0.00834463277904423
0.07059 -0.0097526886767376
0.071223 -0.00642455655491691
0.071856 -2.43024744925192e-05
0.072489 -0.00501650065722355
0.073122 -0.000792332964143447
0.073755 -0.00168836853540286
0.074388 0.0111121396254459
0.075021 0.00227978899446026
0.075654 -0.00719258704456784
0.076287 -0.0124407953905158
0.07692 0.00407186013697909
0.077553 -0.00604054131009145
0.078186 -0.000280312637709495
0.078811 0.00432787030019607
0.079436 0.0102161040541865
0.080061 0.0136722412576157
0.080686 0.00624794652432339
0.081311 0.00176776866802631
0.081936 -0.00220038886183681
0.082561 -0.00604054131009145
0.083186 0.00675996685075734
0.083811 -0.00591253622848296
0.084436 0.010088098972578
0.085061 0.0120081751967053
0.085686 -0.0120567801456904
0.086311 0.0018957737496348
0.086936 0.00355983981054514
0.087561 0.00880804815649315
0.088186 0.0113681497886629
0.088811 -0.00412046508596413
0.089436 -0.00591253622848296
0.090061 0.00176776866802631
0.090687 0.000999738178375384
0.091312 0.00535191095306397
0.091937 -0.0013043532905774
0.092562 0.0114961548702714
0.093187 -0.0097526886767376
0.093812 -0.00463248541239808
0.094437 0.00868004307488466
0.095062 -0.00591253622848296
0.095687 -0.0110327394928225
0.096312 -0.0013043532905774
0.096937 -0.0148728919410771
0.097562 -0.00412046508596413
0.098187 0.000999738178375384
0.098812 0.0113681497886629
0.099437 -0.00898465818708667
0.100062 0.00304781948411119
0.100687 0.0172563835426533
0.101312 -2.43024744925192e-05
0.101937 -0.00732059212617633
0.102562 -0.015896932593945
0.103188 -0.00757660228939331
0.103813 0.00368784489215363
0.104438 -0.00399246000435564
0.105063 -0.00143235837218589
0.105688 0.000359712770332945
0.106313 0.00804001766684222
0.106938 -0.0107767293296055
0.107563 0.00215178391285178
0.108188 -0.00796061753421877
0.108813 -0.00348043967792169
0.109438 0.00432787030019607
0.110063 -0.00770460737100179
0.110688 -0.00680857179974238
0.111313 0.00407186013697909
0.111938 -0.000408317719317983
0.112563 -0.00156036345379437
0.113188 -0.00770460737100179
0.113813 -0.00514450573883204
0.114438 0.00624794652432339
0.115063 0.00765600242201676
0.115689 0.00355983981054514
0.116314 -0.0045044803307896
0.116939 -0.0150008970226856
0.117564 -0.00168836853540286
0.118189 -0.00117634820896891
0.118814 -0.0132088258801668
0.119439 -2.43024744925192e-05
0.120064 -0.00309642443309623
0.120689 0.00599193636110641
0.121314 -0.00796061753421877
0.121939 -0.0107767293296055
0.122564 -0.00693657688135087
0.123189 -0.00412046508596413
0.123814 0.00151175850480934
0.124439 0.0619301570240156
0.125064 0.00816802274845071
0.125689 -0.151454314017334
0.126314 -0.162078735790838
0.126939 0.0925233715284442
0.127564 0.360053992090184
0.12819 0.166894323942976
0.128822 -0.366502851119593
0.129455 -0.539821731617486
0.130088 0.0117521650334884
0.130721 0.747781384282294
0.131354 0.762885983912095
0.131987 -0.0276734001019259
0.13262 -0.920636849402737
0.133252 -1
0.133885 -0.0615947467281752
0.134518 0.918028142821583
0.135151 0.935436833920337
0.135784 0.0829229904078076
0.136417 -0.782263356183962
0.13705 -0.834233419317008
0.137682 -0.0335616338559164
0.138315 0.693635234761903
0.138948 0.60249561665666
0.139581 -0.0756753057051089
0.140214 -0.593327855729834
0.140847 -0.423337107353762
0.14148 0.217968351504762
0.142112 0.501115592022738
0.142745 0.156013892006254
0.143378 -0.359718581794344
0.144011 -0.470826992630511
0.144644 0.03108093235637
0.145277 0.598911474371622
0.14591 0.502267637757214
0.146542 -0.116636931819825
0.147175 -0.629809303988253
0.147808 -0.484651541444228
0.148441 0.230384844420786
0.149074 0.685314904457352
0.149707 0.411768045060013
0.15034 -0.255266435201817
0.150972 -0.640177715598541
0.151605 -0.330533423187608
0.152238 0.387063064309575
0.152871 0.605951753860089
0.153504 0.171118491636056
0.154137 -0.447658072859375
0.15477 -0.603440257176905
0.155403 -0.0557065129741848
0.156035 0.487803063535455
0.156668 0.418296304222046
0.157301 -0.016408952920379
0.157934 -0.337573702676075
0.158567 -0.301220259499265
0.1592 -0.00885665310547818
0.159833 0.234993027358691
0.160465 0.2833789482067
0.161098 0.236401083256385
0.161731 0.037097171191969
0.162364 -0.302884325560175
0.162997 -0.550062138146165
0.16363 -0.457002443816794
0.164263 -0.016152942757162
0.164895 0.450169569542559
0.165528 0.55423770089026
0.166161 0.222960549687493
0.166794 -0.218784986943398
0.167427 -0.406824451826267
0.16806 -0.251298277671954
0.168693 0.0853550869583689
0.169325 0.352245682112066
0.169958 0.31128405599735
0.170591 0.0507937149240772
0.171224 -0.208416575333111
0.171857 -0.278051339728128
0.17249 -0.167582954300003
0.173123 0.0272407799081154
0.173755 0.152685759884434
0.174388 0.149741643007438
0.175021 0.0625701824320581
0.175654 -0.0340736541823503
0.176287 -0.085147681744137
0.17692 -0.104860464311844
0.177553 -0.0722191685016797
0.178186 -0.00527251082044052
0.17881 0.036841161028752
0.179435 0.047337577720648
0.18006 0.0456735116597377
0.180685 -0.0128248106353413
0.18131 -0.0449540861190718
0.181935 -0.0655629042580383
0.18256 -0.0290814559996193
0.183185 -0.00488849557561506
0.18381 0.00483989062663002
0.184435 0.0132882260127902
0.18506 -0.000664327882534959
0.185685 -0.00412046508596413
0.18631 0.00202377883124329
0.186935 -0.000920338045751935
0.18756 0.00586393127949792
0.188185 0.00317582456571968
0.18881 -0.000280312637709495
0.189435 -0.00757660228939331
0.19006 -0.015896932593945
0.190685 -0.00437647524918111
0.19131 0.01584832764496
0.191935 0.00688797193236583
0.19256 0.0145682768288751
0.193185 0.00343183472893665
0.19381 0.0123921904415308
0.194435 0.0118801701150969
0.19506 0.00675996685075734
0.195685 -0.00322442951470472
0.19631 0.00752799734040827
0.196935 -0.00578453114687448
0.19756 0.00445587538180456
0.198185 -0.00719258704456784
0.19881 0.00739999225879978
0.199435 -0.00988069375834609
0.20006 0.00407186013697909
0.200685 -0.00642455655491691
0.20131 -0.00565652606526599
0.201935 0.00586393127949792
0.20256 -0.00309642443309623
0.203185 0.00778400750362524
0.20381 -0.00821662769743575
0.204435 -0.00834463277904423
0.20506 0.00215178391285178
0.205685 -0.0115447598192564
0.20631 -0.00309642443309623
0.206935 -0.0107767293296055
0.20756 -0.00501650065722355
0.208185 0.00637595160593188
0.20881 0.00151175850480934
0.209435 -0.00706458196295935
0.21006 -0.0121847852272989
0.210685 -0.00616854639169994
0.21131 0.00458388046341305
0.211935 -0.00706458196295935
0.21256 0.00355983981054514
0.213185 -0.000152307556101007
0.21381 0.0140562565024411
0.214435 -0.0132088258801668
0.21506 -0.0118007699824734
0.215685 0.00151175850480934
0.21631 -0.00834463277904423
0.216935 -0.00744859720778482
0.21756 0.00368784489215363
0.218185 -0.00770460737100179
0.21881 -0.00501650065722355
0.219435 -0.00680857179974238
0.22006 0.00394385505537061
0.220685 -0.0130808207985583
0.22131 -0.015896932593945
0.221935 0.00663196176914885
0.22256 -0.0013043532905774
0.223185 0.0191764597667807
0.22381 -0.00706458196295935
0.224435 -0.00565652606526599
0.22506 0.000103702607115969
0.225685 0.00151175850480934
0.22631 -0.00412046508596413
0.226935 -0.0165369580019875
0.22756 0.0102161040541865
0.228185 0.0132882260127902
0.228817 0.00701597701397432
0.22945 -0.0033524345963132
0.230083 0.00586393127949792
0.230716 0.00727198717719129
0.231349 0.0166163581346109
0.231982 0.0104721142174035
0.232615 0.00547991603467246
0.233248 -0.00565652606526599
0.233881 -0.00156036345379437
0.234514 -0.00463248541239808
0.235147 0.00215178391285178
0.23578 0.000231707688724457
0.236413 0.0118801701150969
0.237045 0.00407186013697909
0.237678 0.00381584997376212
0.238311 0.0122641853599223
0.238944 0.00125574834159236
0.239577 -0.00936867343191214
0.24021 0.00560792111628095
0.240843 0.00381584997376212
0.241476 -0.00821662769743575
0.242109 0.0209685309092995
0.242742 0.0102161040541865
0.243375 0.010600119299012
0.244008 0.00215178391285178
0.244641 0.00061572293354992
0.245273 0.0162323428897854
0.245906 -0.00220038886183681
0.246539 0.00586393127949792
0.247172 0.00919206340131861
0.247805 0.00599193636110641
0.248438 0.0071439820955828
0.249071 0.0129042107679648
0.249704 0.00816802274845071
0.250337 -0.000664327882534959
0.25097 -0.00629655147330843
0.251603 -0.0151289021042941
};
\addlegendentry{S1}
\addplot [semithick, darkorange25512714]
table {%
0 -0.191182169930715
0.000631 -0.143878906864874
0.001263 -0.173980983361318
0.001894 0.0496344420408375
0.002526 -0.163230241755446
0.003157 0.0496344420408375
0.003788 -0.18903202160954
0.00442 -0.158929945113096
0.005051 0.0883371118219799
0.005682 -0.00196911766735233
0.006314 -0.0987257921202082
0.006945 -0.18903202160954
0.007577 -0.148179203507223
0.008208 0.0818866668584561
0.008839 0.294751350654739
0.009471 -0.122377423653129
0.010102 0.150691413136043
0.010734 -0.135278313580176
0.011365 -0.00841956263087605
0.011996 -0.0127198592732252
0.012628 0.107688446712551
0.013259 -0.0428219357696693
0.01389 0.00233117897499683
0.014522 -0.0299210458426218
0.015153 0.0689857769314087
0.015785 0.182943637953661
0.016416 0.161442154741915
0.017047 0.133490226566646
0.017679 0.346354910362929
0.01831 -0.0428219357696693
0.018941 0.129189929924297
0.019573 0.488264699560451
0.020204 0.0840368151796307
0.020836 0.309802388902961
0.021467 0.0431839970773138
0.022098 0.16359230306309
0.02273 0.217346011092454
0.023361 0.146391116493693
0.023992 0.062535331967885
0.024624 0.103388150070202
0.025255 0.29045105401239
0.025887 0.292601202333565
0.026518 0.0947875567855036
0.027149 0.180793489632487
0.027781 0.256048680873597
0.028412 0.26249912583712
0.029044 0.137790523208995
0.029675 0.0990878534278527
0.030306 0.197994676201883
0.030938 0.0861869635008053
0.031569 0.172192896347788
0.0322 0.0689857769314087
0.032832 0.228096752698327
0.033463 0.260348977515946
0.034095 0.0818866668584561
0.034726 0.210895566128931
0.035357 0.101238001749027
0.035989 0.114138891676075
0.03662 0.180793489632487
0.037251 0.101238001749027
0.037883 -0.0449720840908438
0.038514 0.221646307734804
0.039146 0.062535331967885
0.039777 -0.227734691390683
0.040408 0.0603851836467104
0.04104 0.200144824523058
0.041671 0.217346011092454
0.042303 0.281850460727692
0.042935 0.182943637953661
0.043568 -0.0041192659885269
0.044201 0.0431839970773138
0.044834 0.170042748026614
0.045467 0.0883371118219799
0.0461 0.124889633281948
0.046733 0.294751350654739
0.047366 0.0496344420408375
0.047999 -0.156779796791922
0.048632 0.0560848870043612
0.049265 0.167892599705439
0.049898 0.0388837004349646
0.050531 -0.11377683036843
0.051164 -0.0041192659885269
0.051797 0.114138891676075
0.05243 0.00878162393852055
0.053063 0.124889633281948
0.053696 -0.0514225290543676
0.054329 0.223796456055978
0.054962 -0.0277708975214472
0.055595 0.193694379559534
0.056228 0.092637408464329
0.056861 0.124889633281948
0.057494 0.101238001749027
0.058127 0.0990878534278527
0.05876 -0.0664735673025896
0.059393 -0.105176237083732
0.060026 0.178643341311312
0.060659 -0.0557228256967167
0.061292 0.0754362218949324
0.061925 0.00018103065382225
0.062558 -0.0041192659885269
0.063191 0.0324332554714409
0.063824 0.114138891676075
0.064457 0.077586370216107
0.06509 0.327003575472358
0.065723 0.161442154741915
0.066356 0.0646854802890595
0.066989 0.0904872601431544
0.067621 0.174343044668963
0.068254 0.142090819851344
0.068887 0.0646854802890595
0.06952 -0.107326385404907
0.070153 0.092637408464329
0.070786 0.434510991531087
0.071419 0.0711359252525832
0.072052 0.107688446712551
0.072685 -0.0213204525579235
0.073318 -0.0922753471566845
0.073951 0.0195323655443934
0.074584 0.213045714450105
0.075217 0.1119887433549
0.07585 -0.00626941430970148
0.076483 -0.154629648470747
0.077116 -0.255686619565952
0.077749 -0.227734691390683
0.078382 0.101238001749027
0.079015 0.0474842937196629
0.079648 0.101238001749027
0.080281 0.206595269486582
0.080914 0.245297939267724
0.081547 -0.126677720295478
0.08218 -0.0686237156237642
0.082813 0.021682513865568
0.083446 0.1119887433549
0.084079 0.0431839970773138
0.084712 0.0969377051066782
0.085345 -0.184731724967191
0.085978 0.109838595033726
0.086611 -0.201932911536588
0.087244 0.0969377051066782
0.087877 -0.122377423653129
0.08851 -0.122377423653129
0.089143 0.109838595033726
0.089776 -0.0127198592732252
0.090409 0.101238001749027
0.091042 -0.00196911766735233
0.091675 -0.049272380733193
0.092308 -0.0772243089084625
0.092934 0.0302831071502663
0.093561 -0.152479500149573
0.094188 -0.0879750505143354
0.094814 -0.0987257921202082
0.095441 0.0861869635008053
0.096068 0.18724393459601
0.096695 -0.0449720840908438
0.097321 0.0453341453984883
0.097948 0.0539347386831866
0.098575 -0.0535726773755421
0.099201 -0.126677720295478
0.099828 0.0044813272961714
0.100455 -0.19333231825189
0.101082 -0.00841956263087605
0.101708 -0.0105697109520506
0.102335 -0.109476533726081
0.102962 -0.0535726773755421
0.103589 -0.173980983361318
0.104215 -0.0707738639449388
0.104842 -0.0471222324120184
0.105469 0.0238326621867426
0.106095 -0.0944254954778591
0.106722 -0.0944254954778591
0.107349 -0.00626941430970148
0.107976 -0.0772243089084625
0.108602 -0.0922753471566845
0.109229 -0.182581576646017
0.109856 -0.259986916208301
0.110483 0.0324332554714409
0.111109 -0.107326385404907
0.111736 -0.111626682047256
0.112363 -0.150329351828398
0.112989 -0.0277708975214472
0.113616 -0.232034988033032
0.114243 -0.0277708975214472
0.11487 0.0259828105079172
0.115496 0.253898532552422
0.116123 -0.24063558131773
0.11675 0.135640374887821
0.117377 0.0324332554714409
0.118003 0.0044813272961714
0.11863 0.118439188318424
0.119257 0.284000609048866
0.119883 -0.0277708975214472
0.12051 0.152841561457217
0.121137 -0.0922753471566845
0.121764 -0.115926978689605
0.12239 0.0195323655443934
0.123017 -0.0621732706602404
0.123644 -0.0944254954778591
0.12427 -0.339542404091761
0.124897 -0.0127198592732252
0.125524 0.062535331967885
0.126151 -0.201932911536588
0.126777 0.348505058684104
0.127404 0.309802388902961
0.128031 0.00663147561734598
0.128658 -0.180431428324842
0.129284 -0.176131131682493
0.129911 0.103388150070202
0.130538 0.430210694888737
0.131164 0.357105651968802
0.131791 -0.330941810807062
0.132418 -0.780322809932549
0.133045 -0.388995815478776
0.133671 0.367856393574675
0.134298 1
0.134925 0.552769149195688
0.135552 -0.11377683036843
0.136178 -0.97813645548061
0.136805 -0.754521030078454
0.137432 0.150691413136043
0.138058 0.72263086656848
0.138685 0.681778048466163
0.139312 -0.0729240122661133
0.139939 -0.930833192414769
0.140565 -0.416947743654045
0.141192 0.021682513865568
0.141819 0.535567962626292
0.142446 0.475363809633404
0.143077 -0.195482466573064
0.143708 -0.3954462604423
0.144339 0.0668356286102341
0.14497 0.425910398246388
0.145601 0.410859359998166
0.146232 -0.290088992704745
0.146863 -0.485752489931632
0.147495 -0.498653379858679
0.148126 0.223796456055978
0.148757 0.69467893839321
0.149388 0.41730980496169
0.150019 -0.414797595332871
0.15065 -0.625512130807979
0.151281 -0.24063558131773
0.151913 0.303351943939437
0.152544 0.391508025107595
0.153175 0.0840368151796307
0.153806 -0.279338251098873
0.154437 -0.509404121464552
0.155068 -0.154629648470747
0.155699 0.522667072699244
0.156331 0.718330569926131
0.156962 -0.034221342484971
0.157593 -0.588959609348012
0.158224 -0.483602341610457
0.158855 -0.064323418981415
0.159486 0.0496344420408375
0.160117 0.206595269486582
0.160749 0.189394082917185
0.16138 -0.421248040296395
0.162011 -0.595410054311535
0.162642 -0.167530538397795
0.163273 0.0861869635008053
0.163904 0.430210694888737
0.164535 0.436661139852261
0.165167 -0.158929945113096
0.165798 -0.277188102777698
0.166429 -0.363194035624681
0.16706 -0.408347150369347
0.167691 -0.150329351828398
0.168322 0.361405948611151
0.168953 0.307652240581787
0.169584 0.197994676201883
0.170216 -0.0105697109520506
0.170847 -0.182581576646017
0.171478 -0.535205901318647
0.172109 -0.0944254954778591
0.17274 0.0453341453984883
0.173371 0.120589336639598
0.174002 0.271099719121819
0.174634 -0.0879750505143354
0.175265 -0.0750741605872879
0.175896 -0.176131131682493
0.176527 -0.219134098105984
0.177158 -0.0127198592732252
0.177789 0.0388837004349646
0.17842 -0.034221342484971
0.179052 -0.156779796791922
0.179683 -0.178281280003668
0.180314 -0.296539437668269
0.180945 -0.212683653142461
0.181576 -0.0944254954778591
0.182207 0.174343044668963
0.182838 -0.167530538397795
0.18347 -0.120227275331954
0.184101 -0.0772243089084625
0.184732 -0.350293145697634
0.185363 0.18724393459601
0.185994 -0.0793744572296371
0.186625 -0.148179203507223
0.187256 0.249598235910073
0.187888 0.0947875567855036
0.188519 -0.120227275331954
0.18915 0.0324332554714409
0.189781 -0.0600231223390659
0.190412 -0.229884839711857
0.191043 -0.128827868616652
0.191674 -0.0213204525579235
0.192306 0.00663147561734598
0.192932 0.03673355211379
0.193559 0.118439188318424
0.194186 0.0173822172232189
0.194813 -0.0729240122661133
0.19544 -0.115926978689605
0.196067 -0.251386322923603
0.196693 -0.0815246055508116
0.19732 -0.300839734310618
0.197947 -0.272887806135349
0.198574 -0.130978016937827
0.199201 -0.100875940441383
0.199828 -0.0277708975214472
0.200455 -0.0686237156237642
0.201081 0.105538298391376
0.201708 -0.163230241755446
0.202335 -0.0944254954778591
0.202962 0.197994676201883
0.203589 -0.126677720295478
0.204216 -0.100875940441383
0.204843 -0.31589077255884
0.205469 -0.105176237083732
0.206096 0.0109317722596951
0.206723 -0.197632614894239
0.20735 -0.167530538397795
0.207977 -0.171830835040144
0.208604 -0.0320711941637964
0.20923 -0.111626682047256
0.209857 0.021682513865568
0.210484 0.0324332554714409
0.211111 -0.0836747538719862
0.211738 0.00233117897499683
0.212365 0.0840368151796307
0.212992 -0.115926978689605
0.213618 -0.100875940441383
0.214245 0.0689857769314087
0.214872 -0.11377683036843
0.215499 0.0539347386831866
0.216126 0.109838595033726
0.216753 -0.135278313580176
0.21738 -0.0621732706602404
0.218006 -0.242785729638905
0.218633 -0.0299210458426218
0.21926 -0.0729240122661133
0.219887 -0.118077127010779
0.220514 -0.0320711941637964
0.221141 -0.0729240122661133
0.221767 0.206595269486582
0.222394 -0.0213204525579235
0.223021 0.0173822172232189
0.223648 0.385057580144071
0.224275 -0.103026088762557
0.224902 -0.0535726773755421
0.225529 0.146391116493693
0.226155 0.150691413136043
0.226782 0.0990878534278527
0.227409 -0.0664735673025896
0.228036 0.092637408464329
0.228663 0.00878162393852055
0.22929 0.0668356286102341
0.229917 -0.11377683036843
0.230543 0.0109317722596951
0.23117 0.133490226566646
0.231797 0.26679942247947
0.232424 0.206595269486582
0.233051 0.217346011092454
0.233678 0.172192896347788
0.234304 -0.049272380733193
0.234931 0.0044813272961714
0.235558 0.0754362218949324
0.236185 0.197994676201883
0.236812 0.0668356286102341
0.237439 0.0539347386831866
0.238066 0.372156690217024
0.238692 0.256048680873597
0.239319 0.0818866668584561
0.239946 0.281850460727692
0.240573 0.18724393459601
0.2412 0.185093786274836
0.241827 0.062535331967885
0.242454 0.0345834037926155
0.243085 0.193694379559534
0.243716 0.193694379559534
0.244347 0.13994067153017
0.244978 0.281850460727692
0.245609 -0.105176237083732
0.24624 -0.0299210458426218
0.246871 -0.0686237156237642
0.247502 0.16359230306309
0.248133 0.271099719121819
0.248764 0.230246901019502
0.249395 0.0044813272961714
0.250026 0.189394082917185
0.250657 -0.0234706008790981
0.251289 0.00663147561734598
};
\addlegendentry{S2}
\end{axis}

\end{tikzpicture}

%% file: Figures/normalized_event_filtered.tex
\begin{tikzpicture}

\definecolor{darkgrey176}{RGB}{176,176,176}
\definecolor{darkorange25512714}{RGB}{255,127,14}
\definecolor{lightgrey204}{RGB}{204,204,204}
\definecolor{steelblue31119180}{RGB}{31,119,180}

\begin{axis}[
height=4.3cm,
width=4.3cm,
legend cell align={left},
legend style={fill opacity=0.8, draw opacity=1, text opacity=1, at={(0.45,0.35)}, draw=lightgrey204},
tick align=outside,
tick pos=left,
x grid style={darkgrey176},
xlabel={Time in Seconds},
xmin=-0.01258015, xmax=0.26418315,
xtick style={color=black},
y grid style={darkgrey176},
ymin=-1.1, ymax=1.1,
ytick style={color=black}
]
\addplot [semithick, steelblue31119180]
table {%
0.000335 1.91452977327136e-05
0.000968 0.0125157700156641
0.001601 0.0110587039563466
0.002234 0.000559928112826706
0.002867 -0.00493559694619732
0.0035 -0.00262553630865389
0.004133 -0.000464556689793115
0.004766 -0.00201965660548824
0.005399 -0.00320909698090013
0.006032 -0.00151413291683826
0.006665 0.00104191564298858
0.007298 0.0029173340362674
0.007931 0.00401059104772299
0.008563 0.00305706594370936
0.009196 -0.000394415253726689
0.009829 -0.00337442001100143
0.010462 -0.00325462536231868
0.011095 -0.00145740980813191
0.011728 -6.68066419431121e-05
0.012361 0.00135177002042144
0.012994 0.0025327462724303
0.013627 0.00121411915619997
0.01426 -0.00173846627115943
0.014893 -0.00178591122631809
0.015526 0.00136941675227486
0.016159 0.00206596147729189
0.016792 -0.00137476194406467
0.017425 -0.00270054110120493
0.018058 0.00135513483034087
0.01869 0.00425740124447795
0.019323 0.000312920605093876
0.019956 -0.00517507237482676
0.020589 -0.00419508918839203
0.021222 0.00138993233406389
0.021855 0.00389308742282255
0.022488 0.00206106693615059
0.023121 0.000742011152921372
0.023754 0.0011688408967637
0.024387 6.85144436970865e-05
0.02502 -0.00274678792146523
0.025653 -0.00368912022864203
0.026286 -0.00206528636349345
0.026919 -0.000472003775943573
0.027552 0.00120296147224643
0.028185 0.00481275391891044
0.028817 0.00701841305846081
0.02945 0.00204288836998333
0.030083 -0.0070734815295692
0.030716 -0.00940849462716157
0.031349 -0.00188134483403027
0.031982 0.00537104811489961
0.032615 0.00434129619807792
0.033248 0.000237655590406871
0.033881 0.000453055135485496
0.034514 0.00211432590484124
0.035147 -0.000350033383198188
0.03578 -0.00378330057558581
0.036413 -0.002775433995822
0.037045 -0.000222233801014335
0.037678 -0.000727022108009552
0.038311 -0.000514182301312622
0.038944 0.00455180551445726
0.039577 0.00775572595017281
0.04021 0.0013655454412821
0.040843 -0.008025885930329
0.041476 -0.00797051368561577
0.042109 0.000209607877056848
0.042742 0.00451420744967834
0.043375 0.00194452736636176
0.044008 0.000367597166846544
0.044641 0.00221617843719214
0.045273 0.00142185191623454
0.045906 -0.00267888467861443
0.046539 -0.00294818313217195
0.047172 0.00141593075740705
0.047805 0.00202832761053029
0.048438 -0.00318318529429655
0.049071 -0.00471299827333032
0.049704 0.00205211654650145
0.050337 0.00736703746855174
0.05097 0.00282173782324266
0.051603 -0.0042738628662836
0.052236 -0.0032503034338463
0.052869 0.00200644867503866
0.053501 0.000865790048413982
0.054134 -0.00456326658450245
0.054767 -0.00318527455325191
0.0554 0.00454839281420411
0.056033 0.00628571243076421
0.056666 -0.000867859505734304
0.057299 -0.00487325226394293
0.057932 0.00051608132688517
0.058565 0.00481392825510068
0.059198 -0.000922553632934362
0.059831 -0.00803751250602427
0.060464 -0.00423961689188686
0.061097 0.00601652137045941
0.061729 0.00855440673678425
0.062362 0.00116050942661087
0.062995 -0.00513472955886952
0.063628 -0.00384405120801406
0.064261 0.00032466566372741
0.064894 0.00207053185543523
0.065527 0.0016056515670711
0.06616 -0.000379532124552264
0.066793 -0.0037835431153334
0.067426 -0.00444412464480883
0.068059 0.00165625839235236
0.068692 0.00879391388532922
0.069325 0.00674160448577302
0.069957 -0.00301224925799458
0.07059 -0.00786748812537258
0.071223 -0.00344412505398785
0.071856 0.000516174652101845
0.072489 -0.00184759675990219
0.073122 -0.00248328509890399
0.073755 0.00455796225021394
0.074388 0.0101881793220038
0.075021 0.00476688492155387
0.075654 -0.00505668835512074
0.076287 -0.00700901475878703
0.07692 -0.00249697758025023
0.077553 -0.00134239275529207
0.078186 -0.00341555783787681
0.078811 -0.000504064923348065
0.079436 0.0067550899603492
0.080061 0.00861991492799624
0.080686 0.00245935282581909
0.081311 -0.00385746240961857
0.081936 -0.00533097723496471
0.082561 -0.0046541213363107
0.083186 -0.00297413585154552
0.083811 0.002013384635192
0.084436 0.00695601592140051
0.085061 0.00412598792508311
0.085686 -0.00467784724029251
0.086311 -0.00722220856527209
0.086936 0.00104498548154417
0.087561 0.008756587132098
0.088186 0.00519168258836644
0.088811 -0.00432762954804165
0.089436 -0.00767775414974054
0.090061 -0.00309060884074487
0.090687 0.00182680603372562
0.091312 0.00290427250200725
0.091937 0.00215476981757834
0.092562 0.000930783878329591
0.093187 -0.00105044758935902
0.093812 -0.00151090056034604
0.094437 0.00160713949086894
0.095062 0.00409080322463684
0.095687 0.000235662607066204
0.096312 -0.00645036048783487
0.096937 -0.00624858427979207
0.097562 0.00149327559917583
0.098187 0.00587607345220221
0.098812 0.00139192079782033
0.099437 -0.00279327525654846
0.100062 0.00188423697702225
0.100687 0.00797304547590254
0.101312 0.00322193535571994
0.101937 -0.00874382816856042
0.102562 -0.012422623621664
0.103188 -0.00335796728473174
0.103813 0.00626976774938558
0.104438 0.00671684953360331
0.105063 0.00299416714728471
0.105688 0.00301653451419211
0.106313 0.00391777828336343
0.106938 -0.00105413703567941
0.107563 -0.00846751437161759
0.108188 -0.00841253295813285
0.108813 -0.000709175853672246
0.109438 0.00374200806060872
0.110063 0.000557708444702476
0.110688 -0.000499317953780358
0.111313 0.00688242861404728
0.111938 0.011384872938496
0.112563 0.00134986254099998
0.113188 -0.0123333682890235
0.113813 -0.0102682725680813
0.114438 0.00317181562227831
0.115063 0.00582842412463757
0.115689 -0.00458558843804332
0.116314 -0.0073310633251803
0.116939 0.00313881047343016
0.117564 0.00824949586425076
0.118189 0.00240655527365804
0.118814 0.00265198246009652
0.119439 0.0102430764124134
0.120064 0.00418828557570125
0.120689 -0.0134677431524928
0.121314 -0.0152382659872017
0.121939 -0.00516655143483363
0.122564 -0.0138123292483369
0.123189 -0.0194832330784032
0.123814 0.0319700619999665
0.124439 0.0921679564723017
0.125064 0.0268928335097925
0.125689 -0.140332237828015
0.126314 -0.154949510435434
0.126939 0.111024883731291
0.127564 0.344964296048729
0.12819 0.127147090958084
0.128822 -0.413184641737318
0.129455 -0.597217036850375
0.130088 -0.0363365169699251
0.130721 0.746403673497784
0.131354 0.827508018441862
0.131987 -0.0127964179356394
0.13262 -0.948033054743677
0.133252 -0.977317058199167
0.133885 -0.0241382830625502
0.134518 0.963025244495055
0.135151 1
0.135784 0.0784352007517282
0.136417 -0.85076675467789
0.13705 -0.878071817458255
0.137682 -0.05798828850784
0.138315 0.705124713557166
0.138948 0.65012891807928
0.139581 -0.0604412798575713
0.140214 -0.593050984199102
0.140847 -0.393423188208095
0.14148 0.236818113465422
0.142112 0.537554354314858
0.142745 0.168900429710852
0.143378 -0.419608819388958
0.144011 -0.52172878599863
0.144644 0.0062641798859512
0.145277 0.570123457263739
0.14591 0.510559819786145
0.146542 -0.137133668479683
0.147175 -0.6660679667455
0.147808 -0.482952753465049
0.148441 0.229252502395105
0.149074 0.706892668818034
0.149707 0.434407696081943
0.15034 -0.297854010367612
0.150972 -0.703233540819609
0.151605 -0.347709534638482
0.152238 0.370829614926538
0.152871 0.654403556112189
0.153504 0.20019320788282
0.154137 -0.453792295355435
0.15477 -0.559065850582408
0.155403 -0.0288039217132217
0.156035 0.498158771971616
0.156668 0.448487432011984
0.157301 -0.0503217778975143
0.157934 -0.426924775181396
0.158567 -0.367985553622903
0.1592 -0.0549679535362175
0.159833 0.20442233805094
0.160465 0.311069275439448
0.161098 0.290649374486689
0.161731 0.11389316490272
0.162364 -0.212863540096291
0.162997 -0.478716153090196
0.16363 -0.400608810903234
0.164263 0.0341848122946233
0.164895 0.476563159578425
0.165528 0.539276220424918
0.166161 0.174961716834928
0.166794 -0.297651812988591
0.167427 -0.503140843613034
0.16806 -0.318544100441242
0.168693 0.0709902980834712
0.169325 0.355359078208525
0.169958 0.343673037338858
0.170591 0.0896879005217783
0.171224 -0.181006696375265
0.171857 -0.26993656978571
0.17249 -0.154532356530778
0.173123 0.0337860967672333
0.173755 0.150200069636469
0.174388 0.144990389116324
0.175021 0.0610198731695433
0.175654 -0.0320164337276274
0.176287 -0.086903575922754
0.17692 -0.08905753201584
0.177553 -0.0487106474147929
0.178186 0.00928083705587223
0.17881 0.0565403423659867
0.179435 0.071458162620664
0.18006 0.0471638213086657
0.180685 -0.00245283878778393
0.18131 -0.0457799184974621
0.181935 -0.0541298316681083
0.18256 -0.0268625391083298
0.183185 0.00837535475474814
0.18381 0.024609277054587
0.184435 0.0197191800351136
0.18506 0.00877114193095841
0.185685 0.00160381520397688
0.18631 -0.00253811337586043
0.186935 -0.00442228684141604
0.18756 -0.00221699630271098
0.188185 0.000756261614364208
0.18881 -0.00242161369283476
0.189435 -0.00988716935006713
0.19006 -0.0102366693452308
0.190685 0.000940721487715703
0.19131 0.01246082284265
0.191935 0.0121330225143288
0.19256 0.00299018043429216
0.193185 -0.00300828222687823
0.19381 -0.00232763337751886
0.194435 -0.00162497475211381
0.19506 -0.00453525244778251
0.195685 -0.00631836899059557
0.19631 -0.00290871512572291
0.196935 0.0024005624451761
0.19756 0.00459117920268226
0.198185 0.00395868376183511
0.19881 0.00349746957548516
0.199435 0.00271913547259601
0.20006 -0.000973586759076002
0.200685 -0.00585338189269935
0.20131 -0.00615516090823876
0.201935 -0.000517255099121045
0.20256 0.00441833189508456
0.203185 0.00279335567921482
0.20381 -0.00158456000979948
0.204435 -0.0011421929762744
0.20506 0.00291094580671449
0.205685 0.00231688811359356
0.20631 -0.00365002382656162
0.206935 -0.00534690707158121
0.20756 0.00162089242703387
0.208185 0.00777240285308588
0.20881 0.00380861824244787
0.209435 -0.00494860717874622
0.21006 -0.00720090461574264
0.210685 -0.00237587602388674
0.21131 0.00144783642291812
0.211935 0.001800413258412
0.21256 0.00310874013647497
0.213185 0.00571759348121167
0.21381 0.00413941518585869
0.214435 -0.00284322594982846
0.21506 -0.00845367329679174
0.215685 -0.00668997116829045
0.21631 9.00519374514124e-05
0.216935 0.00461776704172444
0.21756 0.00354308315164176
0.218185 0.000948571815693622
0.21881 0.00217265602397805
0.219435 0.00493706846635264
0.22006 0.00142032396876519
0.220685 -0.00821662919414776
0.22131 -0.0117678367906092
0.221935 -0.00199131003150771
0.22256 0.00968680616306259
0.223185 0.00833169283343
0.22381 -0.00159505051925544
0.224435 -0.00389363465308416
0.22506 0.00252702713587474
0.225685 0.00319875878255369
0.22631 -0.00515323033207312
0.226935 -0.0078686390861545
0.22756 0.00186057795078094
0.228185 0.00965844118560681
0.228817 0.00287102738963172
0.22945 -0.00877439724669995
0.230083 -0.00847205092571559
0.230716 0.00278589348401011
0.231349 0.0101413487836807
0.231982 0.00674667900620634
0.232615 -0.000754719410799722
0.233248 -0.00500939543601042
0.233881 -0.00552014020123662
0.234514 -0.00426504861214474
0.235147 -0.00178165254878871
0.23578 0.00137104412305374
0.236413 0.00377759922183365
0.237045 0.00461838042178556
0.237678 0.00396312358631753
0.238311 0.00197272499209154
0.238944 -0.00102616863197105
0.239577 -0.00420999086335554
0.24021 -0.00624446825734329
0.240843 -0.00528429250180636
0.241476 -0.000472472553180775
0.242109 0.00535752168915224
0.242742 0.00693175387823218
0.243375 0.00283158611091398
0.244008 -0.00190308800810783
0.244641 -0.00225112680070791
0.245273 0.000223196054904535
0.245906 0.000575972855895575
0.246539 -0.00200033324343744
0.247172 -0.00373386808730575
0.247805 -0.002185693816255
0.248438 0.000950642539160827
0.249071 0.00284550368249847
0.249704 0.00272188112172219
0.250337 0.00163176565303647
0.25097 0.000668439654034669
0.251603 6.41959258395557e-05
};
\addlegendentry{S1}
\addplot [semithick, darkorange25512714]
table {%
0 -0.000361747187285082
0.000631 -0.0329733591917773
0.001263 -0.0136022121583624
0.001894 0.0405093056758666
0.002526 0.0484750578322059
0.003157 -0.0250910663427153
0.003788 -0.0873299019911082
0.00442 -0.0262479487979282
0.005051 0.101898259990462
0.005682 0.113404289699034
0.006314 -0.0388597399093409
0.006945 -0.168342219748877
0.007577 -0.101007621238218
0.008208 0.0841696858556555
0.008839 0.166219365054445
0.009471 0.0691442729114633
0.010102 -0.0717153119254943
0.010734 -0.10590133434307
0.011365 -0.0335685378891904
0.011996 0.0389700069481971
0.012628 0.0428162827351285
0.013259 0.00313408548290902
0.01389 -0.025004677494598
0.014522 -0.0256964754088536
0.015153 -0.00771356342554163
0.015785 0.0310251100779119
0.016416 0.0712908541046016
0.017047 0.04781193320306
0.017679 -0.0569126099935888
0.01831 -0.128566789690747
0.018941 -0.0524640286434628
0.019573 0.0884242910255255
0.020204 0.102224525108862
0.020836 -0.0170771654541288
0.021467 -0.0715615620673762
0.022098 0.0165056285583332
0.02273 0.0759140320786199
0.023361 -0.0125870928270063
0.023992 -0.10163691719615
0.024624 -0.0322028383928785
0.025255 0.0842509242010908
0.025887 0.0561191340846782
0.026518 -0.0589595986971955
0.027149 -0.0622960729200558
0.027781 0.0419098075951482
0.028412 0.0712829004027809
0.029044 -0.010216704527296
0.029675 -0.0518410489009438
0.030306 -0.000342894416549809
0.030938 0.0240708059051449
0.031569 -0.0260523235102023
0.0322 -0.0371263087830605
0.032832 0.0362775185018751
0.033463 0.0689353070435886
0.034095 -0.00838460355332243
0.034726 -0.073348748462181
0.035357 -0.0255404659695889
0.035989 0.0399606827333193
0.03662 0.0167673479084234
0.037251 -0.0169879709966708
0.037883 0.0375847662086635
0.038514 0.0819792316720373
0.039146 -0.0197565906828299
0.039777 -0.162849856104084
0.040408 -0.125120253480861
0.04104 0.0756523929145942
0.041671 0.189874528206442
0.042303 0.102575097125506
0.042935 -0.0392525660401801
0.043568 -0.0940793730278132
0.044201 -0.091401191845407
0.044834 -0.0716557421489815
0.045467 0.00994116249210959
0.0461 0.121911529406666
0.046733 0.114026246131911
0.047366 -0.0348362984837278
0.047999 -0.122120374340521
0.048632 -0.0239483227391158
0.049265 0.0934422483897126
0.049898 0.0415197950072536
0.050531 -0.0826520352954399
0.051164 -0.0712541009014646
0.051797 0.0470740201134158
0.05243 0.0778713760756104
0.053063 -0.00533262447786592
0.053696 -0.0475968682874053
0.054329 0.000493881681935629
0.054962 0.0238270762128388
0.055595 -0.0192145818951691
0.056228 -0.0253888666280671
0.056861 0.038782872531178
0.057494 0.0557831952798396
0.058127 -0.0292030153280494
0.05876 -0.0922377271773108
0.059393 -0.0321583831102851
0.060026 0.0637561981231039
0.060659 0.0721496027714495
0.061292 0.0192256009118889
0.061925 -0.0107636185363469
0.062558 -0.029370351258605
0.063191 -0.066830599821203
0.063824 -0.0568245664817473
0.064457 0.0411202672253969
0.06509 0.106948734521097
0.065723 0.0291974906571576
0.066356 -0.0798501674515028
0.066989 -0.0436841677419902
0.067621 0.073152472990158
0.068254 0.0560223050160556
0.068887 -0.0928908513775012
0.06952 -0.135256592566897
0.070153 0.0245672926590859
0.070786 0.181506541119396
0.071419 0.132177396648819
0.072052 -0.0576768337736666
0.072685 -0.182989560991937
0.073318 -0.14686291549375
0.073951 0.00337089232605854
0.074584 0.158429953931613
0.075217 0.196278535676673
0.07585 0.0655965274468982
0.076483 -0.121990084675715
0.077116 -0.185924573579839
0.077749 -0.0947944491154551
0.078382 0.0198873232468347
0.079015 0.0745206687168101
0.079648 0.108309797669504
0.080281 0.125887213641805
0.080914 0.0464105830561262
0.081547 -0.117226607414929
0.08218 -0.194340997504989
0.082813 -0.0823850851624948
0.083446 0.0873819435767402
0.084079 0.126549160197541
0.084712 0.038775316477284
0.085345 -0.0254792375829237
0.085978 -0.000370913164134944
0.086611 0.0310673282879687
0.087244 -0.00987333931895878
0.087877 -0.0797300753447356
0.08851 -0.0804511533693682
0.089143 0.00268788304669424
0.089776 0.0777447265407804
0.090409 0.0654615100226204
0.091042 0.00358507083246971
0.091675 -0.0136296979059353
0.092308 0.019177934772463
0.092934 0.0132213082318746
0.093561 -0.0592275256504969
0.094188 -0.0998951664138737
0.094814 -0.0314879206164841
0.095441 0.0740499012420223
0.096068 0.0963926083274541
0.096695 0.0342793852226604
0.097321 -0.0182501490219863
0.097948 -0.0207768802805405
0.098575 -0.0183301507109483
0.099201 -0.0411385578872208
0.099828 -0.0545446138773448
0.100455 -0.0185239423996142
0.101082 0.0500883404704783
0.101708 0.0881201753693076
0.102335 0.049654220315727
0.102962 -0.0345291114618605
0.103589 -0.0745429221587243
0.104215 -0.029207278742153
0.104842 0.0298063508032918
0.105469 0.0181632515401873
0.106095 -0.0296693397965716
0.106722 -0.0177728436791429
0.107349 0.0456732279702699
0.107976 0.0555440865346966
0.108602 -0.0150666639634326
0.109229 -0.0678321320702408
0.109856 -0.0333930527829466
0.110483 0.035893113360429
0.111109 0.0607187758515744
0.111736 0.0292894135286025
0.112363 -0.0287026040889263
0.112989 -0.076780783683046
0.113616 -0.0664253038209893
0.114243 0.0171382716157608
0.11487 0.0986381445635113
0.115496 0.0868085814060757
0.116123 -0.000292988853949209
0.11675 -0.0660530020709501
0.117377 -0.068802828339385
0.118003 -0.0365918895398872
0.11863 0.0122477245440325
0.119257 0.0607683140042947
0.119883 0.0522678624717497
0.12051 -0.0270745371491578
0.121137 -0.0747590595941806
0.121764 -0.00721142459872039
0.12239 0.0869355142726933
0.123017 0.0691980080104137
0.123644 -0.0310983639682314
0.12427 -0.0791709515590478
0.124897 -0.0559688557089666
0.125524 -0.0354342965875817
0.126151 0.00317428111157939
0.126777 0.116973457399809
0.127404 0.181462638917821
0.128031 0.00380096318937958
0.128658 -0.295223651194423
0.129284 -0.299913711736745
0.129911 0.132320947923957
0.130538 0.542658020233267
0.131164 0.368592441671661
0.131791 -0.310760164970777
0.132418 -0.77671705773087
0.133045 -0.466639750299167
0.133671 0.378866128336257
0.134298 0.939236248466632
0.134925 0.633737653135411
0.135552 -0.299166518079192
0.136178 -1
0.136805 -0.787782644568633
0.137432 0.160991165798964
0.138058 0.939364466887759
0.138685 0.800528588279086
0.139312 -0.07167827698153
0.139939 -0.766143773462411
0.140565 -0.624977592709883
0.141192 0.0894986608757959
0.141819 0.558454568925356
0.142446 0.331083296467079
0.143077 -0.23345962576794
0.143708 -0.433689948991979
0.144339 -0.0344555099612765
0.14497 0.47011450170237
0.145601 0.428052660758221
0.146232 -0.183571906579298
0.146863 -0.6764967821018
0.147495 -0.442517595882187
0.148126 0.294325666595877
0.148757 0.716366509816473
0.149388 0.374292461391448
0.150019 -0.318520440111833
0.15065 -0.613944268799706
0.151281 -0.254350860833487
0.151913 0.336198698767474
0.152544 0.545877773066124
0.153175 0.164714188524712
0.153806 -0.420602120587218
0.154437 -0.580749082873554
0.155068 -0.10304395700716
0.155699 0.513160078967984
0.156331 0.584949562823825
0.156962 0.0489617698071662
0.157593 -0.485728077968811
0.158224 -0.489270479117821
0.158855 -0.0538969707653833
0.159486 0.357728224232708
0.160117 0.42802523563221
0.160749 0.150235281989545
0.16138 -0.261443775654372
0.162011 -0.479990903121541
0.162642 -0.283063388150023
0.163273 0.184054882454853
0.163904 0.493347507210834
0.164535 0.383725012647925
0.165167 0.0102281604936567
0.165798 -0.304030716593459
0.166429 -0.399333563573066
0.16706 -0.289628238869766
0.167691 -0.0296644123166125
0.168322 0.274163713588126
0.168953 0.429306115891404
0.169584 0.299353611511062
0.170216 -0.0297306001507686
0.170847 -0.311447503681375
0.171478 -0.361553864078716
0.172109 -0.184911477843089
0.17274 0.0740782651163028
0.173371 0.247063526554913
0.174002 0.237348879374425
0.174634 0.0757233347852391
0.175265 -0.100836761742311
0.175896 -0.161793517072762
0.176527 -0.0912903489155253
0.177158 0.0236990738841164
0.177789 0.0944740314803009
0.17842 0.087637995584991
0.179052 0.00808257052891253
0.179683 -0.102724257351568
0.180314 -0.149241539477155
0.180945 -0.0520870746540556
0.181576 0.125112988262531
0.182207 0.201711047727853
0.182838 0.0936612135955258
0.18347 -0.072220759015237
0.184101 -0.12905813301548
0.184732 -0.0790845315769997
0.185363 -0.036758740287951
0.185994 -0.0314584745798382
0.186625 0.00468979383609864
0.187256 0.0819771180622562
0.187888 0.118467311069118
0.188519 0.0705070952169282
0.18915 -0.00799355675747484
0.189781 -0.0584655567096774
0.190412 -0.0825471706231206
0.191043 -0.0885827936417587
0.191674 -0.0525767588529495
0.192306 0.0266951970152399
0.192932 0.0911280343839258
0.193559 0.0872257207012982
0.194186 0.0325983751120174
0.194813 -0.0161380432105833
0.19544 -0.0335212706258401
0.196067 -0.0321027512116012
0.196693 -0.0264588362840908
0.19732 -0.0236240333401157
0.197947 -0.025601729831504
0.198574 -0.0132057984127921
0.199201 0.0372482433445707
0.199828 0.0897591364441368
0.200455 0.0602916366708878
0.201081 -0.0497756104962279
0.201708 -0.105302077893626
0.202335 -0.0193796243922531
0.202962 0.0874301075879124
0.203589 0.0482052169489009
0.204216 -0.0824136463850424
0.204843 -0.0978131758325503
0.205469 0.0368534295336741
0.206096 0.127153130981346
0.206723 0.0516937610885901
0.20735 -0.0676608095604986
0.207977 -0.0765352150533979
0.208604 -0.00574761388629079
0.20923 0.0274468645730995
0.209857 0.00854425127285197
0.210484 0.000440940242813696
0.211111 0.0164354674281901
0.211738 0.0225057621232894
0.212365 0.00853305032166018
0.212992 -0.0170223845039594
0.213618 -0.0491726666767313
0.214245 -0.0553689713723966
0.214872 0.00474603465351519
0.215499 0.0874478595160867
0.216126 0.0855050477600447
0.216753 -0.0111936511219119
0.21738 -0.0802618057837139
0.218006 -0.0479999772092676
0.218633 0.0095072582384986
0.21926 0.0118575590639041
0.219887 -0.00477397319645964
0.220514 0.00898270059143221
0.221141 0.0198951718236783
0.221767 0.0060670477841002
0.222394 0.0160157702083348
0.223021 0.047740248221568
0.223648 0.0119584135263738
0.224275 -0.0912424969755325
0.224902 -0.116511153424954
0.225529 -0.00278572536588082
0.226155 0.102797118542508
0.226782 0.0744317859770546
0.227409 0.000475228979354302
0.228036 0.00647089566195449
0.228663 0.0377984811476275
0.22929 -0.0189970158361434
0.229917 -0.117826662333472
0.230543 -0.120246949368765
0.23117 -0.00602735008583716
0.231797 0.114118950005407
0.232424 0.144183994842028
0.233051 0.0743690343567991
0.233678 -0.0387267212469544
0.234304 -0.106672851384243
0.234931 -0.0805905305877497
0.235558 -0.0141235899003981
0.236185 0.00843396363425417
0.236812 -0.00211352497852628
0.237439 0.022807878367067
0.238066 0.0696970208108282
0.238692 0.0611457265500974
0.239319 0.000643584528707459
0.239946 -0.0302874869853881
0.240573 -0.0177133550501845
0.2412 -0.0306359469220458
0.241827 -0.0812632688072109
0.242454 -0.0864973432451378
0.243085 0.00538487177895643
0.243716 0.131409577715827
0.244347 0.177671598028691
0.244978 0.0856202253513583
0.245609 -0.0898166353045203
0.24624 -0.201285403805498
0.246871 -0.138287192692903
0.247502 0.03325994988925
0.248133 0.129826183479555
0.248764 0.0784287780611695
0.249395 -0.00817466414497173
0.250026 -0.020997359339189
0.250657 0.00410370644912734
0.251289 0.00104265696058273
};
\addlegendentry{S2}
\end{axis}

\end{tikzpicture}

%% file: Figures/cwt_peaks.tex
\begin{tikzpicture}

\definecolor{darkgrey176}{RGB}{176,176,176}
\definecolor{darkorange25512714}{RGB}{255,127,14}
\definecolor{lightgrey204}{RGB}{204,204,204}
\definecolor{steelblue31119180}{RGB}{31,119,180}

\begin{axis}[
height=4.3cm,
width=4.3cm,
legend cell align={left},
legend style={fill opacity=0.8, draw opacity=1, text opacity=1, at={(0.45,0.35)}, draw=lightgrey204}, 
tick align=outside,
tick pos=left,
x grid style={darkgrey176},
xlabel={Time in seconds},
xmin=0.08727425, xmax=0.22314875,
xtick style={color=black},
y grid style={darkgrey176},
ymin=-1.12219337926582, ymax=1.10105682758409,
ytick style={color=black}
]
\addplot [semithick, steelblue31119180]
table {%
0.095062 0.00428618378057816
0.095687 0.00556666468542194
0.096312 0.00350102621211968
0.096937 -0.00171615049744895
0.097562 -0.00664630624781795
0.098187 -0.00813622685319493
0.098812 -0.00621822488033935
0.099437 -0.0026318274663214
0.100062 0.00212088996238558
0.100687 0.00807175817250498
0.101312 0.0126151790664384
0.101937 0.0110223141019596
0.102562 0.0017775992891007
0.103188 -0.0100577367041838
0.103813 -0.016633534741102
0.104438 -0.0141121295563389
0.105063 -0.00485851204578505
0.105688 0.00591458257130085
0.106313 0.0137321665551584
0.106938 0.0157360298558435
0.107563 0.0109761301390949
0.108188 0.00148163308961257
0.108813 -0.00817715372161113
0.109438 -0.013910563466883
0.110063 -0.0144151315367924
0.110688 -0.010154706567678
0.111313 -0.00178767443177669
0.111938 0.00855613751585547
0.112563 0.0158108391485573
0.113188 0.0151426266009733
0.113813 0.00735466612569172
0.114438 -0.00131170436846137
0.115063 -0.00595714681853379
0.115689 -0.00756362147727944
0.116314 -0.00951294254483293
0.116939 -0.0119027527893402
0.117564 -0.0115188269738
0.118189 -0.00591077268783802
0.118814 0.00458595349267503
0.119439 0.0173495892471612
0.120064 0.0274665398855441
0.120689 0.0281027326171311
0.121314 0.0137659633084984
0.121939 -0.0144500964139111
0.122564 -0.0440507631977205
0.123189 -0.0516454800896265
0.123814 -0.0194228782615562
0.124439 0.0344125164703605
0.125064 0.0507535279111926
0.125689 -0.00922209864352064
0.126314 -0.0840558303368432
0.126939 -0.0413221228633477
0.127564 0.150424823819816
0.12819 0.297887006302497
0.128822 0.137301401786729
0.129455 -0.317544812568468
0.130088 -0.650672549179835
0.130721 -0.410804561222363
0.131354 0.342742953938321
0.131987 0.955661147108489
0.13262 0.772100656384526
0.133252 -0.15504106492286
0.133885 -1.02113655168173
0.134518 -1.01652664533713
0.135151 -0.129639643570246
0.135784 0.825684205883099
0.136417 1
0.13705 0.314291648950473
0.137682 -0.52081384707733
0.138315 -0.750263066169146
0.138948 -0.291995132656388
0.139581 0.283304261847562
0.140214 0.413399751611637
0.140847 0.102207828947099
0.14148 -0.189808163756675
0.142112 -0.125471466160206
0.142745 0.142983643249597
0.143378 0.210254472061395
0.144011 -0.059546112691922
0.144644 -0.351812623864028
0.145277 -0.274031045530918
0.14591 0.156469156609129
0.146542 0.488702154622175
0.147175 0.32887125643822
0.147808 -0.199503502634298
0.148441 -0.555899999233243
0.149074 -0.355196304232433
0.149707 0.208498155112005
0.15034 0.557066332762476
0.150972 0.339900074802373
0.151605 -0.19829213664406
0.152238 -0.490020118691838
0.152871 -0.256394236938989
0.153504 0.21018121753713
0.154137 0.389399946079863
0.15477 0.107350783873955
0.155403 -0.293385586476272
0.156035 -0.352171736502046
0.156668 0.00768003037860704
0.157301 0.41563808552422
0.157934 0.454460144021368
0.158567 0.0669685654253081
0.1592 -0.422481044489611
0.159833 -0.619192609341173
0.160465 -0.367808610474387
0.161098 0.167146618219582
0.161731 0.631802580813895
0.162364 0.702794707811077
0.162997 0.293861883010385
0.16363 -0.353692806987283
0.164263 -0.79554798623583
0.164895 -0.695844632279588
0.165528 -0.104174513412614
0.166161 0.554452197914254
0.166794 0.801163448613406
0.167427 0.476197455165503
0.16806 -0.149713073915777
0.168693 -0.60700675981632
0.169325 -0.59380399745548
0.169958 -0.179155940496106
0.170591 0.293684543411619
0.171224 0.492105891770775
0.171857 0.327918434862473
0.17249 -0.0237195278358127
0.173123 -0.295362357952841
0.173755 -0.327209470108145
0.174388 -0.14821039020439
0.175021 0.0879255891795066
0.175654 0.229090754775196
0.176287 0.213935286322122
0.17692 0.0820197225132919
0.177553 -0.073648531939223
0.178186 -0.16696023191651
0.17881 -0.159359322197893
0.179435 -0.06960157847576
0.18006 0.0441272586759839
0.180685 0.118653621995714
0.18131 0.120093289107628
0.181935 0.0596594942043789
0.18256 -0.0185622406390533
0.183185 -0.069129003786602
0.18381 -0.0729476895912006
0.184435 -0.0422451865907242
0.18506 -0.00391672983197243
0.185685 0.0212933745177175
0.18631 0.02818734193143
0.186935 0.023289070629671
0.18756 0.0156464113920799
0.188185 0.00980646017142608
0.18881 0.0043380264352394
0.189435 -0.00413569120450798
0.19006 -0.0150679432350355
0.190685 -0.0224030162555638
0.19131 -0.0198267448504389
0.191935 -0.00752377144839426
0.19256 0.00731711638816082
0.193185 0.0167768434668262
0.19381 0.0180052838673953
0.194435 0.0130351127919589
0.19506 0.00501592225485212
0.195685 -0.00373778415952569
0.19631 -0.0109836293851618
0.196935 -0.0141775398385897
0.19756 -0.0118768772796493
0.198185 -0.00502211189849337
0.19881 0.00347447127160247
0.199435 0.0100806810531758
0.20006 0.0119182733628174
0.200685 0.00810838890015019
0.20131 0.000843669312219553
0.201935 -0.00548620433896326
0.20256 -0.00746557585959255
0.203185 -0.00524404366161219
0.20381 -0.00160286476885015
0.204435 0.00118661411005992
0.20506 0.00266847452177383
0.205685 0.00275750254558121
0.20631 0.000921298145047996
0.206935 -0.00213832893106438
0.20756 -0.00346277953391237
0.208185 -0.000588432905518225
0.20881 0.00476659939559171
0.209435 0.00751052301037112
0.21006 0.00431480423446757
0.210685 -0.00296219418443629
0.21131 -0.00922641601084504
0.211935 -0.0104475153293414
0.21256 -0.00582733697461597
0.213185 0.00250145137805894
0.21381 0.0103744488647296
0.214435 0.0131359419734887
0.21506 0.00854720535781174
0.215685 -0.000927051188167723
0.21631 -0.00952465121237999
0.216935 -0.0126197257926888
0.21756 -0.0097046451432019
0.218185 -0.00321899596173181
0.21881 0.00427531852615298
0.219435 0.010673064554268
0.22006 0.0130918245386853
0.220685 0.00875421615143659
0.22131 -0.00133475227829067
0.221935 -0.0109685122787159
0.22256 -0.0133971960186604
0.223185 -0.0077274647263565
0.22381 0.000708049379166878
0.224435 0.00629977485999242
0.22506 0.00763014690981455
0.225685 0.00594576257028419
0.22631 0.00221825946518654
};
\addlegendentry{S1}
\addplot [thick, darkorange25512714]
table {%
0.094814 -0.113791130564574
0.095441 -0.133413275951564
0.096068 -0.0757589146718325
0.096695 0.0215002765001079
0.097321 0.0984409807042794
0.097948 0.119589353710208
0.098575 0.0859405206318465
0.099201 0.0174017495563594
0.099828 -0.0593806072808245
0.100455 -0.110430852130144
0.101082 -0.10418117853905
0.101708 -0.0380878243439005
0.102335 0.0475143454144405
0.102962 0.0955448996086264
0.103589 0.0799368215481318
0.104215 0.0252088001322431
0.104842 -0.0236303286537271
0.105469 -0.0442905354530202
0.106095 -0.044140207300217
0.106722 -0.0331253993515341
0.107349 -0.00930950127511768
0.107976 0.0232803233811287
0.108602 0.0407124923461983
0.109229 0.0218931443686579
0.109856 -0.0185262160794237
0.110483 -0.0370164927845998
0.111109 -0.00756100668958622
0.111736 0.0454976156557828
0.112363 0.0670121498358685
0.112989 0.0244634596248213
0.113616 -0.0564734702230626
0.114243 -0.108861985941015
0.11487 -0.0821444172001768
0.115496 0.0100712528583221
0.116123 0.0983662589733624
0.11675 0.11660499710728
0.117377 0.0565715080269788
0.118003 -0.0301544026947533
0.11863 -0.0789377508154084
0.119257 -0.0640740542818347
0.119883 -0.0150751931519344
0.12051 0.0148496871672409
0.121137 0.0024052773618135
0.121764 -0.0219280861642954
0.12239 -0.0099131704761568
0.123017 0.0444678226820047
0.123644 0.0907922276597302
0.12427 0.0720973069750147
0.124897 -0.0156568328773898
0.125524 -0.111726266122877
0.126151 -0.131855810251511
0.126777 -0.0365086728626498
0.127404 0.112406693425717
0.128031 0.169023314009109
0.128658 0.0334797546402667
0.129284 -0.199418134595049
0.129911 -0.269278903100915
0.130538 -0.0131368825169887
0.131164 0.387349197793215
0.131791 0.505886427538244
0.132418 0.105037419497356
0.133045 -0.535760593517991
0.133671 -0.807873364810629
0.134298 -0.360950083188013
0.134925 0.491377436147131
0.135552 1
0.136178 0.670193923650627
0.136805 -0.246419432339765
0.137432 -0.955994040482416
0.138058 -0.850113574611167
0.138685 -0.0618695284009119
0.139312 0.686136606458883
0.139939 0.779227721891648
0.140565 0.255396646585235
0.141192 -0.319736835569481
0.141819 -0.462915630781156
0.142446 -0.208823153574487
0.143077 0.0456692438643269
0.143708 0.0437495521882718
0.144339 -0.0724131967842908
0.14497 -0.00970904322746151
0.145601 0.256191763464711
0.146232 0.411698279256762
0.146863 0.181135889499866
0.147495 -0.298932270875798
0.148126 -0.571519567338146
0.148757 -0.344112406271188
0.149388 0.176129430934763
0.150019 0.485534166425758
0.15065 0.309765247456755
0.151281 -0.12108113782652
0.151913 -0.336447442458981
0.152544 -0.129873144326261
0.153175 0.235838888775606
0.153806 0.32728598557572
0.154437 0.0148853320031721
0.155068 -0.381684349426275
0.155699 -0.430812776562489
0.156331 -0.0506102327162918
0.156962 0.396873071631951
0.157593 0.476208531881268
0.158224 0.117485109582192
0.158855 -0.32672521612748
0.159486 -0.43777360690224
0.160117 -0.1244335509981
0.160749 0.312140234252722
0.16138 0.465081902394015
0.162011 0.181716099311586
0.162642 -0.30847459696368
0.163273 -0.591063112223938
0.163904 -0.414643504571488
0.164535 0.104717873533846
0.165167 0.585394973424699
0.165798 0.682115087891782
0.166429 0.329890809735826
0.16706 -0.237288200960811
0.167691 -0.661187093392429
0.168322 -0.684930042937678
0.168953 -0.300697624431025
0.169584 0.250642020343219
0.170216 0.630025078320641
0.170847 0.616614146742625
0.171478 0.247095914481621
0.172109 -0.222062927585145
0.17274 -0.498411197059112
0.173371 -0.445474037331271
0.174002 -0.150605311225205
0.174634 0.16284824652523
0.175265 0.300931179066001
0.175896 0.220698979018666
0.176527 0.0314063035239272
0.177158 -0.10792428171044
0.177789 -0.107948819770665
0.17842 -0.00130872365992425
0.179052 0.0974140599460481
0.179683 0.0871057192631875
0.180314 -0.0371409606696037
0.180945 -0.175462234717636
0.181576 -0.20304972481227
0.182207 -0.0788048885507421
0.182838 0.114927824724805
0.18347 0.242556707896825
0.184101 0.226866381194366
0.184732 0.0951149611718136
0.185363 -0.0670740603041115
0.185994 -0.184883922966265
0.186625 -0.220904009815593
0.187256 -0.167321378615867
0.187888 -0.0430245725602048
0.188519 0.102269409038168
0.18915 0.202660796386482
0.189781 0.210745839562122
0.190412 0.124023244295079
0.191043 -0.01556861939622
0.191674 -0.142616818237158
0.192306 -0.197897839943063
0.192932 -0.158180518202721
0.193559 -0.0506987371285852
0.194186 0.0636198022153267
0.194813 0.130257739619793
0.19544 0.132087331288402
0.196067 0.0873934828857319
0.196693 0.0246058414510846
0.19732 -0.0373000091007843
0.197947 -0.0869134394561565
0.198574 -0.108470529026375
0.199201 -0.0837222408832067
0.199828 -0.0142509812230337
0.200455 0.0618412495719608
0.201081 0.092093709148834
0.201708 0.0604486814442568
0.202335 0.007502361996629
0.202962 -0.0151568912988275
0.203589 -0.00586672280865484
0.204216 -0.00751984524258827
0.204843 -0.0399584283058107
0.205469 -0.0648455263785116
0.206096 -0.0361179751738651
0.206723 0.0327562715799459
0.20735 0.0785214587373993
0.207977 0.060902990197641
0.208604 0.00258586243745983
0.20923 -0.045197371190898
0.209857 -0.0554128662946823
0.210484 -0.0350512384967263
0.211111 -0.001011553312156
0.211738 0.0345112481014647
0.212365 0.0579773713376918
0.212992 0.0511627423836304
0.213618 0.00593669811479465
0.214245 -0.0568431477216655
0.214872 -0.0914936916980641
0.215499 -0.0631110880248236
0.216126 0.0146898136762232
0.216753 0.0840732364054301
0.21738 0.0938823344909462
0.218006 0.045161286480785
0.218633 -0.0178419851989313
0.21926 -0.0551587940307139
0.219887 -0.060380742392027
0.220514 -0.0486053614946722
0.221141 -0.0306532403794201
0.221767 -0.00402212263315377
0.222394 0.0365315135481691
0.223021 0.0813754986847071
0.223648 0.0991296398895572
0.224275 0.0580086472367051
0.224902 -0.0347134041795479
0.225529 -0.121858632657067
0.226155 -0.14100887267662
};
\addlegendentry{S2}
\addplot [semithick, blue, mark=x, mark size=3, mark options={solid}, only marks, forget plot]
table {%
0.12819 0.297887006302497
};
\addplot [semithick, blue, mark=x, mark size=3, mark options={solid}, only marks, forget plot]
table {%
0.131987 0.955661147108489
};
\addplot [semithick, blue, mark=x, mark size=3, mark options={solid}, only marks, forget plot]
table {%
0.136417 1
};
\addplot [semithick, blue, mark=x, mark size=3, mark options={solid}, only marks, forget plot]
table {%
0.140214 0.413399751611637
};
\addplot [semithick, blue, mark=x, mark size=3, mark options={solid}, only marks, forget plot]
table {%
0.143378 0.210254472061395
};
\addplot [semithick, blue, mark=x, mark size=3, mark options={solid}, only marks, forget plot]
table {%
0.146542 0.488702154622175
};
\addplot [semithick, blue, mark=x, mark size=3, mark options={solid}, only marks, forget plot]
table {%
0.15034 0.557066332762476
};
\addplot [semithick, blue, mark=x, mark size=3, mark options={solid}, only marks, forget plot]
table {%
0.154137 0.389399946079863
};
\addplot [semithick, blue, mark=x, mark size=3, mark options={solid}, only marks, forget plot]
table {%
0.157934 0.454460144021368
};
\addplot [semithick, blue, mark=x, mark size=3, mark options={solid}, only marks, forget plot]
table {%
0.162364 0.702794707811077
};
\addplot [semithick, blue, mark=x, mark size=3, mark options={solid}, only marks, forget plot]
table {%
0.166794 0.801163448613406
};
\addplot [semithick, blue, mark=x, mark size=3, mark options={solid}, only marks, forget plot]
table {%
0.171224 0.492105891770775
};
\addplot [semithick, blue, mark=x, mark size=3, mark options={solid}, only marks, forget plot]
table {%
0.175654 0.229090754775196
};
\addplot [semithick, red, mark=x, mark size=3, mark options={solid}, only marks, forget plot]
table {%
0.131791 0.505886427538244
};
\addplot [semithick, red, mark=x, mark size=3, mark options={solid}, only marks, forget plot]
table {%
0.135552 1
};
\addplot [semithick, red, mark=x, mark size=3, mark options={solid}, only marks, forget plot]
table {%
0.139939 0.779227721891648
};
\addplot [semithick, red, mark=x, mark size=3, mark options={solid}, only marks, forget plot]
table {%
0.146232 0.411698279256762
};
\addplot [semithick, red, mark=x, mark size=3, mark options={solid}, only marks, forget plot]
table {%
0.150019 0.485534166425758
};
\addplot [semithick, red, mark=x, mark size=3, mark options={solid}, only marks, forget plot]
table {%
0.153806 0.32728598557572
};
\addplot [semithick, red, mark=x, mark size=3, mark options={solid}, only marks, forget plot]
table {%
0.157593 0.476208531881268
};
\addplot [semithick, red, mark=x, mark size=3, mark options={solid}, only marks, forget plot]
table {%
0.16138 0.465081902394015
};
\addplot [semithick, red, mark=x, mark size=3, mark options={solid}, only marks, forget plot]
table {%
0.165798 0.682115087891782
};
\addplot [semithick, red, mark=x, mark size=3, mark options={solid}, only marks, forget plot]
table {%
0.170216 0.630025078320641
};
\addplot [semithick, red, mark=x, mark size=3, mark options={solid}, only marks, forget plot]
table {%
0.175265 0.300931179066001
};
\addplot [semithick, red, mark=x, mark size=3, mark options={solid}, only marks, forget plot]
table {%
0.18347 0.242556707896825
};
\addplot [semithick, red, mark=x, mark size=3, mark options={solid}, only marks, forget plot]
table {%
0.189781 0.210745839562122
};
\addplot [semithick, red, mark=x, mark size=3, mark options={solid}, only marks, forget plot]
table {%
0.245609 0.271635820211377
};
\end{axis}

\end{tikzpicture}

%% file: Figures/raw_prediction.tex
\begin{tikzpicture}

\definecolor{darkgray176}{RGB}{176,176,176}
\definecolor{goldenrod}{RGB}{218,165,32}
\definecolor{lightgray204}{RGB}{204,204,204}
\definecolor{royalblue}{RGB}{65,105,225}
\definecolor{sienna}{RGB}{160,82,45}
\definecolor{lightgray204}{RGB}{204,204,204}
\definecolor{limegreen}{RGB}{50,205,50}

\begin{axis}[
height = 6cm,
width = 6cm,
legend cell align={left},
legend style={fill opacity=0.8, draw opacity=1, text opacity=1, draw=lightgray204},
tick align=outside,
tick pos=left,
x grid style={darkgray176},
xlabel={TDoA Sensor Pair 1-2 (\(\displaystyle \mu\) sec)},
every x tick scale label/.style={
    yshift=-2em,
    xshift= 4.5cm,
},
xmin=0, xmax=31000,
xtick style={color=black},
y grid style={darkgray176},
ylabel={TDoA Sensor Pair 1-3 (\(\displaystyle \mu\) sec)},
ymin=0, ymax=31000,
ytick style={color=black}
]
\addplot [draw=royalblue, fill=royalblue,, mark=*, only marks, opacity=0.3, mark size=1pt]
table{%
x  y
2895 5967
2927 6151
3431 5820
2876 5921
3551 5891
3078 5734
3405 5805
2947 5578
6619 9267
2616 28260
3520 4908
7370 12422
3077 4884
3529 6607
2838 5694
3198 6217
2814 5213
3493 6000
3409 5855
3497 5894
4121 5693
7558 9573
3337 5634
8048 13029
3025 4892
3501 4544
2818 5825
2918 6496
3489 5666
2264 5356
2902 5973
3528 6056
2915 5166
2783 5822
2851 5199
3511 5996
2957 5264
3589 5084
3397 5150
2884 6050
2958 6065
3592 6976
3512 6457
3727 6230
2960 5498
3586 6191
8000 10635
3607 6038
3559 5976
7257 11252
7296 11267
6722 16736
2348 6224
3454 6347
2977 6266
3689 6136
2995 6234
3517 5962
4022 5950
3743 6031
3129 5208
3908 5960
4516 6587
3658 5971
4212 5838
9212 12890
14729 18018
9264 12884
9790 12722
9311 13584
9323 13513
9352 13179
8705 8892
2918 9651
9330 26894
7944 17972
1704 5270
7837 11593
14191 18060
9825 12841
9893 12920
14149 17753
8958 12556
9189 13479
8511 12379
4054 8168
4767 7760
8562 12381
8862 8889
7845 11938
8088 8144
9311 13126
14871 18043
9805 13246
9185 12988
9190 13041
9382 13107
9212 12836
2205 12828
9191 12824
8570 17616
7959 11738
8421 11447
2181 11479
9208 13262
9932 12983
9217 12985
14925 18094
9797 13571
9278 12906
8743 13117
8616 12890
9254 12897
14252 17313
8582 11554
2447 6136
4297 17381
8579 12395
9131 12874
14142 17243
9827 12987
9491 12647
8503 12256
8645 12371
4088 7811
4689 8459
8518 13060
8695 12364
2420 9261
8169 9055
3573 7849
9074 12700
9058 12714
10307 13306
9690 13305
2342 5294
9093 12380
8574 8770
9104 12597
8563 12922
9323 12469
8089 11874
7946 13625
9206 13124
8757 13123
9952 13109
14380 18178
9204 12805
9194 12608
8618 13000
8579 16091
8577 12384
3583 7814
1352 4617
8139 11732
4380 10581
2806 7404
55123 55900
9904 17953
9295 13061
9981 13197
9212 12910
9287 16968
4895 8779
9259 13151
8288 12667
7855 8780
1093 11130
9296 12392
8670 12870
9128 12790
9118 17932
8623 12637
8532 12984
9122 12694
8536 12444
8567 8713
4082 4295
8649 17522
9308 12330
9277 12286
4816 17313
3574 7825
9268 12821
8496 12689
9698 12644
9321 13247
8820 13268
8458 12270
8459 12056
8069 8883
4434 7687
1592 11562
957 7281
7333 10957
9197 12881
3666 7625
9831 13001
9330 13045
9802 12891
9099 13415
4640 8295
2158 12692
7916 8106
9196 13228
1784 8122
4216 20049
2621 8832
2861 7944
9057 12930
9058 12859
9117 13371
9766 12691
9105 12309
8684 12529
8606 12793
10591 12253
15854 18030
8875 12123
4031 7703
870 16590
3614 12986
8175 12708
9950 13654
8769 12575
9345 13201
9346 13248
9318 13170
8444 8759
8583 11854
9504 12997
4233 7429
8177 8863
4124 7372
3602 12259
14725 17086
14135 16534
14151 16517
14302 17502
9825 13338
9256 13122
4789 5205
8693 11762
3544 12998
8175 11801
8055 11744
8019 11742
9129 12794
9851 13643
14837 17349
14873 17913
2217 12702
8489 12180
1737 8512
11141 14950
1627 8664
8890 13279
8443 12207
1095 17428
1769 8472
8054 17629
9983 12591
14885 17599
9858 17982
9314 12482
9209 12369
9242 13125
2233 5847
9398 13122
4801 8068
1891 5146
2324 2445
2757 12526
9772 12773
14404 17446
9430 13227
9191 17408
9940 13048
9349 13063
8510 12213
8557 12057
8516 12192
8271 11979
2872 12771
9130 12652
1231 7578
3474 13409
9838 12676
10454 12964
10438 12734
4370 13178
1821 12978
9221 12878
1068 1675
2208 5232
2230 26191
8571 12369
8116 8695
1721 17469
9194 13170
10539 13482
14490 17061
9901 13072
9338 13125
2367 9189
1892 12542
8571 12298
2317 5883
3453 12231
4111 8400
1572 1599
4082 4430
9319 13194
15113 17599
15024 18164
9795 13590
9950 13579
9857 12889
2245 9055
2379 12473
9244 12490
8512 44865
9309 9462
1742 8696
1793 4246
2339 17626
14774 17388
10525 13122
9709 12292
9172 12986
8616 9237
1662 12642
897 8375
1611 8430
10395 12527
1516 4554
1310 5140
2212 12169
};
\addlegendentry{Stiff 330 N/m}
\addplot [draw=goldenrod, fill=goldenrod, mark=*, only marks, opacity=0.3, mark size=1pt]
table{%
x  y
1566 7030
1729 7246
998 6501
887 6425
1572 6356
877 1763
1792 2533
1087 2184
1868 6817
1159 6404
902 7522
869 1712
1485 11853
14231 15456
9833 11088
9961 10582
5460 6460
9978 10719
2216 6466
1706 6180
1722 6730
1091 6139
1039 5907
1176 6250
1522 5859
1755 10331
9669 10775
9050 10797
9674 10618
1605 5973
985 6011
9331 10585
1498 5829
1580 5764
470 5712
1641 10374
2278 6552
1573 5559
1626 5931
9648 10709
9152 10291
9758 10941
9766 10189
9692 10813
1586 5844
2125 6518
1078 6113
1716 6753
1089 5569
1340 5905
1553 6192
1809 6405
9152 10574
14143 15324
9938 10563
9119 10374
1553 5984
4790 6078
1607 6568
1485 5988
1500 6566
8378 9828
1111 6606
1812 5721
1696 6408
14227 15427
14229 14995
9134 10579
9130 10567
5914 10570
1138 6763
1519 6003
1041 5789
1071 5518
394 4946
1536 5910
1092 5905
989 5866
9772 10891
9924 10506
9829 11086
9853 10543
6116 11200
1031 5878
1095 5855
1691 6659
1744 6396
1350 5702
1658 10221
1667 6324
1003 6380
14820 16059
9320 10842
9210 10981
4920 6292
5573 6144
5481 10627
1083 6096
1718 6186
5466 6662
1213 6251
1651 6077
910 5833
985 5941
9287 10506
9137 10467
9782 11052
9168 10240
9229 10322
5329 10249
1550 0
1831 5984
1411 6094
535 5174
1681 6212
758 5751
1723 6817
14139 15233
9901 10774
4630 5823
5385 6291
9696 10690
5256 6349
1553 6079
1578 6278
1069 6070
1424 5776
2212 5864
1622 5952
491 5975
9079 10876
9659 10877
9754 10874
5278 6222
5463 10424
1614 6037
5443 10998
1448 5725
1047 5657
1417 5632
1617 5722
5987 10645
1238 6339
14855 15637
9963 11256
5567 6600
1116 6592
1688 6274
6214 11075
9356 10053
1383 6186
2239 6290
13644 15597
1093 5589
1086 5499
456 5527
3929 8834
9951 11147
9721 10906
9767 10329
9679 10872
1086 6310
1536 5591
1321 6161
1708 6292
1870 6036
5679 10294
1473 6160
1524 9858
9738 10341
9281 10771
9864 10926
1691 6536
5928 10854
5955 10786
1091 6146
888 5678
1347 6050
1407 10668
1076 5524
1261 6023
1899 6707
9870 10487
9914 10380
9088 10357
9057 10162
4914 5561
1171 6206
4140 5537
1038 6188
1505 6162
1616 10181
879 5559
839 5562
1020 5834
9345 10095
9747 10798
9117 10586
9198 9936
5387 6128
5452 5908
5292 10232
5397 5735
1526 5794
8902 10731
6123 9962
857 5663
5332 10812
9773 10550
9904 11056
4948 5613
5801 10774
5586 10700
1821 6392
1146 6184
810 5986
1821 6214
1146 10789
5392 9841
507 5597
951 5874
9835 10437
10426 11074
9102 9755
9128 10201
5382 5775
437 5852
4838 9970
4769 5548
1635 9893
846 10315
1047 5519
1603 10387
1548 9767
9931 11129
9878 11333
9820 11075
10134 10838
1185 5998
9365 10036
888 5804
2178 6401
1222 6317
1648 6112
1630 5944
979 5921
360 5292
14060 15381
9976 11043
9830 10638
4910 6177
5678 6183
1483 9970
1077 5994
1060 6016
1093 5832
1039 5901
1571 6460
1729 6538
1134 5522
};
\addlegendentry{Stiff 410 N/m}
\addplot [draw=sienna, fill=sienna, mark=*, only marks, opacity=0.3, mark size=1pt]
table{%
x  y
5585 6514
5912 10040
5286 10148
5414 10252
5910 10182
5520 5594
5910 15079
5448 10341
5349 6543
5461 31489
5590 6828
5549 6812
5278 5763
5996 10814
6029 10902
6646 10798
5267 5348
4875 4955
5374 16116
5460 28011
5334 10804
5449 14708
5415 10380
5428 7200
2375 6188
5056 6350
5443 5452
5407 5534
5551 5714
6075 6192
6119 10029
6036 20976
1638 11550
5303 10786
4889 9905
5022 5867
5360 10848
1661 6384
1956 6523
6104 11094
5913 10121
5381 10408
5265 5614
5668 5690
6080 33795
6192 11180
6111 11019
5949 11068
1789 1893
5581 6271
1655 9880
4648 9298
1146 1313
5547 5961
1858 6203
6004 9877
6196 10118
6043 16152
1803 11914
5574 5810
4811 10440
1491 5509
1628 4630
1252 6869
2274 7102
6027 6328
6118 6657
5725 6035
5711 5824
5440 9855
5128 15860
4736 14663
5378 11159
1436 10404
5283 5990
1709 10646
5481 6716
1626 6052
5586 5796
6060 6566
1597 5764
5338 9718
6320 9952
5448 16215
5459 11935
1018 7108
5428 10397
5471 11290
1655 6799
5513 6117
2404 25665
9949 10503
6238 10885
5724 5978
5047 5237
5987 16476
5456 11750
1203 7617
5646 11365
5342 10950
5772 6931
5514 10628
5698 7501
1826 6861
5850 6159
10407 10547
5279 5303
6420 17620
5440 12396
6083 12397
6029 11649
5398 7696
5983 11184
1060 7121
1676 6561
5483 5997
5350 6537
6148 6201
6796 10595
6055 9875
5546 8787
10493 21252
5320 11283
5583 7902
5440 11444
5493 11213
5979 9916
5356 10755
6026 16573
6049 9874
6117 11122
5477 6316
6040 6550
4836 4957
5519 16527
5378 15546
5454 11150
5478 11203
5392 11094
1613 5254
5692 10697
5548 6851
6059 6795
10479 10999
5394 6001
5489 5765
6031 9771
5352 16574
5282 11555
5392 8539
5283 11501
5576 14797
5477 6287
5344 10794
5479 6675
5325 5884
6168 11142
5974 6216
10440 10478
5343 9593
5566 16239
5458 11669
5976 11533
5957 12064
4874 6631
2131 5845
1669 7286
5378 10451
5354 15523
11040 11055
6081 6447
5931 10055
6102 9908
6011 8962
5977 12092
5528 11767
5936 11500
5430 7455
5834 10386
5362 6540
4843 10183
5318 10247
5994 10920
5397 6033
5483 6060
5451 9798
5436 16622
5515 11082
6004 11534
5394 11140
5889 11479
5801 6820
1095 4053
5505 10524
5405 7172
6113 11020
5374 6262
5365 5643
6249 10667
5386 20455
5315 11809
5219 11152
6012 11901
4830 10977
1721 4875
1623 9762
5379 6606
5341 6443
6072 11074
6304 11336
6019 6420
5659 10020
5443 5570
4904 11946
5309 11356
5608 11870
5448 7370
5805 5922
1770 6838
5565 6840
5489 6814
6021 6182
6174 11207
6031 10453
6144 10500
5606 27612
5581 5644
5527 7411
5474 7262
5319 7312
5191 6421
5905 6752
5389 7075
1797 7095
6608 6945
5514 6235
5427 5593
6016 18022
6192 27536
6189 12454
5413 11707
5752 11397
5530 7885
9819 11578
5380 6795
5391 6880
5326 6466
5425 5941
9798 10232
6150 6314
5405 27355
6017 26534
5877 16474
6021 11431
6031 11384
5593 11297
9842 11263
1566 6518
5426 9701
5376 6504
5858 6157
5980 6476
5953 10796
5685 5736
5527 19972
5985 12653
5290 7778
5334 7726
5869 10780
1538 6431
1647 7241
5560 6886
5600 6574
10425 11057
5373 5898
5384 5837
5987 10362
5545 26796
5982 15849
5974 11488
5448 10965
4679 6980
5752 7115
5996 7274
6021 7151
2293 7059
6113 10938
5561 6356
6143 6299
6195 10698
5591 21055
5428 11751
6214 11823
6111 11824
5504 7436
1580 6532
4860 7290
1984 6386
1595 3199
};
\addlegendentry{Stiff 510 N/m}
\addplot [thick, limegreen]
table {%
0 6000
31000 6000
};
\addlegendentry{PCS}

\addplot [royalblue, forget plot]
table {%
8500 -9.09494701772928e-13
8500 31000
};
\addplot [royalblue, forget plot]
table {%
0 12749
31000 12749
};
\addplot [goldenrod, forget plot]
table {%
7639 -9.09494701772928e-13
7639 31000
};
\addplot [goldenrod, forget plot]
table {%
0 11458
31000 11458
};
\addplot [sienna, forget plot]
table {%
6850 -9.09494701772928e-13
6850 31000
};
\addplot [sienna, forget plot]
table {%
0 10276
31000 10276
};
\addplot [thick, limegreen, forget plot]
table {%
20000 -9.09494701772928e-13
20000 31000
};
\addplot [thick, limegreen, forget plot]
table {%
0 30000
31000 30000
};
\addplot [thick, limegreen, forget plot]
table {%
5000 -9.09494701772928e-13
5000 31000
};
\end{axis}

\end{tikzpicture}

%% file: Figures/corridor_prediction.tex
\begin{tikzpicture}

\definecolor{darkgrey176}{RGB}{176,176,176}
\definecolor{goldenrod}{RGB}{218,165,32}
\definecolor{lightgrey204}{RGB}{204,204,204}
\definecolor{limegreen}{RGB}{50,205,50}
\definecolor{royalblue}{RGB}{65,105,225}
\definecolor{sienna}{RGB}{160,82,45}

\begin{axis}[
height = 6cm,
width = 6cm,
legend cell align={left},
legend style={
  fill opacity=0.8,
  draw opacity=1,
  text opacity=1,
  at={(0.97,0.03)},
  anchor=south east,
  draw=lightgrey204
},
tick align=outside,
tick pos=left,
x grid style={darkgrey176},
xlabel={TDoA Sensor Pair 1-2 (\(\displaystyle \mu\) sec)},
every x tick scale label/.style={
    yshift=-2em,
    xshift= 4.5cm,
},
xmin=0, xmax=31000,
xtick style={color=black},
y grid style={darkgrey176},
ylabel={TDoA Sensor Pair 1-3 (\(\displaystyle \mu\) sec)},
ymin=0, ymax=31000,
ytick style={color=black}
]
\addplot [draw=royalblue, fill=royalblue, mark=*, only marks, opacity=0.3, mark size=1pt]
table{%
x  y
4280 6726
7917 10988
4145 6460
7991 10529
8618 11223
8625 11260
4089 6189
7844 11341
4212 6464
4252 6545
9098 12988
9264 12884
4288 6611
9323 12878
9352 13179
8705 12633
8687 13089
8574 12309
8470 12226
9191 13911
8601 12871
8496 12752
9582 13180
7332 10885
8511 12379
13463 17613
8550 12152
8584 11715
8862 8889
8478 12571
8716 23291
7900 12793
9231 13009
8606 13049
7982 13079
9382 12479
9816 13449
9195 13156
7959 11738
8421 11447
8680 12921
8708 12979
9217 12985
9309 13088
9164 12946
9910 12904
9248 13522
9254 12897
8626 12918
8582 11554
4297 7299
14198 17332
7879 12231
8507 12226
9194 12987
13515 17268
9088 13490
8518 13060
9327 12996
8169 12183
14112 17736
9690 13305
8471 8641
8574 12592
8478 12597
8563 12922
8696 12469
8716 11874
9206 13631
7889 11701
8752 13160
9360 13759
14259 17860
9194 12608
13610 24930
9209 13016
8168 11209
4079 7762
8672 13464
9295 13061
8817 13831
7290 9863
14335 22016
9259 13151
9296 12392
9110 13384
7826 12801
8537 12978
8567 12463
8478 9320
8649 9357
9277 12286
9254 11636
14249 14620
9032 13230
8195 13268
9082 12266
9084 12681
8069 8883
8539 12195
6653 26658
7954 22722
8729 12688
8620 13652
14864 17953
9099 12790
14119 17766
9084 12692
8539 8729
8595 12587
4216 26314
8300 8832
9873 26854
8459 12913
8488 12026
14110 17314
8606 12793
9959 12253
15854 18030
9508 13385
13476 29820
8466 13248
9321 13654
8775 13208
9345 12572
9318 13170
4669 9392
8347 12380
9296 12492
8177 12661
8694 13120
7768 12585
8481 12767
9111 12742
8664 13084
14862 18375
9256 13122
9326 18070
8541 17363
8652 12375
13506 29720
8610 13589
8579 12925
9205 13501
13497 21579
8633 12443
13955 18344
9076 12840
7961 10563
8601 12494
8602 13208
9947 13115
9242 13125
9855 13122
9205 11983
8165 13141
9191 12981
6785 13030
9135 12838
9182 12682
8516 12192
8271 11979
9130 12652
8129 8211
9098 13393
8621 12975
9200 13391
6755 9677
8571 12369
8743 13728
8606 12438
9230 22234
8655 12898
8862 13262
9345 13758
8571 12298
8516 12254
9735 13907
8559 12943
9795 13590
9950 13579
8512 11718
14988 18903
8587 12338
8019 13092
9074 10410
9797 13611
8616 9237
13469 17792
9196 12834
8642 13164
7591 7647
};
\addlegendentry{Stiff 330 N/m}
\addplot [draw=goldenrod, fill=goldenrod, mark=*, only marks, opacity=0.3, mark size=1pt]
table{%
x  y
7150 12058
6574 7460
6898 7442
7407 7654
7808 11688
7787 12478
6819 12287
7872 12122
6745 11219
7379 11759
8642 11196
7919 11531
8075 10640
6765 10965
7834 11454
7303 11071
7452 11210
8439 10224
7853 10816
6726 10719
7941 11006
14272 16638
8547 10570
7224 11497
7338 10940
9061 10815
7282 11544
7188 10951
8618 10522
7031 11596
8469 11258
6854 11408
14183 16533
7418 11262
7249 11051
8413 10998
8424 10970
8382 10453
14303 16724
8706 10751
8034 10777
14134 15579
7172 11195
6801 11796
14130 15469
6060 10808
7961 9910
7339 11262
8444 10911
6684 11561
6728 11125
9859 11176
6749 11208
7351 11567
6136 10896
8642 10753
8306 10737
14844 16957
7878 10819
7378 11262
6749 10963
6223 10726
6851 11191
7365 11747
9246 11226
9223 11058
14539 16402
6014 11079
5933 10856
14879 16647
14775 16519
9160 10419
5954 10874
6422 11112
5580 10219
8590 10614
14638 17106
8623 11212
7992 11622
5956 10887
6651 11354
6545 11323
7166 12045
6616 11142
14809 16316
8585 11085
7120 10837
6783 11640
5941 7150
14723 15941
6564 11286
14913 16139
6511 10788
6120 11353
6474 11312
9193 10747
8015 11113
15017 16310
9329 11036
6173 11014
6725 11308
9363 10686
8865 10623
9183 11365
7987 11198
7345 11216
14267 16200
14254 16213
6661 10412
6809 11087
6533 11521
6565 11545
6086 11310
7186 10607
6961 11172
14292 15683
6312 10296
7798 11222
14751 16168
14158 15903
6786 11360
6669 10896
6691 10907
6553 10851
6580 11411
6110 11165
14134 15770
7680 11302
14270 16227
8790 11101
8133 11123
6101 10567
6581 10973
6151 10561
6827 11235
13519 15555
7785 10593
9752 11442
9069 10624
6487 11210
5421 10875
7942 10441
7181 11384
9826 10916
9189 10799
6565 10833
14525 16387
14845 17140
9130 11277
14410 15075
10215 10777
6218 10706
14998 16446
6818 11227
6475 11026
5393 10466
8658 10648
12873 14642
9315 11263
6712 11134
7232 10799
6062 10845
9219 10595
9137 10536
9143 10526
6742 11214
14162 20464
7033 11114
14943 16398
7475 11654
9371 10669
10356 11500
9052 11471
9167 10502
8508 11015
8474 10940
8551 10325
14256 16175
6674 10697
6203 10608
6327 11231
6148 11057
9164 10433
8574 10894
6734 10969
8598 10995
6791 11179
6837 11450
};
\addlegendentry{Stiff 410 N/m}
\addplot [draw=sienna, fill=sienna, mark=*, only marks, opacity=0.3, mark size=1pt]
table{%
x  y
5910 10696
5918 10780
5414 10252
5910 10182
6543 10639
5349 6543
4961 6199
4922 6812
6029 10902
5999 10811
9250 9330
5374 9837
9210 9596
9708 15178
9825 11013
5428 6567
5513 6188
9424 10756
5475 6405
6093 6214
6119 10029
6032 9645
6099 10822
5567 6860
6546 10754
5381 10408
5933 10678
10101 10123
6080 11086
5564 11180
6111 11019
5540 6332
4955 6271
6077 10424
6280 10625
6056 9945
6043 9887
5609 10654
4811 6045
5014 6869
6067 7102
5529 6040
6356 6666
5754 16486
5467 6124
6060 6566
6661 10828
5338 10348
6320 10585
5393 11483
9875 11289
5423 7432
5513 6749
9824 11745
6238 10885
10147 10401
5352 16476
4832 8004
5579 11993
5342 10950
5772 6931
5698 7501
10018 11283
10850 15739
10504 17460
5396 11649
6031 7699
5982 11817
5350 6537
5614 12264
6168 6192
6726 10543
6172 9413
6098 9880
9695 15658
6232 12323
5979 9916
5989 15187
5392 12773
6139 6351
6040 6550
9836 9957
10568 17190
5378 8630
5454 11150
5697 11330
6059 7425
6139 6721
5394 6001
5491 6390
5406 9771
9784 13497
10345 16618
5477 6287
5513 6059
6047 6261
6052 6082
5566 9944
5342 11533
6010 6089
5354 15523
6081 6447
5378 8962
5977 11462
5306 11500
5430 6831
6467 11019
5440 6830
5483 6060
6076 10423
5436 9689
5515 15513
5371 11534
5394 11140
5889 11479
5166 6820
5505 6118
5405 7172
10431 11328
5620 10667
5386 9776
5914 16184
5853 6759
6012 7484
6304 11336
6032 6442
5664 6260
9279 9437
5935 11982
6240 16281
5448 7370
5565 6840
5489 6814
5642 6610
6174 11207
6081 6131
10575 14931
5581 6270
6191 12474
4850 7262
5319 6690
5191 6421
6022 7708
6147 6868
6060 6226
6016 6097
6192 11209
6222 11836
6045 11710
5530 7251
9819 11578
5380 6795
5391 6256
5951 6464
5315 6428
6042 6437
6150 6314
10459 15364
6017 9544
5877 12061
6064 15863
5593 11297
5937 10879
5980 6476
5953 10796
9902 9933
5966 11974
5334 10858
5239 10780
5961 7064
6193 7519
5600 7210
5466 11556
6008 10898
5987 10362
6170 12304
5350 11443
5974 11488
5312 7613
5752 7115
5386 6526
10623 10790
6189 10721
6195 10698
6223 16643
6117 16815
5588 11196
5504 7436
5394 6364
5395 6142
};
\addlegendentry{Stiff 510 N/m}
\addplot [thick, red, dashed]
table {%
3000 0
23000 30000
};
\addlegendentry{CCS}
\addplot [thick, limegreen]
table {%
0 6000
31000 6000
};
\addlegendentry{PCS}
\addplot [thick, red,dashed, forget plot]
table {%
0 4500
16500 30000
};
\addplot [royalblue, forget plot]
table {%
8500 -9.09494701772928e-13
8500 31000
};
\addplot [royalblue, forget plot]
table {%
0 12749
31000 12749
};
\addplot [goldenrod, forget plot]
table {%
7639 -9.09494701772928e-13
7639 31000
};
\addplot [goldenrod, forget plot]
table {%
0 11458
31000 11458
};
\addplot [sienna, forget plot]
table {%
6850 -9.09494701772928e-13
6850 31000
};
\addplot [sienna, forget plot]
table {%
0 10276
31000 10276
};
\addplot [thick, limegreen, forget plot]
table {%
20000 -9.09494701772928e-13
20000 31000
};

\addplot [thick, limegreen, forget plot]
table {%
0 30000
31000 30000
};
\addplot [thick, limegreen, forget plot]
table {%
4000 -9.09494701772928e-13
4000 31000
};

\addplot [thick, red]
table {%
0 0
20000 30000
};
\end{axis}

\end{tikzpicture}

%% file: Figures/filtered_prediction.tex
\begin{tikzpicture}

\definecolor{darkgrey176}{RGB}{176,176,176}
\definecolor{goldenrod}{RGB}{218,165,32}
\definecolor{lightgrey204}{RGB}{204,204,204}
\definecolor{limegreen}{RGB}{50,205,50}
\definecolor{royalblue}{RGB}{65,105,225}
\definecolor{sienna}{RGB}{160,82,45}

\begin{axis}[
height = 6cm,
width = 6cm,
legend cell align={left},
legend style={
  fill opacity=0.8,
  draw opacity=1,
  text opacity=1,
  at={(0.97,0.03)},
  anchor=south east,
  draw=lightgrey204
},
tick align=outside,
tick pos=left,
x grid style={darkgrey176},
xlabel={TDoA Sensor Pair 1-2 (\(\displaystyle \mu\) sec)},
every x tick scale label/.style={
    yshift=-2em,
    xshift= 4.5cm,
},
xmin=0, xmax=31000,
xtick style={color=black},
y grid style={darkgrey176},
ylabel={TDoA Sensor Pair 1-3 (\(\displaystyle \mu\) sec)},
ymin=0, ymax=31000,
ytick style={color=black}
]
\addplot [draw=royalblue, fill=royalblue, mark=*, only marks, opacity=0.3, mark size=1pt]
table{%
x  y
4280 6726
7917 10988
4145 6460
7991 10529
8618 11223
8625 11260
4089 6189
7844 11341
4212 6464
4252 6545
9098 12988
9264 12884
4288 6611
9323 12878
9352 13179
8705 12633
8687 13089
8574 12309
8470 12226
9191 13911
8601 12871
8496 12752
9582 13180
7332 10885
8511 12379
13463 17613
8550 12152
8584 11715
8478 12571
7900 12793
9231 13009
8606 13049
7982 13079
9382 12479
9816 13449
9195 13156
7959 11738
8421 11447
8680 12921
8708 12979
9217 12985
9309 13088
9164 12946
9910 12904
9248 13522
9254 12897
8626 12918
8582 11554
4297 7299
14198 17332
7879 12231
8507 12226
9194 12987
13515 17268
9088 13490
8518 13060
9327 12996
8169 12183
14112 17736
9690 13305
8574 12592
8478 12597
8563 12922
8696 12469
8716 11874
9206 13631
7889 11701
8752 13160
9360 13759
14259 17860
9194 12608
9209 13016
8168 11209
4079 7762
8672 13464
9295 13061
8817 13831
7290 9863
14335 22016
9259 13151
9296 12392
9110 13384
7826 12801
8537 12978
8567 12463
8478 9320
8649 9357
9277 12286
9254 11636
9032 13230
8195 13268
9082 12266
9084 12681
8069 8883
8539 12195
8729 12688
8620 13652
9099 12790
14119 17766
9084 12692
8595 12587
8300 8832
8459 12913
8488 12026
14110 17314
8606 12793
9959 12253
9508 13385
8466 13248
9321 13654
8775 13208
9345 12572
9318 13170
4669 9392
8347 12380
9296 12492
8177 12661
8694 13120
7768 12585
8481 12767
9111 12742
8664 13084
14862 18375
9256 13122
8652 12375
8610 13589
8579 12925
9205 13501
13497 21579
8633 12443
13955 18344
9076 12840
7961 10563
8601 12494
8602 13208
9947 13115
9242 13125
9855 13122
9205 11983
8165 13141
9191 12981
6785 13030
9135 12838
9182 12682
8516 12192
8271 11979
9130 12652
8129 8211
9098 13393
8621 12975
9200 13391
6755 9677
8571 12369
8743 13728
8606 12438
8655 12898
8862 13262
9345 13758
8571 12298
8516 12254
9735 13907
8559 12943
9795 13590
9950 13579
8512 11718
14988 18903
8587 12338
8019 13092
9074 10410
9797 13611
8616 9237
13469 17792
9196 12834
8642 13164
7591 7647
};
\addlegendentry{Stiff 330 N/m}
\addplot [draw=goldenrod, fill=goldenrod, mark=*, only marks, opacity=0.3, mark size=1pt]
table{%
x  y
7150 12058
6574 7460
6898 7442
7407 7654
7808 11688
7787 12478
6819 12287
7872 12122
6745 11219
7379 11759
8642 11196
7919 11531
8075 10640
6765 10965
7834 11454
7303 11071
7452 11210
8439 10224
7853 10816
6726 10719
7941 11006
8547 10570
7224 11497
7338 10940
9061 10815
7282 11544
7188 10951
8618 10522
7031 11596
8469 11258
6854 11408
7418 11262
7249 11051
8413 10998
8424 10970
8382 10453
8706 10751
8034 10777
7172 11195
6801 11796
6060 10808
7961 9910
7339 11262
8444 10911
6684 11561
6728 11125
9859 11176
6749 11208
7351 11567
6136 10896
8642 10753
8306 10737
7878 10819
7378 11262
6749 10963
6223 10726
6851 11191
7365 11747
9246 11226
9223 11058
6014 11079
5933 10856
9160 10419
5954 10874
6422 11112
5580 10219
8590 10614
8623 11212
7992 11622
5956 10887
6651 11354
6545 11323
7166 12045
6616 11142
8585 11085
7120 10837
6783 11640
5941 7150
6564 11286
6511 10788
6120 11353
6474 11312
9193 10747
8015 11113
9329 11036
6173 11014
6725 11308
9363 10686
8865 10623
9183 11365
7987 11198
7345 11216
6661 10412
6809 11087
6533 11521
6565 11545
6086 11310
7186 10607
6961 11172
6312 10296
7798 11222
6786 11360
6669 10896
6691 10907
6553 10851
6580 11411
6110 11165
7680 11302
8790 11101
8133 11123
6101 10567
6581 10973
6151 10561
6827 11235
7785 10593
9752 11442
9069 10624
6487 11210
5421 10875
7942 10441
7181 11384
9826 10916
9189 10799
6565 10833
9130 11277
6218 10706
6818 11227
6475 11026
5393 10466
8658 10648
9315 11263
6712 11134
7232 10799
6062 10845
9219 10595
9137 10536
9143 10526
6742 11214
14162 20464
7033 11114
7475 11654
9371 10669
9052 11471
9167 10502
8508 11015
8474 10940
8551 10325
6674 10697
6203 10608
6327 11231
6148 11057
9164 10433
8574 10894
6734 10969
8598 10995
6791 11179
6837 11450
};
\addlegendentry{Stiff 410 N/m}
\addplot [draw=sienna, fill=sienna, mark=*, only marks, opacity=0.3, mark size=1pt]
table{%
x  y
5910 10696
5918 10780
5414 10252
5910 10182
6543 10639
5349 6543
4961 6199
4922 6812
6029 10902
5999 10811
5374 9837
9708 15178
9825 11013
5428 6567
5513 6188
9424 10756
5475 6405
6093 6214
6119 10029
6032 9645
6099 10822
5567 6860
6546 10754
5381 10408
5933 10678
6080 11086
5564 11180
6111 11019
5540 6332
4955 6271
6077 10424
6280 10625
6056 9945
6043 9887
5609 10654
4811 6045
5014 6869
6067 7102
5529 6040
6356 6666
5467 6124
6060 6566
6661 10828
5338 10348
6320 10585
5393 11483
9875 11289
5423 7432
5513 6749
9824 11745
6238 10885
4832 8004
5579 11993
5342 10950
5772 6931
5698 7501
10018 11283
10850 15739
10504 17460
5396 11649
6031 7699
5982 11817
5350 6537
5614 12264
6168 6192
6726 10543
6172 9413
6098 9880
9695 15658
6232 12323
5979 9916
6139 6351
6040 6550
10568 17190
5378 8630
5454 11150
5697 11330
6059 7425
6139 6721
5394 6001
5491 6390
5406 9771
9784 13497
10345 16618
5477 6287
5513 6059
6047 6261
6052 6082
5566 9944
5342 11533
6010 6089
6081 6447
5378 8962
5977 11462
5306 11500
5430 6831
6467 11019
5440 6830
5483 6060
6076 10423
5436 9689
5371 11534
5394 11140
5889 11479
5166 6820
5505 6118
5405 7172
5620 10667
5386 9776
5853 6759
6012 7484
6304 11336
6032 6442
5664 6260
5935 11982
5448 7370
5565 6840
5489 6814
5642 6610
6174 11207
6081 6131
10575 14931
5581 6270
6191 12474
4850 7262
5319 6690
5191 6421
6022 7708
6147 6868
6060 6226
6016 6097
6192 11209
6222 11836
6045 11710
5530 7251
9819 11578
5380 6795
5391 6256
5951 6464
5315 6428
6042 6437
6150 6314
10459 15364
6017 9544
5877 12061
5593 11297
5937 10879
5980 6476
5953 10796
5966 11974
5334 10858
5239 10780
5961 7064
6193 7519
5600 7210
5466 11556
6008 10898
5987 10362
6170 12304
5350 11443
5974 11488
5312 7613
5752 7115
5386 6526
6189 10721
6195 10698
5588 11196
5504 7436
5394 6364
5395 6142
};
\addlegendentry{Stiff 510 N/m}
\addplot [thick, red, dashed]
table {%
3000 0
23000 30000
};
\addlegendentry{CCS}
\addplot [thick, limegreen]
table {%
0 6000
31000 6000
};
\addlegendentry{PCS}
\addplot [thick, red,dashed, forget plot]
table {%
0 4500
16500 30000
};
\addplot [royalblue, forget plot]
table {%
8500 -9.09494701772928e-13
8500 31000
};
\addplot [royalblue, forget plot]
table {%
0 12749
31000 12749
};
\addplot [goldenrod, forget plot]
table {%
7639 -9.09494701772928e-13
7639 31000
};
\addplot [goldenrod, forget plot]
table {%
0 11458
31000 11458
};
\addplot [sienna, forget plot]
table {%
6850 -9.09494701772928e-13
6850 31000
};
\addplot [sienna, forget plot]
table {%
0 10276
31000 10276
};
\addplot [thick, limegreen, forget plot]
table {%
20000 -9.09494701772928e-13
20000 31000
};

\addplot [thick, limegreen, forget plot]
table {%
0 30000
31000 30000
};
\addplot [thick, limegreen, forget plot]
table {%
4000 -9.09494701772928e-13
4000 31000
};

\addplot [thick, red]
table {%
0 0
20000 30000
};
\end{axis}

\end{tikzpicture}

%% file: Evaluation.tex
\section{System Evaluation}
\label{sec:eval}
We evaluate \name through three sets of experiments: (1) Sensing capability characterization on multiple tissue simulants prepared using ballistic gel to control stiffness level and ranges; (2) On-body data collection with multiple users across different muscle groups (thigh, trapezius, and forearm); (3) Continuous monitoring experiment on a user during a workout session.  

\subsection{Prototype Implementation}
We customize printable circuit boards (PCBs) for reliable, comfortable data collection. 
The sensing platform is built around the ESP32-S3 Mini N8 microcontroller unit (MCU) \cite{ESP32s3-2026Feb}, which performs actuator control, signal acquisition, and wireless communication.
The platform is designed to interface with two actuators and six IMU sensors, enabling data collection across various muscle groups.
For example, the symmetric design allows simultaneous data collection on both sides of the trapezius muscles.
A GD-01 bone conduction speaker \cite{Industries2026Feb} with a MAX98357A amplifier \cite{MAX98357A_2026Feb} is adopted as the shear wave generating actuator. 
BMI270 triaxial Inertial Measurement Units (IMUs) \cite{Bosh} are used for capturing shear wave signals.
The system is driven by a 5V Lithium-Ion battery and supports wireless communication via the User Datagram Protocol (UDP). 

Given the expected shear-wave velocity range in skeletal muscle (2–8 m/s) \cite{gennisson2010viscoelastic} and the typical length of target muscles (10–20 cm) \cite{shier1996human}, we design the sensor spacing to provide sufficient propagation distance for reliable TDoA estimation while maintaining compatibility with wearable deployment on common muscle groups.
The distances between the bone-conduction transducer and three IMU sensors are $d_1 = 2.5$ cm, $d_2 = 6.5$ cm, and $d_3 = 8.5$ cm, respectively.
Based on prior literature, shear waves at frequencies between {$200$} and {$300$} Hz \cite{tsuchida2020application} is ideal for vibroacoustic muscle sensing.
The bone-conduction transducer (BCT) generates excitation impulses at frequencies ranging from 200 Hz to 350 Hz in 10 Hz increments.
The signal frequency determines how deeply the shear wave will travel, allowing \name to effectively capture stiffness.
In each trial, the BCT produces $16$ pulses, each a 45-millisecond impulse, as the excitation signal.
Excitation signals are separated by 200 ms intervals to ensure one actuation to completely dissipate and do not overlap with succeeding signals.

\subsection{Experiment Design}\label{sec:exp_design}
We conduct a series of experiments on both tissue simulants and human subjects. This section describes the data collection and ground-truth collection used throughout the evaluation, as well as the evaluation metrics.

\begin{figure*}
    \centering
    \includegraphics[width=0.9\linewidth]{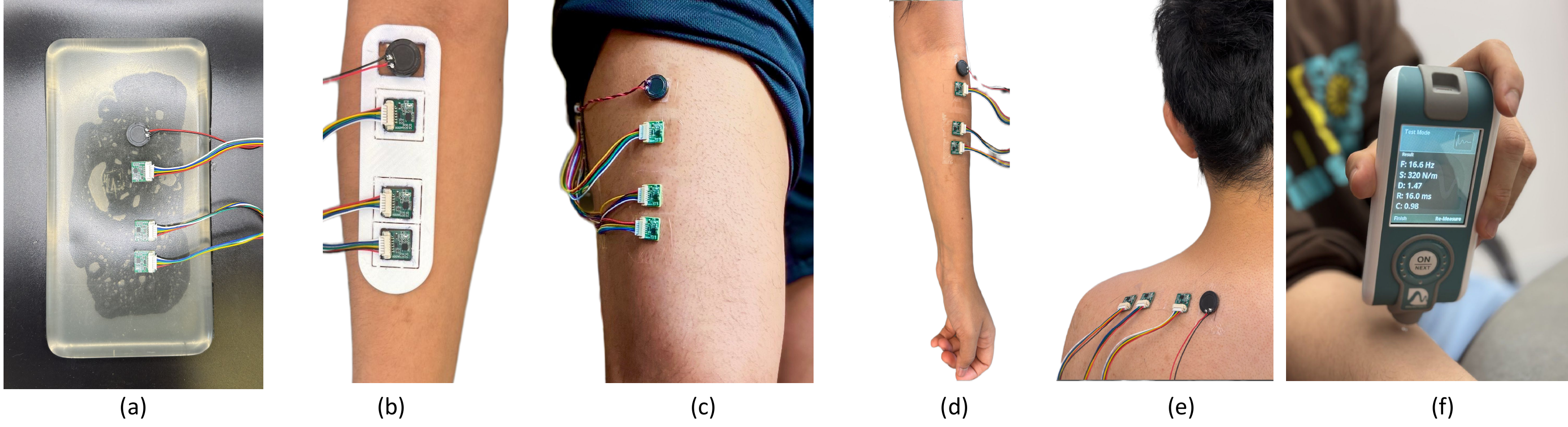}
    \caption{Experiment settings. (a) shows the system installation on the ballistic gel for controlled characterization of sensing capability. (b) depicts the 3D-printed installation guide tool used during on-body system installation to ensure consistent sensor placement.
    (c-e) present the data collection deployment for different muscles (thigh, forearm, and trapezius).
    (f) shows the ground truth collection using Myoton Pro \cite{Myoton026Feb}.}
    \label{fig:deployment}
\end{figure*}

\paragraph{Data Collection Procedures}
We use the following data collection procedure for all experiments in the subsequent sections. 
First, the system will be installed on the surface of the target soft tissue.
Figure \ref{fig:deployment} shows the different deployments of \name system across tissue simulant gels and different human muscles.

For experiments with tissue simulant gels, we use tissue simulants of three different stiffnesses. Each gel block has a length, width, and height of $220$ mm, $120$ mm, and $25$ mm, respectively.
The three tissue simulants samples (Sim Gel 1,2,3) are fabricated with the ground truth stiffness measured of $393 $ N/m (or $31.5$ kPa),  $593.4$ N/m, and $713$ N/m, which simulate stiffness to that of resting muscle,  moderately stiffer muscle, and high stiffness muscle that is prone to injury, respectively. 
We collected $200+$ trials for each simulant gel sample. 

All human-subject recruitment, enrollment, and data collection procedures were conducted in accordance with an Institutional Review Board (IRB)-approved protocol.
We use a flexible 3D-printed stencil to minimize distance variance across deployments caused by human error, which is demonstrated in Figure \ref{fig:deployment} (b).
\name actuator and sensors are deployed in a line aligning with the underlying muscle fibers' direction.
To measure resting muscle stiffness, data were collected while participants maintained relaxed standing or seated postures and avoided voluntary contraction of the target muscle. 
To ensure participant comfort and compliance, participants were allowed to select the muscle groups to be measured. 
In total, 14 participants were enrolled in the study, including 11 males and 3 females. 
For each muscle group investigated, measurements were collected from at least 5 participants, as summarized in Table \ref{tab:participants}.

\begin{table}[!t]
\centering
\caption{Human-subject data collection summary.}
\vspace{-1ex}
\label{tab:participants}
\begin{tabular}{lcc}
\toprule
Experiment & Participant ID & Measurements \# \\
\midrule
Thigh  & 1-5 (5M) & 2000 \\
Trapezius  & 2,6-9 (2M+3F)& 2000 \\
Forearm & 1,10-13 (5M) & 2000 \\
Cross-day (forearm) & 11 & 1700\\
Deployment quality &  3,5,6,11,12,14 & 500\\
\midrule
Total & - & 8300\\
\bottomrule
\end{tabular}
\end{table}


\paragraph{Ground Truth Collection Procedures}
We utilize the Myoton Pro\cite{Myoton026Feb} palpation device to collect ground truth stiffness to verify and evaluate \name's stiffness estimation capability.
Figure \ref{fig:deployment} (f) demonstrates an example session of the ground truth collection procedure. 
A single stiffness reading from the Myoton Pro may be subject to measurement variability due to placement angle and pressure. 
To ensure an accurate stiffness ground truth, we collect 5 or more stiffness readings from the Myoton Pro and calculate the aggregated mean stiffness for each data collection instance.

\paragraph{Evaluation Metrics}
We compare the output of \name system, which is the stiffness estimated with the aggregated TDoA estimation from multiple measurement events. 
Because \name rejects uncertain measurements based on physical knowledge and signal quality assessment to ensure estimation accuracy, we consider (1) estimation error, and (2) rejection rate, as the major evaluation metrics that reflects the system's performance. 
We report the error percentage calculated as $\frac{|S_{est} - S_{GT}|}{S_{GT}}$ to reflect practical estimation quality -- the estimation error within $30\%$ is considered good and acceptable for non-clinical use. \cite{warneke2025discussion}.

\subsection{Characterization with Tissue Simulant }
To systematically characterize the capability of our low-cost active vibroacoustic elasticity sensing approach, we first conduct controlled experiments using ballistic gel made tissue simulants \cite{Breeze2016Mar,Wermer2020Jan}. 
Ballistic gels have been widely used as soft-tissue simulants because their mechanical properties can approximate those of biological tissues and can be reproducibly controlled across experiments \cite{Andrade2025Jan,Appleby-Thomas2016Aug,Caldwell2019Oct}. As a result, they provide a well-controlled environment for evaluating sensing performance before deployment on human subjects.
We simulate different tissue stiffness levels using ballistic gel with varying concentrations of gelatin and water.

\begin{figure}
    \centering
    \input{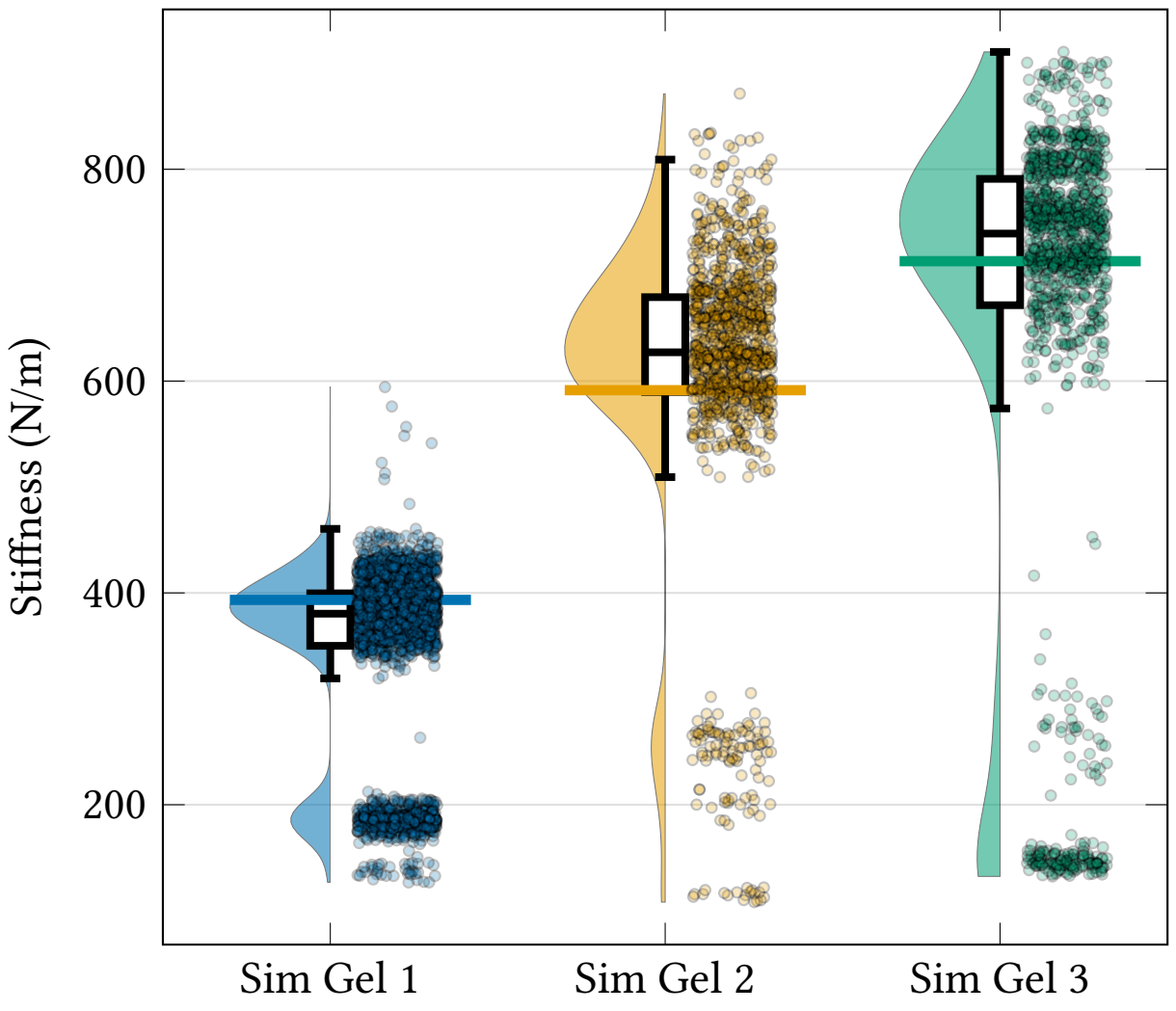}
    \caption{Stiffness estimation distribution for three simulant ballistic gels with different stiffness levels. Each data point is calculated with one trial of TDoA measurement.}
    \label{fig:jelly_pred}
\end{figure}

\begin{figure}
    \centering
    \input{Figures/seed_accuracy_jelly}
    \caption{Stiffness estimation across ballistic gel samples with different stiffness levels. The x-axis depicts the number of TDoA trials aggregated for a single stiffness estimate.}
    \label{fig:jelly_error}
\end{figure}
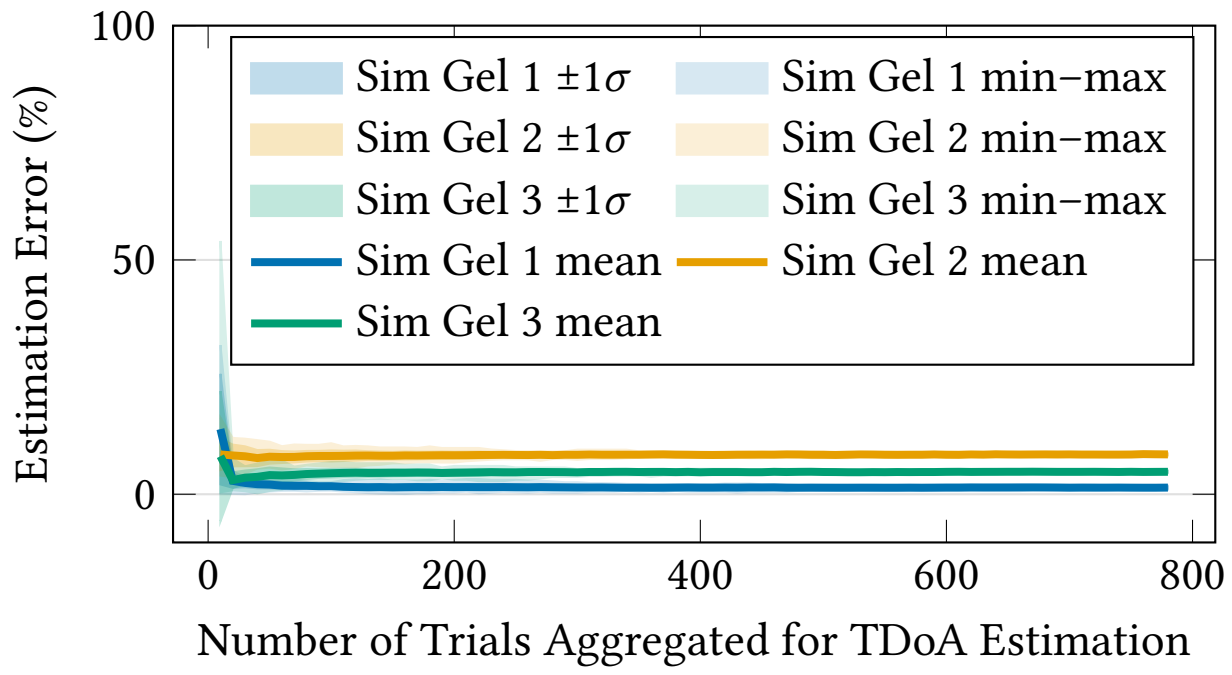

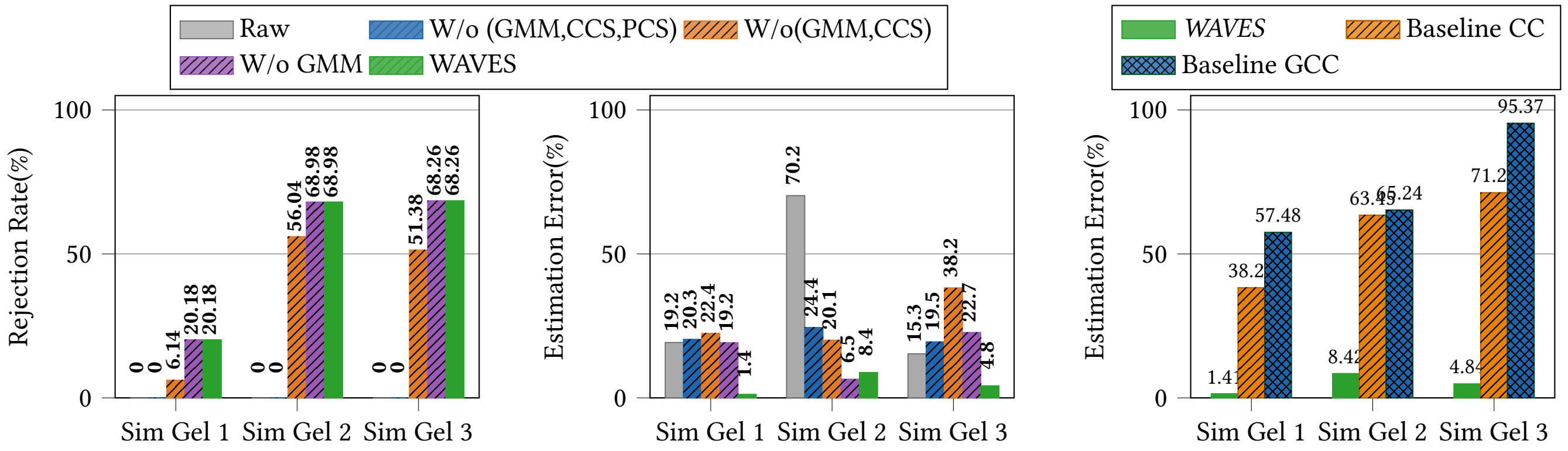
\begin{figure*}[t]
    \centering
    \begin{subfigure}[t]{0.32\textwidth}
        \centering
        \begin{minipage}[t][0.22\textheight][t]{\linewidth}
            \centering
            \input{Figures/jelly_baselines_reject}
        \end{minipage}
        \caption{Rejection rate for different screening and refinement strategies.}
        \label{fig:gel_rejection}
    \end{subfigure}
    \hfill
    \begin{subfigure}[t]{0.32\textwidth}
        \centering
        \begin{minipage}[t][0.22\textheight][t]{\linewidth}
            \centering
            \input{Figures/jelly_baselines}
        \end{minipage}
        \caption{Stiffness estimation error across different baseline methods.}
        \label{fig:gel_error}
    \end{subfigure}
    \hfill
    \begin{subfigure}[t]{0.32\textwidth}
        \centering
        \begin{minipage}[t][0.22\textheight][t]{\linewidth}
            \centering
            \input{Figures/tdoa_methods}
        \end{minipage}
        \caption{Stiffness estimation error comparison of \name\ with CC and GCC-PHAT baselines.}
        \label{fig:tdoa_methods}
    \end{subfigure}

    \caption{Evaluation of stiffness estimation performance with rejection rate and estimation error across different baselines as well as different stiffness simulant gels. }
    \label{fig:gel_evaluation}
\end{figure*}

\subsubsection{Overall Stiffness Estimation Performance}
We first evaluate whether \name can accurately distinguish simulant gel samples with different stiffness levels.
Figure \ref{fig:jelly_pred} shows the distribution of predicted stiffness across three different tissue simulant gels.
The x-axis shows the different simulant gels, while the y-axis shows the stiffness in N/m.
The ground-truth stiffness measurements for Sim Gel 1, 2, and 3 are shown as horizontal blue, yellow, and green lines, with values of $ 393$ N/m, $593.4$ N/m, and $713$ N/m, respectively. 
Three forms of analysis are shown for \name output.
First, each data point represents an estimate from a single active sensing trial.
Built upon that, the box plot shows that mean stiffness is close to ground truth stiffness, and the half violin plot illustrates that the estimated stiffness values are concentrated around the mean with a low stiffness estimation outlier.
The GMM applied to the distribution, as illustrated by the violin plot, reports estimated stiffness values of 387.81 N/m, 641.44 N/m, and 747.51 N/m, which correspond to estimation errors of 1.42\%, 8.46\%, and 4.8\%, respectively.  
We observe that the estimated stiffness is more concentrated around the ground truth for Sim Gel 1, while the estimated stiffness for Sim Gel 2 and Sim Gel 3 is more spread out.
This indicates that stiffness estimation becomes more error-prone at higher stiffness levels, likely because higher shear-wave velocity produces smaller TDoA values, making the estimation more sensitive to sampling-rate limitations.

We further evaluate the impact of the number of trials on the overall stiffness estimation. 
\name uses the Gaussian Mixture Model(GMM) to find the centroid locus among stiffness estimation from multiple trials of measurements.
We randomly select $N$ different events and predict the overall stiffness from the GMM centroid output.
We repeated this process with 20 different random seed values, varying $N$ from 10 to 750 in increments of 10 trials per estimation.
Figure \ref{fig:jelly_error} shows how the number of trials used for estimation impacts the error in stiffness estimation.
We observe that the error is highest for N = 10, as the variation among the estimations is higher, as shown by the min-max estimation.
The mean of the error is 10\%, with a minimum error of 0.2\% and the maximum error of 54\%.
However, this changes rapidly as the number of trials increases.
We observe that the mean stiffness estimation stabilizes when the number of trials reaches 50, and the mean estimation error for all three tissue-simulant gels converges to within 10\% of the ground-truth stiffness.
For the on-body experiment in Section \ref{sec:muscle_stiffness}, we adopt $N \approx 100$, which translates to approximately 1-2 minutes of data collection.




\subsubsection{Ablation Study}
We conduct the ablation study to understand the effectiveness of each modularized design and define the following baselines.

\textbf{\textit{Baseline 1 [w/o GMM]}}: This baseline uses all the in-trial screening and refinement strategies except using GMM to find the centroid locus for cross-trial refinement; instead, the mean of the TDoA measurements is used to estimate stiffness.

\textbf{\textit{Baseline 2 [w/o (GMM, CCS)]}}: For this baseline, we modify baseline 1 by removing the Cross-Sensor Consistency Screening (CCS) illustrated by dashed red lines in Figure \ref{fig:estimation_quality} (b).

\textbf{\textit{Baseline 3 [w/o (GMM, CCS, PCS)]}}: This baseline is a modified version of baseline 2 by removing the Physiological Constraint Screening (PCS) marked by green lines in Figure \ref{fig:estimation_quality} (a).

\textbf{\textit{Baseline 4 [Raw, w/o (GMM, CCS, PCS, CCR)]}}: This baseline is a modified version of baseline 3 by removing the Cross-Sensor Consistency Refinement (CCR); instead, the raw estimation is obtained by combining the means of TDoA estimations from TDoA-12 and TDoA-13. 

Figure \ref{fig:gel_rejection} and \ref{fig:gel_error} show the rejection and estimation error percentages, respectively, across all baselines and our \name. 
\name outperforms all the baselines in stiffness estimation with an average error of 4.86\%.
The next-best method is baseline 1, with an average estimation error of 16.13\%.
The rejection rate across \name and baseline 1 is the same, as they both perform the same screening process.
The stiffness estimation errors for baseline 2 and 3 are 26.9\% and 21.4\%, respectively.
This shows that the screening process can effectively identify and reject low-quality data points, thereby improving the overall system's performance.
The baseline 4 performs the worst, with an average estimation error of 34.9\%, while Sim Gel 2 demonstrates an error up to 70.2\%.
On the other hand, the rejection trend is along expectation; with an increase in screening strategies, the rejection rate also increases.

\subsubsection{TDoA Estimation Comparison}
We also evaluated \name's performance against commonly used TDoA estimation methods to further cement the effectiveness of \name's wavelet-based TDoA estimation and physiology-based filtering.
We select Cross Correlation (CC) \cite{bourke1996cross} and Generalized Cross-Correlation with Phase Transform (GCC-PHAT) \cite{Lee2020Jan, Wang, salvati2021time} as two TDoA baselines.
For both baselines, we compute the TDoA for all measurement trials using CC and GCC-PHAT, and take the mean of these TDoAs as the cross-trial aggregation output.
Figure \ref{fig:tdoa_methods} shows the stiffness estimation error across these three different methods.
\name outperforms both baselines across all three stiffness levels with approximately $16 \times$ and $22 \times$ improvement on average.
We also observe that the performance of both the CC and GCC-PHAT baselines degrades with increasing stiffness.
For example, on Sim Gel 3 with a stiffness of $713.2$ N/m, the estimation errors are $71.26\%$ and $95.37\%$ for CC and GCC, respectively. 
This may be because increasing stiffness leads to higher shear-wave velocities and, therefore, smaller TDoA values. 
Combined with signal attenuation in the viscoelastic medium, this makes TDoA estimation with correlation-based methods more error-prone.

\subsection{On-Body Stiffness Estimation Analysis}\label{sec:muscle_stiffness}
\name shows highly accurate stiffness estimation performance on the tissue simulant, which has an ideal, consistent density compared to human muscle.
To evaluate and validate \name's effectiveness in predicting muscle stiffness, we conduct a thorough on-body evaluation as detailed in Section \ref{sec:exp_design}.

\begin{figure}
    \centering
    \input{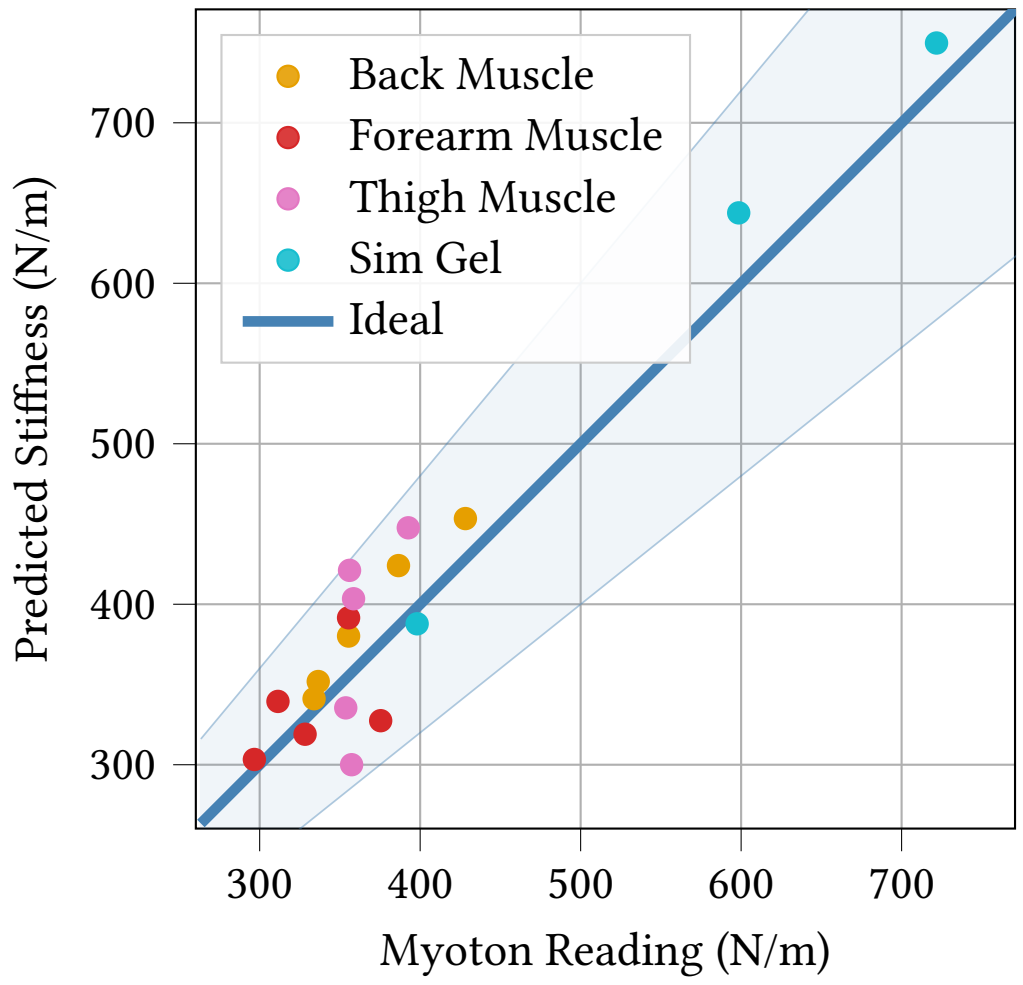}
    \vspace{-2ex}
    \caption{Overall stiffness estimation across different muscles and tissue simulant gel samples.}
    \label{fig:overall_stiff}
    \vspace{-2ex}
\end{figure}

Figure \ref{fig:overall_stiff} shows the results for muscle stiffness estimation across 13 participants and three different muscle groups, along with the stiffness estimation for tissue simulant gels for comparison.
The ground truth is shown along the x-axis, while the stiffness estimation from \name is shown along the y-axis.
The dark blue line at the center shows the ideal scenario with both predicted stiffness and ground truth being the same value, while the $20\%$ error region is marked along this line, indicating the acceptable error range for non-clinical usage \cite{warneke2025discussion}.
The stiffness estimates for the back (trapezius), forearm, and thigh muscles are shown in yellow, red, and pink circles, respectively. 
Each circle represents stiffness estimates for different users. 
For all participants, the stiffness estimation error remains within the $20\%$ range across all three muscle groups.
While the muscle stiffness across different users ranges from $296$ N/m to $428$ N/m, \name consistently tracks inter-subject and inter-muscle variation without performance degradation.

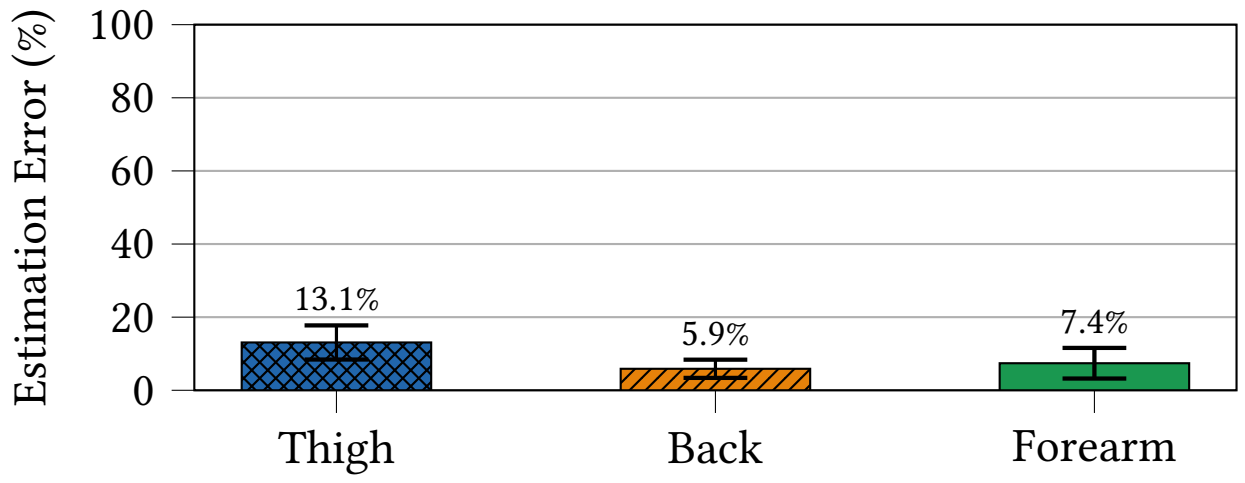
\begin{figure}
    \centering
    \input{Figures/muscle_plot}
        \vspace{-2ex}
    \caption{Stiffness estimation errors across three muscle groups. }
    \label{fig:muscle_stiff}
    \vspace{-2ex}
\end{figure}

\subsubsection{Cross Muscle Analysis}
Because of physiological differences among muscles, such as the presence of subcutaneous fat and proximity to bones, signal propagation across human muscles can vary.  
To evaluate \name's robustness to these physiological differences, we tested \name across three different muscle groups, as discussed in Section \ref{sec:muscle_stiffness}.
By evaluating data from five participants per muscle group, we account for inter-subject and intra-subject physiological variability. 
Figure \ref{fig:muscle_stiff} shows the stiffness estimation error among the thigh — vastus lateralis, back — trapezius muscle, and forearm — brachioradialis, respectively.
Similar to Figure \ref{fig:overall_stiff}, we can observe that the muscle stiffness estimation error is under a 20\% range. 
The average stiffness estimation error for the thigh muscle is 13.1\%, the highest among the three muscles.
The average errors for the forearm and back muscles are 7.4\% and 5.9\%, respectively.

\subsubsection{Multiple Day Deployment Variation}
\begin{figure}
    \centering
    \input{Figures/same_sub_multi}
    \vspace{-2ex}
    \caption{Same subject muscle stiffness estimation error across three days.}
    \vspace{-2ex}
    \label{fig:multi_day}
\end{figure}
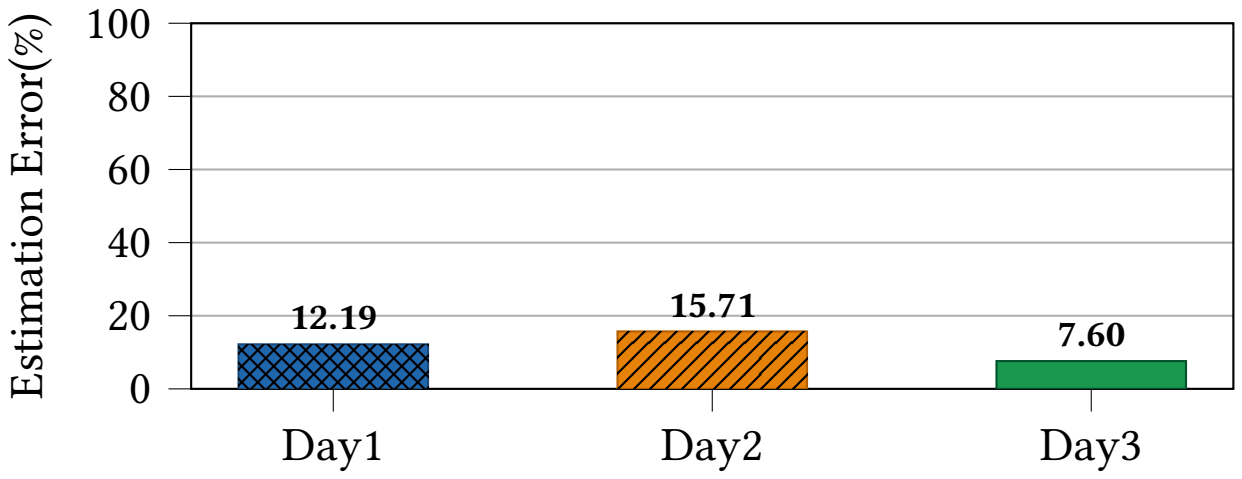

We also tested \name's stiffness estimation over multiple days to validate its robustness across different deployments of the same muscle on the same user.
For this study, we collected data from subject 11's forearm on three different days and compared stiffness estimation errors.
Figure \ref{fig:multi_day} shows the result for stiffness estimation on the same user across the same muscle on different days.
We observe that overall muscle stiffness varies across users and days.
However, the \name's stiffness estimation error is 15.71\% or lower across all 3 days of deployment.
This shows that with consistent deployment, the \name can effectively estimate muscle stiffness.


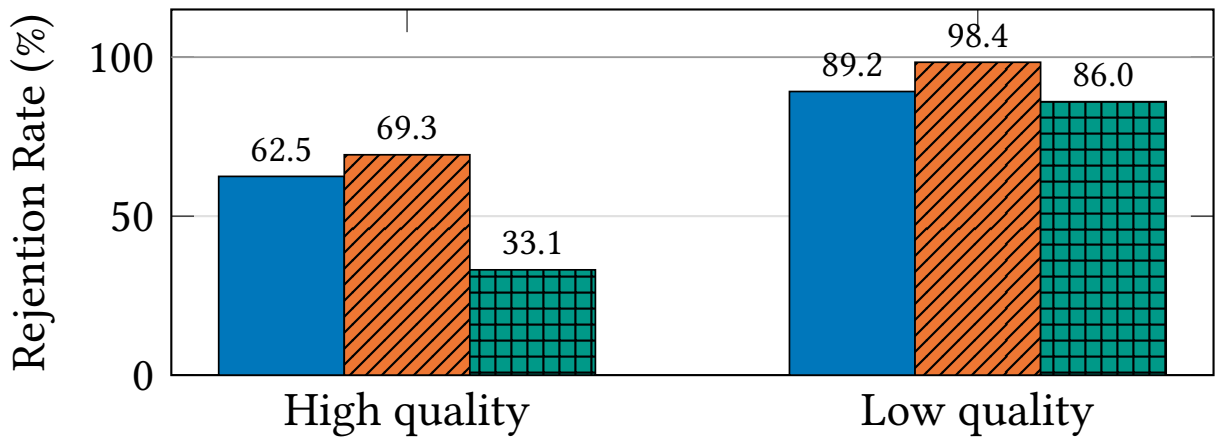
\begin{figure}
    \centering
    \input{Figures/hodogram_comp}
    \vspace{-2ex}
    \caption{Rejection rate comparison at different deployment quality levels, each with three deployment samples.}
    \label{fig:quality_reject}
\end{figure}

\subsubsection{On-body Deployment Quality Comparison}
To evaluate deployment quality, \name relies on the signal's energy concentration along the z-axis (normal to the surface) and the x-axis (along the surface, perpendicular to the signal propagation) to assess the shear-wave components as discussed in Section \ref{sec:sys_quality}.
The $z_H$ ratio is used to verify whether the deployment is valid.
Figure \ref{fig:quality_reject} shows a comparison of rejection rates across six deployments, with high-quality deployments on the left and low-quality on the right.
We observe that in these examples, the rejection rate is lower in high-quality deployments than in low-quality deployments -- the identified high-quality deployments can acquire effective measurement trials $5 \times$ more than low-quality deployments on average.

%% file: Figures/seed_accuracy_jelly.tex
\begin{tikzpicture}

\definecolor{darkcyan0114178}{RGB}{0,114,178}
\definecolor{darkcyan0158115}{RGB}{0,158,115}
\definecolor{darkgrey176}{RGB}{176,176,176}
\definecolor{gainsboro224}{RGB}{224,224,224}
\definecolor{orange2301590}{RGB}{230,159,0}

\begin{axis}[
height = 5cm,
width=\linewidth,
legend cell align={left},
legend columns=2,
legend style={fill opacity=1, draw opacity=1, text opacity=1},
tick pos=both,
x grid style={darkgrey176},
xlabel={Number of Trials Aggregated for TDoA Estimation},
xmin=-28.5, xmax=818.5,
xtick style={color=black},
y grid style={gainsboro224},
ylabel={Estimation Error (\%)},
ymajorgrids,
ymin=-10.29042016393, ymax=100,
ytick style={color=black}
]
\path [draw=darkcyan0114178, fill=darkcyan0114178, opacity=0.25]
(axis cs:10,25.6938748338137)
--(axis cs:10,2.09901782316587)
--(axis cs:20,1.38208266240169)
--(axis cs:30,1.49471698237395)
--(axis cs:40,1.27123923410352)
--(axis cs:50,1.23874831501)
--(axis cs:60,1.08073649496192)
--(axis cs:70,1.02226436123233)
--(axis cs:80,1.16032178468104)
--(axis cs:90,1.14753193211389)
--(axis cs:100,1.20611938304499)
--(axis cs:110,0.996254735442953)
--(axis cs:120,0.892759334093025)
--(axis cs:130,0.847200599593395)
--(axis cs:140,0.867528851348779)
--(axis cs:150,0.842825343201284)
--(axis cs:160,0.853911916681532)
--(axis cs:170,0.871475542065817)
--(axis cs:180,0.838491601369735)
--(axis cs:190,0.909550371161643)
--(axis cs:200,0.876262598846769)
--(axis cs:210,0.912223808547355)
--(axis cs:220,0.86126408563724)
--(axis cs:230,0.881432572313467)
--(axis cs:240,0.900727961151496)
--(axis cs:250,0.878379004011382)
--(axis cs:260,0.888806316159963)
--(axis cs:270,0.922675370322899)
--(axis cs:280,0.913815875551857)
--(axis cs:290,0.902998661028771)
--(axis cs:300,0.906279501893834)
--(axis cs:310,0.93897980190523)
--(axis cs:320,0.949751368429728)
--(axis cs:330,0.947628839425557)
--(axis cs:340,0.957903043190103)
--(axis cs:350,0.964660104442902)
--(axis cs:360,0.96140388087669)
--(axis cs:370,0.975135276318248)
--(axis cs:380,1.02747267469023)
--(axis cs:390,1.02372091064068)
--(axis cs:400,1.00907541508423)
--(axis cs:410,1.0510095100656)
--(axis cs:420,1.03817979079034)
--(axis cs:430,1.00189918782626)
--(axis cs:440,1.05500850961688)
--(axis cs:450,1.06078875304296)
--(axis cs:460,1.11538348982128)
--(axis cs:470,1.04597396499691)
--(axis cs:480,1.07543115323889)
--(axis cs:490,1.08022955740533)
--(axis cs:500,1.06122541685594)
--(axis cs:510,1.04329867268616)
--(axis cs:520,1.05676231966734)
--(axis cs:530,1.07475520155626)
--(axis cs:540,1.06866048219889)
--(axis cs:550,1.06043136752831)
--(axis cs:560,1.06778961859345)
--(axis cs:570,1.08999486121372)
--(axis cs:580,1.06778961859345)
--(axis cs:590,1.08249206510101)
--(axis cs:600,1.09880902877601)
--(axis cs:610,1.11583013772428)
--(axis cs:620,1.15179904532767)
--(axis cs:630,1.13839173119041)
--(axis cs:640,1.12509185083132)
--(axis cs:650,1.13768268879998)
--(axis cs:660,1.14678036906773)
--(axis cs:670,1.17377176098464)
--(axis cs:680,1.16944896340683)
--(axis cs:690,1.15469121214041)
--(axis cs:700,1.12368087690536)
--(axis cs:710,1.13062205867355)
--(axis cs:720,1.11000418444565)
--(axis cs:730,1.14105742973735)
--(axis cs:740,1.16202310495351)
--(axis cs:750,1.12346448625057)
--(axis cs:760,1.12376402170461)
--(axis cs:770,1.13546781868416)
--(axis cs:780,1.16845184246971)
--(axis cs:780,1.6635165912618)
--(axis cs:780,1.6635165912618)
--(axis cs:770,1.65218142068862)
--(axis cs:760,1.66272024762087)
--(axis cs:750,1.68128696256801)
--(axis cs:740,1.67782226005962)
--(axis cs:730,1.6768107363102)
--(axis cs:720,1.70786324695753)
--(axis cs:710,1.7038167540128)
--(axis cs:700,1.67540339099582)
--(axis cs:690,1.70037855720873)
--(axis cs:680,1.72159427644461)
--(axis cs:670,1.73267936025746)
--(axis cs:660,1.75915028016462)
--(axis cs:650,1.74935526693271)
--(axis cs:640,1.73737307242612)
--(axis cs:630,1.74189530738868)
--(axis cs:620,1.74998564477789)
--(axis cs:610,1.73079405684901)
--(axis cs:600,1.747605689154)
--(axis cs:590,1.72855794570828)
--(axis cs:580,1.7102837339361)
--(axis cs:570,1.72195784585907)
--(axis cs:560,1.7102837339361)
--(axis cs:550,1.71663885149575)
--(axis cs:540,1.72552439773499)
--(axis cs:530,1.72367486583884)
--(axis cs:520,1.72219948868182)
--(axis cs:510,1.73695512992539)
--(axis cs:500,1.73588451898197)
--(axis cs:490,1.73422971680956)
--(axis cs:480,1.71583073345165)
--(axis cs:470,1.74655919268822)
--(axis cs:460,1.8033340625681)
--(axis cs:450,1.83038514205168)
--(axis cs:440,1.84446296257597)
--(axis cs:430,1.8336997429522)
--(axis cs:420,1.82934304076113)
--(axis cs:410,1.77173518849412)
--(axis cs:400,1.81731468490337)
--(axis cs:390,1.85703340939783)
--(axis cs:380,1.80718063965529)
--(axis cs:370,1.8096153964737)
--(axis cs:360,1.80737270761589)
--(axis cs:350,1.84051629062628)
--(axis cs:340,1.83165771186123)
--(axis cs:330,1.8958439003859)
--(axis cs:320,1.95612580517257)
--(axis cs:310,1.94911773783981)
--(axis cs:300,1.97756374462136)
--(axis cs:290,2.02856625151619)
--(axis cs:280,2.05813247726762)
--(axis cs:270,2.10761894501757)
--(axis cs:260,2.07346678039542)
--(axis cs:250,2.12099084604447)
--(axis cs:240,2.1490485739564)
--(axis cs:230,2.13914199842003)
--(axis cs:220,2.16558815750544)
--(axis cs:210,2.13484543773634)
--(axis cs:200,2.14675241638139)
--(axis cs:190,2.16250571388014)
--(axis cs:180,2.17629318536502)
--(axis cs:170,2.18261908644343)
--(axis cs:160,2.15168380614608)
--(axis cs:150,2.08616493732387)
--(axis cs:140,2.16919890978055)
--(axis cs:130,2.13743451668845)
--(axis cs:120,2.16391498219901)
--(axis cs:110,2.18424630990598)
--(axis cs:100,2.23295261648074)
--(axis cs:90,2.2735999670446)
--(axis cs:80,2.42857127298571)
--(axis cs:70,2.55597734693041)
--(axis cs:60,2.5778660539615)
--(axis cs:50,2.87969029230338)
--(axis cs:40,3.01318951677956)
--(axis cs:30,3.32309304935103)
--(axis cs:20,4.32701172903648)
--(axis cs:10,25.6938748338137)
--cycle;
\addlegendimage{area legend, draw=darkcyan0114178, fill=darkcyan0114178, opacity=0.25}
\addlegendentry{Sim Gel 1 ±1$\sigma$}

\path [draw=darkcyan0114178, fill=darkcyan0114178, opacity=0.16]
(axis cs:10,31.8230745405349)
--(axis cs:10,0.18045832960569)
--(axis cs:20,0.0955778492941811)
--(axis cs:30,0.00890340874678649)
--(axis cs:40,0.350253657847344)
--(axis cs:50,0.586113955495711)
--(axis cs:60,0.718445475196071)
--(axis cs:70,0.752657823373059)
--(axis cs:80,0.599080354673216)
--(axis cs:90,0.753859262792159)
--(axis cs:100,0.777575423090494)
--(axis cs:110,0.577849173520392)
--(axis cs:120,0.437547061233399)
--(axis cs:130,0.259579266403039)
--(axis cs:140,0.144447036555936)
--(axis cs:150,0.136668294183549)
--(axis cs:160,0.0476750362279541)
--(axis cs:170,0.221923916117098)
--(axis cs:180,0.221923916117098)
--(axis cs:190,0.395717535550376)
--(axis cs:200,0.395717535550376)
--(axis cs:210,0.569057479094511)
--(axis cs:220,0.221923916117098)
--(axis cs:230,0.357185699407815)
--(axis cs:240,0.532309153517408)
--(axis cs:250,0.469172827302035)
--(axis cs:260,0.357185699407815)
--(axis cs:270,0.532309153517408)
--(axis cs:280,0.357185699407815)
--(axis cs:290,0.357185699407815)
--(axis cs:300,0.357185699407815)
--(axis cs:310,0.532309153517408)
--(axis cs:320,0.548870332562721)
--(axis cs:330,0.548870332562721)
--(axis cs:340,0.706971340607415)
--(axis cs:350,0.548870332562721)
--(axis cs:360,0.706971340607415)
--(axis cs:370,0.548870332562721)
--(axis cs:380,0.72357867743059)
--(axis cs:390,0.548870332562721)
--(axis cs:400,0.548870332562721)
--(axis cs:410,0.548870332562721)
--(axis cs:420,0.548870332562721)
--(axis cs:430,0.373700399467338)
--(axis cs:440,0.548870332562721)
--(axis cs:450,0.548870332562721)
--(axis cs:460,0.548870332562721)
--(axis cs:470,0.548870332562721)
--(axis cs:480,0.72357867743059)
--(axis cs:490,0.72357867743059)
--(axis cs:500,0.72357867743059)
--(axis cs:510,0.72357867743059)
--(axis cs:520,0.72357867743059)
--(axis cs:530,0.72357867743059)
--(axis cs:540,0.678082993585201)
--(axis cs:550,0.678082993585201)
--(axis cs:560,0.678082993585201)
--(axis cs:570,0.678082993585201)
--(axis cs:580,0.678082993585201)
--(axis cs:590,0.678082993585201)
--(axis cs:600,0.678082993585201)
--(axis cs:610,0.678082993585201)
--(axis cs:620,0.678082993585201)
--(axis cs:630,0.678082993585201)
--(axis cs:640,0.678082993585201)
--(axis cs:650,0.678082993585201)
--(axis cs:660,0.678082993585201)
--(axis cs:670,0.857798635323215)
--(axis cs:680,0.857798635323215)
--(axis cs:690,0.857798635323215)
--(axis cs:700,0.857798635323215)
--(axis cs:710,0.857798635323215)
--(axis cs:720,0.857798635323215)
--(axis cs:730,0.920616182008125)
--(axis cs:740,0.920616182008125)
--(axis cs:750,0.920616182008125)
--(axis cs:760,0.920616182008125)
--(axis cs:770,0.920616182008125)
--(axis cs:780,0.920616182008125)
--(axis cs:780,1.8398651168645)
--(axis cs:780,1.8398651168645)
--(axis cs:770,1.8398651168645)
--(axis cs:760,1.8398651168645)
--(axis cs:750,1.8398651168645)
--(axis cs:740,1.8398651168645)
--(axis cs:730,1.8398651168645)
--(axis cs:720,2.01908607990247)
--(axis cs:710,2.01908607990247)
--(axis cs:700,2.01908607990247)
--(axis cs:690,2.01908607990247)
--(axis cs:680,2.01908607990247)
--(axis cs:670,2.01908607990247)
--(axis cs:660,2.01908607990247)
--(axis cs:650,2.0078202222421)
--(axis cs:640,2.0078202222421)
--(axis cs:630,2.0078202222421)
--(axis cs:620,2.0078202222421)
--(axis cs:610,2.0078202222421)
--(axis cs:600,2.0078202222421)
--(axis cs:590,2.0078202222421)
--(axis cs:580,2.0078202222421)
--(axis cs:570,2.0078202222421)
--(axis cs:560,2.0078202222421)
--(axis cs:550,2.0078202222421)
--(axis cs:540,2.0078202222421)
--(axis cs:530,2.0078202222421)
--(axis cs:520,2.0078202222421)
--(axis cs:510,2.0078202222421)
--(axis cs:500,2.0078202222421)
--(axis cs:490,2.0078202222421)
--(axis cs:480,1.94408027673434)
--(axis cs:470,2.11255138636262)
--(axis cs:460,2.11255138636262)
--(axis cs:450,2.11255138636262)
--(axis cs:440,2.28058869018088)
--(axis cs:430,2.28058869018088)
--(axis cs:420,2.28058869018088)
--(axis cs:410,2.11255138636262)
--(axis cs:400,2.11255138636262)
--(axis cs:390,2.13615272579138)
--(axis cs:380,2.13615272579138)
--(axis cs:370,2.13615272579138)
--(axis cs:360,2.13615272579138)
--(axis cs:350,2.13615272579138)
--(axis cs:340,2.1978166576792)
--(axis cs:330,2.31730571561835)
--(axis cs:320,2.37605863762369)
--(axis cs:310,2.37605863762369)
--(axis cs:300,2.55381379902838)
--(axis cs:290,2.55381379902838)
--(axis cs:280,2.7310839130938)
--(axis cs:270,2.90592310134107)
--(axis cs:260,2.90592310134107)
--(axis cs:250,3.06480273946467)
--(axis cs:240,3.06480273946467)
--(axis cs:230,3.06480273946467)
--(axis cs:220,3.22329272299632)
--(axis cs:210,3.22329272299632)
--(axis cs:200,3.06480273946467)
--(axis cs:190,2.90592310134107)
--(axis cs:180,3.06480273946467)
--(axis cs:170,2.90592310134107)
--(axis cs:160,2.90592310134107)
--(axis cs:150,2.90592310134107)
--(axis cs:140,3.16607000949294)
--(axis cs:130,3.01201811118751)
--(axis cs:120,3.16607000949294)
--(axis cs:110,3.00755620590901)
--(axis cs:100,2.54333845872473)
--(axis cs:90,2.54333845872473)
--(axis cs:80,2.79478789656928)
--(axis cs:70,3.24486395052571)
--(axis cs:60,3.11673916765042)
--(axis cs:50,3.352731934007)
--(axis cs:40,3.70553447591453)
--(axis cs:30,3.70144843095232)
--(axis cs:20,4.87918984370414)
--(axis cs:10,31.8230745405349)
--cycle;
\addlegendimage{area legend, draw=darkcyan0114178, fill=darkcyan0114178, opacity=0.16}
\addlegendentry{Sim Gel 1 min–max}

\path [draw=orange2301590, fill=orange2301590, opacity=0.25]
(axis cs:10,11.8095005611728)
--(axis cs:10,5.16421918208307)
--(axis cs:20,5.97089718724841)
--(axis cs:30,5.97970108656125)
--(axis cs:40,6.01135467215551)
--(axis cs:50,6.53413654703597)
--(axis cs:60,6.67369161401647)
--(axis cs:70,6.8053601998284)
--(axis cs:80,6.99193601123379)
--(axis cs:90,7.08403694957572)
--(axis cs:100,6.9946526520521)
--(axis cs:110,7.09563652240581)
--(axis cs:120,7.183939431774)
--(axis cs:130,7.34364195916386)
--(axis cs:140,7.33043653246903)
--(axis cs:150,7.38335937329921)
--(axis cs:160,7.42407731169427)
--(axis cs:170,7.45125942945586)
--(axis cs:180,7.51026661284017)
--(axis cs:190,7.52417334492252)
--(axis cs:200,7.56180560684361)
--(axis cs:210,7.56782321487865)
--(axis cs:220,7.63199607813902)
--(axis cs:230,7.73576715820854)
--(axis cs:240,7.83445973761222)
--(axis cs:250,7.81442274690745)
--(axis cs:260,7.84284743377286)
--(axis cs:270,7.88061932091776)
--(axis cs:280,7.87180687956066)
--(axis cs:290,7.96303702096857)
--(axis cs:300,7.89542770045612)
--(axis cs:310,7.97379425983467)
--(axis cs:320,7.94999128155506)
--(axis cs:330,7.92283520995191)
--(axis cs:340,7.95636428593583)
--(axis cs:350,7.99748641817136)
--(axis cs:360,8.05768963733675)
--(axis cs:370,8.11430979653045)
--(axis cs:380,8.06071259551269)
--(axis cs:390,7.99729950108832)
--(axis cs:400,8.02109138455341)
--(axis cs:410,8.01317898468933)
--(axis cs:420,8.00615442581224)
--(axis cs:430,8.05546521792503)
--(axis cs:440,8.09563513963852)
--(axis cs:450,8.12132504176578)
--(axis cs:460,8.11712455067392)
--(axis cs:470,8.17754906502316)
--(axis cs:480,8.17120198258037)
--(axis cs:490,8.07414896188529)
--(axis cs:500,8.07433518628048)
--(axis cs:510,8.08297247631895)
--(axis cs:520,8.15348991779575)
--(axis cs:530,8.24455341394078)
--(axis cs:540,8.2305957228529)
--(axis cs:550,8.16029310849881)
--(axis cs:560,8.15652872295159)
--(axis cs:570,8.17872440622685)
--(axis cs:580,8.18036368488244)
--(axis cs:590,8.22329033386239)
--(axis cs:600,8.17433754217803)
--(axis cs:610,8.1564357584124)
--(axis cs:620,8.28892026318759)
--(axis cs:630,8.23446418576281)
--(axis cs:640,8.26001048797974)
--(axis cs:650,8.28236281676292)
--(axis cs:660,8.23275315879073)
--(axis cs:670,8.25313098091982)
--(axis cs:680,8.28784129634422)
--(axis cs:690,8.2560580617682)
--(axis cs:700,8.28784129634422)
--(axis cs:710,8.23112819290849)
--(axis cs:720,8.23843960514461)
--(axis cs:730,8.21019357221546)
--(axis cs:740,8.23861484003495)
--(axis cs:750,8.24299805315592)
--(axis cs:760,8.38003169443299)
--(axis cs:770,8.35773197548651)
--(axis cs:780,8.34075240576806)
--(axis cs:780,8.66223412660042)
--(axis cs:780,8.66223412660042)
--(axis cs:770,8.6928675033504)
--(axis cs:760,8.73840213005247)
--(axis cs:750,8.70578664761051)
--(axis cs:740,8.67581249407365)
--(axis cs:730,8.70398857725825)
--(axis cs:720,8.6998849404866)
--(axis cs:710,8.70692604254434)
--(axis cs:700,8.70898528637588)
--(axis cs:690,8.70697854375354)
--(axis cs:680,8.70898528637588)
--(axis cs:670,8.71839306896834)
--(axis cs:660,8.70523397314564)
--(axis cs:650,8.69053077735508)
--(axis cs:640,8.77030211658497)
--(axis cs:630,8.66994554527316)
--(axis cs:620,8.68339571585065)
--(axis cs:610,8.68034225604402)
--(axis cs:600,8.69555522845154)
--(axis cs:590,8.74228243267582)
--(axis cs:580,8.69289470122049)
--(axis cs:570,8.68858950756544)
--(axis cs:560,8.67653927849744)
--(axis cs:550,8.70713225960804)
--(axis cs:540,8.71755613047675)
--(axis cs:530,8.73748308162713)
--(axis cs:520,8.728631749924)
--(axis cs:510,8.68619222147995)
--(axis cs:500,8.76370775444576)
--(axis cs:490,8.77468886214685)
--(axis cs:480,8.77931574861745)
--(axis cs:470,8.80691836958496)
--(axis cs:460,8.76441892305928)
--(axis cs:450,8.79549386411016)
--(axis cs:440,8.78678626883352)
--(axis cs:430,8.80624709819151)
--(axis cs:420,8.77884393690864)
--(axis cs:410,8.73839124216243)
--(axis cs:400,8.76289993865049)
--(axis cs:390,8.8548154191004)
--(axis cs:380,8.82501050896848)
--(axis cs:370,8.89628318076437)
--(axis cs:360,8.89975844612097)
--(axis cs:350,8.88053534465302)
--(axis cs:340,8.92903725957862)
--(axis cs:330,8.95293236940925)
--(axis cs:320,8.92576547482998)
--(axis cs:310,8.97036450221976)
--(axis cs:300,8.94836549026327)
--(axis cs:290,8.87991557662465)
--(axis cs:280,8.82245292467625)
--(axis cs:270,8.94470503122725)
--(axis cs:260,8.88428608615221)
--(axis cs:250,8.94845251172172)
--(axis cs:240,8.99394889984823)
--(axis cs:230,9.03332064267457)
--(axis cs:220,9.10996162012845)
--(axis cs:210,9.10192930471297)
--(axis cs:200,9.08282571207481)
--(axis cs:190,9.12802720492055)
--(axis cs:180,9.08685313777339)
--(axis cs:170,9.08009724536756)
--(axis cs:160,9.13206315686678)
--(axis cs:150,9.04124912878493)
--(axis cs:140,9.09539528670947)
--(axis cs:130,9.17555218654521)
--(axis cs:120,9.30861165781313)
--(axis cs:110,9.27949444632305)
--(axis cs:100,9.33031356131946)
--(axis cs:90,9.24132063333818)
--(axis cs:80,9.22834791655476)
--(axis cs:70,9.11100976325554)
--(axis cs:60,9.19728344152159)
--(axis cs:50,9.48183535545469)
--(axis cs:40,9.41583224175306)
--(axis cs:30,10.2145791948887)
--(axis cs:20,10.5869878137049)
--(axis cs:10,11.8095005611728)
--cycle;
\addlegendimage{area legend, draw=orange2301590, fill=orange2301590, opacity=0.25}
\addlegendentry{Sim Gel 2 ±1$\sigma$}

\path [draw=orange2301590, fill=orange2301590, opacity=0.16]
(axis cs:10,16.3549672607093)
--(axis cs:10,3.98785675882236)
--(axis cs:20,3.62184158129171)
--(axis cs:30,4.45993487211186)
--(axis cs:40,5.46887874675211)
--(axis cs:50,5.69104701890453)
--(axis cs:60,5.2987760484557)
--(axis cs:70,5.61251321361284)
--(axis cs:80,6.15749134406469)
--(axis cs:90,6.00711575755126)
--(axis cs:100,5.85705946284512)
--(axis cs:110,5.85705946284512)
--(axis cs:120,5.7643962345709)
--(axis cs:130,6.16104689095678)
--(axis cs:140,6.28868857944984)
--(axis cs:150,6.60789125178953)
--(axis cs:160,6.60789125178953)
--(axis cs:170,6.81692116182938)
--(axis cs:180,6.81692116182938)
--(axis cs:190,6.77269639516496)
--(axis cs:200,6.77269639516496)
--(axis cs:210,7.05185108290249)
--(axis cs:220,7.20641039767627)
--(axis cs:230,7.37076529656279)
--(axis cs:240,7.46534329101934)
--(axis cs:250,7.46534329101934)
--(axis cs:260,7.46893241128316)
--(axis cs:270,7.14767107095426)
--(axis cs:280,7.52589738166188)
--(axis cs:290,7.52721879887201)
--(axis cs:300,7.34919282333839)
--(axis cs:310,7.52721879887201)
--(axis cs:320,7.52721879887201)
--(axis cs:330,7.52721879887201)
--(axis cs:340,7.6833226241528)
--(axis cs:350,7.77932071716618)
--(axis cs:360,7.77932071716618)
--(axis cs:370,7.7907854522221)
--(axis cs:380,7.85166903811462)
--(axis cs:390,7.77932071716618)
--(axis cs:400,7.77932071716618)
--(axis cs:410,7.77932071716618)
--(axis cs:420,7.77932071716618)
--(axis cs:430,7.7882032913911)
--(axis cs:440,7.70213026152964)
--(axis cs:450,7.70213026152964)
--(axis cs:460,7.89554670759355)
--(axis cs:470,7.89554670759355)
--(axis cs:480,7.89554670759355)
--(axis cs:490,7.70213026152964)
--(axis cs:500,7.70213026152964)
--(axis cs:510,7.70213026152964)
--(axis cs:520,7.70213026152964)
--(axis cs:530,8.03503157228261)
--(axis cs:540,8.03503157228261)
--(axis cs:550,7.77932071716618)
--(axis cs:560,7.77932071716618)
--(axis cs:570,7.77932071716618)
--(axis cs:580,7.77932071716618)
--(axis cs:590,8.03503157228261)
--(axis cs:600,7.77932071716618)
--(axis cs:610,7.77932071716618)
--(axis cs:620,8.03503157228261)
--(axis cs:630,8.11965504034892)
--(axis cs:640,8.03503157228261)
--(axis cs:650,8.03503157228261)
--(axis cs:660,8.03503157228261)
--(axis cs:670,8.03503157228261)
--(axis cs:680,8.11965504034892)
--(axis cs:690,8.11965504034892)
--(axis cs:700,8.11965504034892)
--(axis cs:710,8.11965504034892)
--(axis cs:720,8.11965504034892)
--(axis cs:730,8.11965504034892)
--(axis cs:740,8.11965504034892)
--(axis cs:750,8.11965504034892)
--(axis cs:760,8.11965504034892)
--(axis cs:770,8.11965504034892)
--(axis cs:780,8.11965504034892)
--(axis cs:780,8.80517758737913)
--(axis cs:780,8.80517758737913)
--(axis cs:770,8.80517758737913)
--(axis cs:760,8.84718846734671)
--(axis cs:750,8.80517758737913)
--(axis cs:740,8.80517758737913)
--(axis cs:730,8.84718846734671)
--(axis cs:720,8.84718846734671)
--(axis cs:710,8.84718846734671)
--(axis cs:700,8.84718846734671)
--(axis cs:690,8.84718846734671)
--(axis cs:680,8.84718846734671)
--(axis cs:670,8.84718846734671)
--(axis cs:660,8.84718846734671)
--(axis cs:650,8.80517758737913)
--(axis cs:640,9.1503863500743)
--(axis cs:630,8.80517758737913)
--(axis cs:620,8.80517758737913)
--(axis cs:610,8.80517758737913)
--(axis cs:600,8.80517758737913)
--(axis cs:590,9.07539397892768)
--(axis cs:580,8.84718846734671)
--(axis cs:570,8.84718846734671)
--(axis cs:560,8.84718846734671)
--(axis cs:550,8.84718846734671)
--(axis cs:540,8.84718846734671)
--(axis cs:530,8.84718846734671)
--(axis cs:520,8.84718846734671)
--(axis cs:510,8.84718846734671)
--(axis cs:500,9.1503863500743)
--(axis cs:490,9.1503863500743)
--(axis cs:480,9.1503863500743)
--(axis cs:470,9.1503863500743)
--(axis cs:460,9.08314214464693)
--(axis cs:450,8.84718846734671)
--(axis cs:440,9.08314214464693)
--(axis cs:430,9.08314214464693)
--(axis cs:420,9.08314214464693)
--(axis cs:410,9.03556402714764)
--(axis cs:400,9.03556402714764)
--(axis cs:390,9.08314214464693)
--(axis cs:380,9.1503863500743)
--(axis cs:370,9.1503863500743)
--(axis cs:360,9.1503863500743)
--(axis cs:350,9.1503863500743)
--(axis cs:340,9.49724060078359)
--(axis cs:330,9.49724060078359)
--(axis cs:320,9.49724060078359)
--(axis cs:310,9.49724060078359)
--(axis cs:300,9.49724060078359)
--(axis cs:290,9.1503863500743)
--(axis cs:280,9.1503863500743)
--(axis cs:270,9.1503863500743)
--(axis cs:260,9.19689154317998)
--(axis cs:250,9.19689154317998)
--(axis cs:240,9.36794912466022)
--(axis cs:230,9.52833505545501)
--(axis cs:220,9.86128989858473)
--(axis cs:210,9.86128989858473)
--(axis cs:200,9.86128989858473)
--(axis cs:190,9.86128989858473)
--(axis cs:180,10.1957652751329)
--(axis cs:170,9.86128989858473)
--(axis cs:160,9.96084047568469)
--(axis cs:150,9.96084047568469)
--(axis cs:140,9.96084047568469)
--(axis cs:130,10.1957652751329)
--(axis cs:120,10.2948800193176)
--(axis cs:110,10.1957652751329)
--(axis cs:100,10.8693147901338)
--(axis cs:90,10.5317704578136)
--(axis cs:80,10.5317704578136)
--(axis cs:70,10.6041567757065)
--(axis cs:60,10.2392000625352)
--(axis cs:50,11.2060534624464)
--(axis cs:40,11.5311659092286)
--(axis cs:30,11.8446710335446)
--(axis cs:20,12.0086627141397)
--(axis cs:10,16.3549672607093)
--cycle;
\addlegendimage{area legend, draw=orange2301590, fill=orange2301590, opacity=0.16}
\addlegendentry{Sim Gel  2 min–max}

\path [draw=darkcyan0158115, fill=darkcyan0158115, opacity=0.25]
(axis cs:10,22.0117025508035)
--(axis cs:10,-5.78647240185409)
--(axis cs:20,1.37008332010101)
--(axis cs:30,2.03232693533455)
--(axis cs:40,2.0961006579555)
--(axis cs:50,2.48017763004371)
--(axis cs:60,2.53500421417801)
--(axis cs:70,2.9337684906937)
--(axis cs:80,3.25448639090581)
--(axis cs:90,3.4972707742433)
--(axis cs:100,3.58245117202023)
--(axis cs:110,3.68400824415586)
--(axis cs:120,3.76133899303392)
--(axis cs:130,3.73120195793517)
--(axis cs:140,3.76259934450722)
--(axis cs:150,3.82915107556878)
--(axis cs:160,3.87849233834785)
--(axis cs:170,3.90479957526774)
--(axis cs:180,3.89493477556918)
--(axis cs:190,3.9204815563146)
--(axis cs:200,3.88346722518344)
--(axis cs:210,3.93282552532947)
--(axis cs:220,3.92740505536526)
--(axis cs:230,4.05098287604687)
--(axis cs:240,4.03350835969945)
--(axis cs:250,4.07234599635006)
--(axis cs:260,4.117341373993)
--(axis cs:270,4.21627662205439)
--(axis cs:280,4.22784024904148)
--(axis cs:290,4.21896833655251)
--(axis cs:300,4.17966346766718)
--(axis cs:310,4.26123927466664)
--(axis cs:320,4.30785813260997)
--(axis cs:330,4.34172121360928)
--(axis cs:340,4.34826021415906)
--(axis cs:350,4.33856050385256)
--(axis cs:360,4.3384570236526)
--(axis cs:370,4.36481962879635)
--(axis cs:380,4.3560206928152)
--(axis cs:390,4.40675288747706)
--(axis cs:400,4.31514450010258)
--(axis cs:410,4.38014069974199)
--(axis cs:420,4.41042423455418)
--(axis cs:430,4.40922688528828)
--(axis cs:440,4.43172849755059)
--(axis cs:450,4.44931170217569)
--(axis cs:460,4.51117118926287)
--(axis cs:470,4.46948750462803)
--(axis cs:480,4.53795351380508)
--(axis cs:490,4.54450879522993)
--(axis cs:500,4.46553949279343)
--(axis cs:510,4.4727472299896)
--(axis cs:520,4.45363790072851)
--(axis cs:530,4.44024050484518)
--(axis cs:540,4.44502243589698)
--(axis cs:550,4.46172831544744)
--(axis cs:560,4.53638936566937)
--(axis cs:570,4.52084867016096)
--(axis cs:580,4.54229839143384)
--(axis cs:590,4.54229839143384)
--(axis cs:600,4.60908336751171)
--(axis cs:610,4.6260111925173)
--(axis cs:620,4.55904793642684)
--(axis cs:630,4.5830622880005)
--(axis cs:640,4.60287265325774)
--(axis cs:650,4.60227479682048)
--(axis cs:660,4.60227479682048)
--(axis cs:670,4.62470437023788)
--(axis cs:680,4.59766413972151)
--(axis cs:690,4.57245287635708)
--(axis cs:700,4.5768326194136)
--(axis cs:710,4.5768326194136)
--(axis cs:720,4.59951067677289)
--(axis cs:730,4.57432353751959)
--(axis cs:740,4.52833930304332)
--(axis cs:750,4.59912915211016)
--(axis cs:760,4.55698892913031)
--(axis cs:770,4.58713331633208)
--(axis cs:780,4.61301082696201)
--(axis cs:780,4.95800774590599)
--(axis cs:780,4.95800774590599)
--(axis cs:770,4.95308781155488)
--(axis cs:760,4.9525761760594)
--(axis cs:750,4.97193714637192)
--(axis cs:740,4.978046130767)
--(axis cs:730,4.96805830983582)
--(axis cs:720,4.97366861556356)
--(axis cs:710,4.97394466346489)
--(axis cs:700,4.97394466346489)
--(axis cs:690,4.97811658950069)
--(axis cs:680,4.98370277111729)
--(axis cs:670,4.98745998558197)
--(axis cs:660,4.97929993103905)
--(axis cs:650,4.97929993103905)
--(axis cs:640,4.95310279810476)
--(axis cs:630,4.96402781241501)
--(axis cs:620,4.95724471900762)
--(axis cs:610,4.95147472108699)
--(axis cs:600,4.93787367526251)
--(axis cs:590,4.93651940580307)
--(axis cs:580,4.93651940580307)
--(axis cs:570,4.92717461689107)
--(axis cs:560,4.94256985385134)
--(axis cs:550,4.95262556764121)
--(axis cs:540,5.00039578520293)
--(axis cs:530,4.94116269144444)
--(axis cs:520,4.93652397510717)
--(axis cs:510,4.98633802145749)
--(axis cs:500,5.02443942086793)
--(axis cs:490,5.04244902852538)
--(axis cs:480,5.01827044799303)
--(axis cs:470,5.00312899317499)
--(axis cs:460,5.05288290168704)
--(axis cs:450,4.95672888345574)
--(axis cs:440,4.97444514572214)
--(axis cs:430,5.05869496606533)
--(axis cs:420,5.05752601803491)
--(axis cs:410,5.08787171391833)
--(axis cs:400,5.06192535498264)
--(axis cs:390,5.12322868455016)
--(axis cs:380,5.08683558987898)
--(axis cs:370,5.17000140308823)
--(axis cs:360,5.18058647725578)
--(axis cs:350,5.14247206452023)
--(axis cs:340,5.22090289653758)
--(axis cs:330,5.2139707775039)
--(axis cs:320,5.18634450860054)
--(axis cs:310,5.23551642167776)
--(axis cs:300,5.16053327925916)
--(axis cs:290,5.21012274580751)
--(axis cs:280,5.23154773939919)
--(axis cs:270,5.273336307278)
--(axis cs:260,5.3461484571625)
--(axis cs:250,5.26783324955221)
--(axis cs:240,5.31941069450466)
--(axis cs:230,5.37027987085241)
--(axis cs:220,5.39313699892569)
--(axis cs:210,5.30827467227606)
--(axis cs:200,5.29304674535621)
--(axis cs:190,5.16234854454593)
--(axis cs:180,5.31722504011505)
--(axis cs:170,5.34494122650846)
--(axis cs:160,5.29691867412421)
--(axis cs:150,5.33768255611069)
--(axis cs:140,5.34426149229335)
--(axis cs:130,5.35576703004628)
--(axis cs:120,5.41857472436523)
--(axis cs:110,5.42160611245574)
--(axis cs:100,5.38396782212897)
--(axis cs:90,5.30579326630098)
--(axis cs:80,5.36947352243935)
--(axis cs:70,5.31186891901205)
--(axis cs:60,5.49403980482861)
--(axis cs:50,5.69451740304272)
--(axis cs:40,5.32128120366402)
--(axis cs:30,5.04444039614801)
--(axis cs:20,4.66206267499031)
--(axis cs:10,22.0117025508035)
--cycle;
\addlegendimage{area legend, draw=darkcyan0158115, fill=darkcyan0158115, opacity=0.25}
\addlegendentry{Sim Gel 3 ±1$\sigma$}

\path [draw=darkcyan0158115, fill=darkcyan0158115, opacity=0.16]
(axis cs:10,54.0488100685128)
--(axis cs:10,0.427083612068678)
--(axis cs:20,0.870959986229672)
--(axis cs:30,0.655888308484842)
--(axis cs:40,0.172231982310213)
--(axis cs:50,0.716121369761738)
--(axis cs:60,1.00610479005193)
--(axis cs:70,1.88407110793853)
--(axis cs:80,2.47575693734654)
--(axis cs:90,2.8148130771588)
--(axis cs:100,2.8148130771588)
--(axis cs:110,2.8148130771588)
--(axis cs:120,3.09498981369093)
--(axis cs:130,3.25064019822793)
--(axis cs:140,3.25631952025631)
--(axis cs:150,3.25631952025631)
--(axis cs:160,3.27117491846904)
--(axis cs:170,2.98532227553543)
--(axis cs:180,2.98532227553543)
--(axis cs:190,3.40271661639388)
--(axis cs:200,3.10400413634542)
--(axis cs:210,3.10400413634542)
--(axis cs:220,2.80658418180335)
--(axis cs:230,2.80658418180335)
--(axis cs:240,2.80658418180335)
--(axis cs:250,3.10400413634542)
--(axis cs:260,3.10400413634542)
--(axis cs:270,3.40271661639388)
--(axis cs:280,3.70272912220967)
--(axis cs:290,3.70272912220967)
--(axis cs:300,3.70272912220967)
--(axis cs:310,3.92422322944036)
--(axis cs:320,3.92422322944036)
--(axis cs:330,4.00404920853591)
--(axis cs:340,4.00404920853591)
--(axis cs:350,4.04008433963561)
--(axis cs:360,3.7400446315867)
--(axis cs:370,4.00404920853591)
--(axis cs:380,4.00404920853591)
--(axis cs:390,4.04008433963561)
--(axis cs:400,4.04008433963561)
--(axis cs:410,4.19625356074053)
--(axis cs:420,4.30668448507342)
--(axis cs:430,4.30668448507342)
--(axis cs:440,4.30668448507342)
--(axis cs:450,4.34142760675117)
--(axis cs:460,4.47827059425249)
--(axis cs:470,4.47827059425249)
--(axis cs:480,4.47827059425249)
--(axis cs:490,4.42513590711317)
--(axis cs:500,4.30668448507342)
--(axis cs:510,4.30668448507342)
--(axis cs:520,4.30668448507342)
--(axis cs:530,4.30668448507342)
--(axis cs:540,4.30668448507342)
--(axis cs:550,4.30668448507342)
--(axis cs:560,4.33660396591282)
--(axis cs:570,4.33660396591282)
--(axis cs:580,4.33660396591282)
--(axis cs:590,4.33660396591282)
--(axis cs:600,4.50287258167706)
--(axis cs:610,4.50287258167706)
--(axis cs:620,4.30668448507342)
--(axis cs:630,4.30668448507342)
--(axis cs:640,4.42513590711317)
--(axis cs:650,4.42513590711317)
--(axis cs:660,4.42513590711317)
--(axis cs:670,4.42513590711317)
--(axis cs:680,4.42513590711317)
--(axis cs:690,4.42513590711317)
--(axis cs:700,4.42513590711317)
--(axis cs:710,4.42513590711317)
--(axis cs:720,4.42513590711317)
--(axis cs:730,4.42513590711317)
--(axis cs:740,4.19625356074053)
--(axis cs:750,4.42513590711317)
--(axis cs:760,4.42513590711317)
--(axis cs:770,4.50287258167706)
--(axis cs:780,4.50287258167706)
--(axis cs:780,5.12018491097224)
--(axis cs:780,5.12018491097224)
--(axis cs:770,5.12018491097224)
--(axis cs:760,5.12018491097224)
--(axis cs:750,5.12018491097224)
--(axis cs:740,5.12018491097224)
--(axis cs:730,5.12018491097224)
--(axis cs:720,5.12018491097224)
--(axis cs:710,5.12018491097224)
--(axis cs:700,5.12018491097224)
--(axis cs:690,5.12018491097224)
--(axis cs:680,5.12018491097224)
--(axis cs:670,5.12018491097224)
--(axis cs:660,5.12018491097224)
--(axis cs:650,5.12018491097224)
--(axis cs:640,5.12018491097224)
--(axis cs:630,5.12018491097224)
--(axis cs:620,5.12018491097224)
--(axis cs:610,5.12018491097224)
--(axis cs:600,5.12018491097224)
--(axis cs:590,5.12018491097224)
--(axis cs:580,5.12018491097224)
--(axis cs:570,5.12018491097224)
--(axis cs:560,5.12018491097224)
--(axis cs:550,5.25593596718273)
--(axis cs:540,5.25593596718273)
--(axis cs:530,5.22255838745198)
--(axis cs:520,5.25593596718273)
--(axis cs:510,5.25593596718273)
--(axis cs:500,5.25593596718273)
--(axis cs:490,5.25593596718273)
--(axis cs:480,5.34489416945731)
--(axis cs:470,5.43089428005298)
--(axis cs:460,5.34489416945731)
--(axis cs:450,5.28107712804661)
--(axis cs:440,5.28107712804661)
--(axis cs:430,5.43089428005298)
--(axis cs:420,5.43089428005298)
--(axis cs:410,5.43089428005298)
--(axis cs:400,5.43089428005298)
--(axis cs:390,5.43089428005298)
--(axis cs:380,5.34489416945731)
--(axis cs:370,5.34489416945731)
--(axis cs:360,5.34489416945731)
--(axis cs:350,5.31584896023093)
--(axis cs:340,5.34489416945731)
--(axis cs:330,5.34489416945731)
--(axis cs:320,5.37312397771569)
--(axis cs:310,5.37312397771569)
--(axis cs:300,5.37312397771569)
--(axis cs:290,5.68091639026267)
--(axis cs:280,5.68091639026267)
--(axis cs:270,5.68091639026267)
--(axis cs:260,5.71276862999619)
--(axis cs:250,5.71276862999619)
--(axis cs:240,6.02153535587048)
--(axis cs:230,6.02153535587048)
--(axis cs:220,6.02153535587048)
--(axis cs:210,6.02153535587048)
--(axis cs:200,6.02153535587048)
--(axis cs:190,5.71276862999619)
--(axis cs:180,6.3316568325839)
--(axis cs:170,6.3316568325839)
--(axis cs:160,6.28744514636973)
--(axis cs:150,6.59769425777978)
--(axis cs:140,6.59769425777978)
--(axis cs:130,6.59769425777978)
--(axis cs:120,6.59769425777978)
--(axis cs:110,6.85389331610906)
--(axis cs:100,6.85389331610906)
--(axis cs:90,6.57759724818625)
--(axis cs:80,6.26811665957295)
--(axis cs:70,5.89878350574039)
--(axis cs:60,7.15990383398547)
--(axis cs:50,7.15990383398547)
--(axis cs:40,6.54008401343691)
--(axis cs:30,5.95535866712156)
--(axis cs:20,7.57811442519399)
--(axis cs:10,54.0488100685128)
--cycle;
\addlegendimage{area legend, draw=darkcyan0158115, fill=darkcyan0158115, opacity=0.16}
\addlegendentry{Sim Gel 3 min–max}

\addplot [line width=1.5pt, darkcyan0114178]
table {%
10 13.8964463284898
20 2.85454719571908
30 2.40890501586249
40 2.14221437544154
50 2.05921930365669
60 1.82930127446171
70 1.78912085408137
80 1.79444652883338
90 1.71056594957925
100 1.71953599976287
110 1.59025052267447
120 1.52833715814602
130 1.49231755814092
140 1.51836388056466
150 1.46449514026258
160 1.5027978614138
170 1.52704731425462
180 1.50739239336738
190 1.53602804252089
200 1.51150750761408
210 1.52353462314185
220 1.51342612157134
230 1.51028728536675
240 1.52488826755395
250 1.49968492502793
260 1.48113654827769
270 1.51514715767024
280 1.48597417640974
290 1.46578245627248
300 1.4419216232576
310 1.44404876987252
320 1.45293858680115
330 1.42173636990573
340 1.39478037752567
350 1.40258819753459
360 1.38438829424629
370 1.39237533639597
380 1.41732665717276
390 1.44037716001926
400 1.4131950499938
410 1.41137234927986
420 1.43376141577573
430 1.41779946538923
440 1.44973573609643
450 1.44558694754732
460 1.45935877619469
470 1.39626657884256
480 1.39563094334527
490 1.40722963710745
500 1.39855496791895
510 1.39012690130577
520 1.38948090417458
530 1.39921503369755
540 1.39709243996694
550 1.38853510951203
560 1.38903667626478
570 1.4059763535364
580 1.38903667626478
590 1.40552500540465
600 1.42320735896501
610 1.42331209728664
620 1.45089234505278
630 1.44014351928955
650 1.44351897786634
660 1.45296532461617
670 1.45322556062105
680 1.44552161992572
690 1.42753488467457
700 1.39954213395059
710 1.41721940634318
720 1.40893371570159
730 1.40893408302378
740 1.41992268250657
750 1.40237572440929
760 1.39324213466274
770 1.39382461968639
780 1.41598421686575
};
\addlegendentry{Sim Gel  1 mean}
\addplot [line width=1.5pt, orange2301590]
table {%
10 8.48685987162795
20 8.27894250047666
30 8.09714014072499
40 7.71359345695428
50 8.00798595124533
60 7.93548752776903
70 7.95818498154197
80 8.11014196389428
90 8.16267879145695
100 8.16248310668578
110 8.18756548436443
120 8.24627554479357
130 8.25959707285454
140 8.21291590958925
150 8.21230425104207
160 8.27807023428053
170 8.26567833741171
180 8.29855987530678
190 8.32610027492153
200 8.32231565945921
210 8.33487625979581
220 8.37097884913373
230 8.38454390044155
240 8.41420431873022
250 8.38143762931458
260 8.36356675996253
270 8.41266217607251
280 8.34712990211845
290 8.42147629879661
300 8.42189659535969
310 8.47207938102721
320 8.43787837819252
330 8.43788378968058
340 8.44270077275722
350 8.43901088141219
360 8.47872404172886
370 8.50529648864741
380 8.44286155224058
390 8.42605746009436
400 8.39199566160195
410 8.37578511342588
420 8.39249918136044
430 8.43085615805827
440 8.44121070423602
450 8.45840945293797
460 8.4407717368666
470 8.49223371730406
480 8.47525886559891
490 8.42441891201607
500 8.41902147036312
510 8.38458234889945
520 8.44106083385988
530 8.49101824778395
540 8.47407592666483
550 8.43371268405343
560 8.41653400072451
570 8.43365695689615
580 8.43662919305146
590 8.4827863832691
600 8.43494638531479
610 8.41838900722821
620 8.48615798951912
630 8.45220486551798
640 8.51515630228236
650 8.486446797059
660 8.46899356596819
670 8.48576202494408
680 8.49841329136005
690 8.48151830276087
700 8.49841329136005
710 8.46902711772642
720 8.46916227281561
730 8.45709107473685
740 8.4572136670543
750 8.47439235038321
760 8.55921691224273
770 8.52529973941845
780 8.50149326618424
};
\addlegendentry{Sim Gel  2 mean}
\addplot [line width=1.5pt, darkcyan0158115]
table {%
10 8.11261507447473
20 3.01607299754566
30 3.53838366574128
40 3.70869093080976
50 4.08734751654322
60 4.01452200950331
70 4.12281870485288
80 4.31197995667258
90 4.40153202027214
100 4.4832094970746
110 4.5528071783058
120 4.58995685869958
130 4.54348449399073
140 4.55343041840029
150 4.58341681583974
160 4.58770550623603
170 4.6248704008881
180 4.60607990784212
190 4.54141505043027
200 4.58825698526983
210 4.62055009880277
220 4.66027102714548
230 4.71063137344964
240 4.67645952710206
250 4.67008962295114
260 4.73174491557775
270 4.74480646466619
280 4.72969399422033
290 4.71454554118001
300 4.67009837346317
310 4.7483778481722
320 4.74710132060526
330 4.77784599555659
340 4.78458155534832
350 4.74051628418639
360 4.75952175045419
370 4.76741051594229
380 4.72142814134709
390 4.76499078601361
400 4.68853492754261
410 4.73400620683016
420 4.73397512629455
430 4.7339609256768
440 4.70308682163636
450 4.70302029281572
460 4.78202704547495
470 4.73630824890151
480 4.77811198089906
490 4.79347891187765
500 4.74498945683068
510 4.72954262572354
520 4.69508093791784
530 4.69070159814481
540 4.72270911054996
550 4.70717694154432
560 4.73947960976035
570 4.72401164352601
580 4.73940889861846
590 4.73940889861846
600 4.77347852138711
610 4.78874295680214
620 4.75814632771723
630 4.77354505020775
640 4.77798772568125
650 4.79078736392977
660 4.79078736392977
670 4.80608217790992
680 4.7906834554194
690 4.77528473292888
700 4.77538864143925
710 4.77538864143925
720 4.78658964616822
730 4.7711909236777
740 4.75319271690516
750 4.78553314924104
760 4.75478255259486
770 4.77011056394348
780 4.785509286434
};
\addlegendentry{Sim Gel  3 mean}
\end{axis}

\end{tikzpicture}

%% file: Figures/jelly_baselines_reject.tex
\begin{tikzpicture}

\definecolor{chocolate22412342}{RGB}{224,123,42}
\definecolor{darkgrey170}{RGB}{170,170,170}
\definecolor{darkgrey176}{RGB}{176,176,176}
\definecolor{dimgrey102}{RGB}{102,102,102}
\definecolor{forestgreen4416044}{RGB}{44,160,44}
\definecolor{mediumorchid15589182}{RGB}{155,89,182}
\definecolor{steelblue33102172}{RGB}{33,102,172}

\begin{axis}[
height =5cm,
width = \linewidth,
legend cell align={left},
legend columns=3,
scale only axis=false,
legend style={fill opacity=0.8, draw opacity=1, text opacity=1, at={(0.15,1.3)}, anchor=north west},
tick align=outside,
tick pos=left,
x grid style={darkgrey176},
xmin=-0.5, xmax=2.5,
xtick style={color=black},
xtick={0,1,2},
x tick label style={align=center},
xticklabels={Sim Gel 1, Sim Gel 2, Sim Gel 3}, 
y grid style={darkgrey176},
ylabel={Rejection Rate(\%)},
ymajorgrids,
ymin=0, ymax=105.0,
ytick style={color=black}
]
\draw[draw=dimgrey102,fill=darkgrey170,line width=0.32pt] (axis cs:-0.375,0) rectangle (axis cs:-0.225,0);
\addlegendimage{area legend ,draw=dimgrey102,fill=darkgrey170,line width=0.32pt}
\addlegendentry{Raw}

\draw[draw=dimgrey102,fill=darkgrey170,line width=0.32pt] (axis cs:0.625,0) rectangle (axis cs:0.775,0);
\draw[draw=dimgrey102,fill=darkgrey170,line width=0.32pt] (axis cs:1.625,0) rectangle (axis cs:1.775,0);
\draw[draw=steelblue33102172,fill=steelblue33102172,line width=0.32pt,postaction={pattern=north east lines, pattern color=black}] (axis cs:-0.225,0) rectangle (axis cs:-0.075,0);
\addlegendimage{area legend ,draw=steelblue33102172,fill=steelblue33102172,line width=0.32pt,postaction={pattern=north east lines, pattern color=steelblue33102172}}
\addlegendentry{W/o (GMM,CCS,PCS)}

\draw[draw=steelblue33102172,fill=steelblue33102172,line width=0.32pt,postaction={pattern=north east lines, pattern color=black}] (axis cs:0.775,0) rectangle (axis cs:0.925,0);
\draw[draw=steelblue33102172,fill=steelblue33102172,line width=0.32pt,postaction={pattern=north east lines, pattern color=black}] (axis cs:1.775,0) rectangle (axis cs:1.925,0);
\draw[draw=chocolate22412342,fill=chocolate22412342,line width=0.32pt,postaction={pattern=north east lines, pattern color=black}] (axis cs:-0.075,0) rectangle (axis cs:0.075,6.14);
\addlegendimage{area legend ,draw=chocolate22412342,fill=chocolate22412342,line width=0.32pt,postaction={pattern=north east lines, pattern color=black}}
\addlegendentry{W/o(GMM,CCS)}

\draw[draw=chocolate22412342,fill=chocolate22412342,line width=0.32pt,postaction={pattern=north east lines, pattern color=black}] (axis cs:0.925,0) rectangle (axis cs:1.075,56.04);
\draw[draw=chocolate22412342,fill=chocolate22412342,line width=0.32pt,postaction={pattern=north east lines, pattern color=black}] (axis cs:1.925,0) rectangle (axis cs:2.075,51.38);
\draw[draw=mediumorchid15589182,fill=mediumorchid15589182,line width=0.32pt,postaction={pattern=north east lines, pattern color=black}] (axis cs:0.075,0) rectangle (axis cs:0.225,20.18);
\addlegendimage{area legend ,draw=mediumorchid15589182,fill=mediumorchid15589182,line width=0.32pt,postaction={pattern=north east lines, pattern color=black}}
\addlegendentry{W/o GMM}

\draw[draw=mediumorchid15589182,fill=mediumorchid15589182,line width=0.32pt,postaction={pattern=north east lines, pattern color=black}] (axis cs:1.075,0) rectangle (axis cs:1.225,68.04);
\draw[draw=mediumorchid15589182,fill=mediumorchid15589182,line width=0.32pt,postaction={pattern=north east lines, pattern color=black}] (axis cs:2.075,0) rectangle (axis cs:2.225,68.47);
\draw[draw=forestgreen4416044,fill=forestgreen4416044,line width=0.32pt,postaction={pattern=north east lines, pattern color=forestgreen4416044}] (axis cs:0.225,0) rectangle (axis cs:0.375,20.18);
\addlegendimage{area legend ,draw=forestgreen4416044,fill=forestgreen4416044,line width=0.32pt,postaction={pattern=north east lines, pattern color=forestgreen4416044}}
\addlegendentry{WAVES}

\draw[draw=forestgreen4416044,fill=forestgreen4416044,line width=0.32pt,postaction={pattern=north east lines, pattern color=forestgreen4416044}] (axis cs:1.225,0) rectangle (axis cs:1.375,68.04);
\draw[draw=forestgreen4416044,fill=forestgreen4416044,line width=0.32pt,postaction={pattern=north east lines, pattern color=forestgreen4416044}] (axis cs:2.225,0) rectangle (axis cs:2.375,68.47);
\draw (axis cs:-0.2,10.5) node[
  scale=0.8,,
  anchor=south,
  text=black,
  rotate=90.0
]{\bfseries 0};
\draw (axis cs:0.8,10.5) node[
  scale=0.8,,
  anchor=south,
  text=black,
  rotate=90.0
]{\bfseries 0};
\draw (axis cs:1.8,10.5) node[
  scale=0.8,,
  anchor=south,
  text=black,
  rotate=90.0
]{\bfseries 0};
\draw (axis cs:-0.05,10.5) node[
  scale=0.8,,
  anchor=south,
  text=black,
  rotate=90.0
]{\bfseries 0};
\draw (axis cs:0.95,10.5) node[
  scale=0.8,,
  anchor=south,
  text=black,
  rotate=90.0
]{\bfseries 0};
\draw (axis cs:1.95,10.5) node[
  scale=0.8,,
  anchor=south,
  text=black,
  rotate=90.0
]{\bfseries 0};
\draw (axis cs:0.1,16.64) node[
  scale=0.8,,
  anchor=south,
  text=black,
  rotate=90.0
]{\bfseries 6.14};
\draw (axis cs:1.1,66.54) node[
  scale=0.8,,
  anchor=south,
  text=black,
  rotate=90.0
]{\bfseries 56.04};
\draw (axis cs:2.1,61.88) node[
  scale=0.8,,
  anchor=south,
  text=black,
  rotate=90.0
]{\bfseries 51.38};
\draw (axis cs:0.25,30.68) node[
  scale=0.8,,
  anchor=south,
  text=black,
  rotate=90.0
]{\bfseries 20.18};
\draw (axis cs:1.25,78.54) node[
  scale=0.8,,
  anchor=south,
  text=black,
  rotate=90.0
]{\bfseries 68.98};
\draw (axis cs:2.25,78.97) node[
  scale=0.8,,
  anchor=south,
  text=black,
  rotate=90.0
]{\bfseries 68.26};
\draw (axis cs:0.4,30.68) node[
  scale=0.8,,
  anchor=south,
  text=black,
  rotate=90.0
]{\bfseries 20.18};
\draw (axis cs:1.4,78.54) node[
  scale=0.8,,
  anchor=south,
  text=black,
  rotate=90.0
]{\bfseries 68.98};
\draw (axis cs:2.4,78.97) node[
  scale=0.8,,
  anchor=south,
  text=black,
  rotate=90.0
]{\bfseries 68.26};
\end{axis}

\end{tikzpicture}

%% file: Figures/jelly_baselines.tex
\begin{tikzpicture}

\definecolor{chocolate22412342}{RGB}{224,123,42}
\definecolor{darkgrey170}{RGB}{170,170,170}
\definecolor{darkgrey176}{RGB}{176,176,176}
\definecolor{dimgrey102}{RGB}{102,102,102}
\definecolor{forestgreen4416044}{RGB}{44,160,44}
\definecolor{mediumorchid15589182}{RGB}{155,89,182}
\definecolor{steelblue33102172}{RGB}{33,102,172}

\begin{axis}[
height =5cm,
width = \linewidth,
legend cell align={left},
legend columns=3,
legend style={draw=none, fill=none, fill opacity=0, draw opacity=0, text opacity=0, at={(-0.20,1.3)}, anchor=north west},
tick align=outside,
tick pos=left,
x grid style={darkgrey176},
xmin=-0.5, xmax=2.5,
xtick style={color=black},
xtick={0,1,2},
x tick label style={align=center},
xticklabels={Sim Gel 1, Sim Gel 2, Sim Gel 3}, 
y grid style={darkgrey176},
ylabel={Estimation Error(\%)},
ymajorgrids,
ymin=0, ymax=105.0,
ytick style={color=black}
]
\draw[draw=dimgrey102,fill=darkgrey170,line width=0.32pt] (axis cs:-0.375,0) rectangle (axis cs:-0.225,19.237775573088);
\addlegendimage{area legend,draw=dimgrey102,fill=darkgrey170,line width=0.32pt}
\addlegendentry{Raw}

\draw[draw=dimgrey102,fill=darkgrey170,line width=0.32pt] (axis cs:0.625,0) rectangle (axis cs:0.775,70.2487901007673);
\draw[draw=dimgrey102,fill=darkgrey170,line width=0.32pt] (axis cs:1.625,0) rectangle (axis cs:1.775,15.3060744239661);
\draw[draw=steelblue33102172,fill=steelblue33102172,line width=0.32pt,postaction={pattern=north east lines, pattern color=black}] (axis cs:-0.225,0) rectangle (axis cs:-0.075,20.3455313754192);
\addlegendimage{area legend ,draw=steelblue33102172,fill=steelblue33102172,line width=0.32pt,postaction={pattern=north east lines, pattern color=black}}
\addlegendentry{W/o (GMM,CCS,PCS)}

\draw[draw=steelblue33102172,fill=steelblue33102172,line width=0.32pt,postaction={pattern=north east lines, pattern color=black}] (axis cs:0.775,0) rectangle (axis cs:0.925,24.44474577694);
\draw[draw=steelblue33102172,fill=steelblue33102172,line width=0.32pt,postaction={pattern=north east lines, pattern color=black}] (axis cs:1.775,0) rectangle (axis cs:1.925,19.4678486303485);
\draw[draw=chocolate22412342,fill=chocolate22412342,line width=0.32pt,postaction={pattern=north east lines, pattern color=black}] (axis cs:-0.075,0) rectangle (axis cs:0.075,22.4287371493322);
\addlegendimage{area legend ,draw=chocolate22412342,fill=chocolate22412342,line width=0.32pt,postaction={pattern=north east lines, pattern color=black}}
\addlegendentry{W/o (GMM,CCS)}

\draw[draw=chocolate22412342,fill=chocolate22412342,line width=0.32pt,postaction={pattern=north east lines, pattern color=black}] (axis cs:0.925,0) rectangle (axis cs:1.075,20.0604890557012);
\draw[draw=chocolate22412342,fill=chocolate22412342,line width=0.32pt,postaction={pattern=north east lines, pattern color=black}] (axis cs:1.925,0) rectangle (axis cs:2.075,38.2146756268473);
\draw[draw=mediumorchid15589182,fill=mediumorchid15589182,line width=0.32pt,postaction={pattern=north east lines, pattern color=black}] (axis cs:0.075,0) rectangle (axis cs:0.225,19.1775514556552);
\addlegendimage{area legend ,draw=mediumorchid15589182,fill=mediumorchid15589182,line width=0.32pt,postaction={pattern=north east lines, pattern color=black}}
\addlegendentry{W/o GMM}

\draw[draw=mediumorchid15589182,fill=mediumorchid15589182,line width=0.32pt,postaction={pattern=north east lines, pattern color=black}] (axis cs:1.075,0) rectangle (axis cs:1.225,6.49824899637788);
\draw[draw=mediumorchid15589182,fill=mediumorchid15589182,line width=0.32pt,postaction={pattern=north east lines, pattern color=black}] (axis cs:2.075,0) rectangle (axis cs:2.225,22.7331572379831);
\draw[draw=forestgreen4416044,fill=forestgreen4416044,line width=0.32pt,postaction={pattern=north east lines, pattern color=forestgreen4416044}] (axis cs:0.225,0) rectangle (axis cs:0.375,1.23963024785066);
\addlegendimage{area legend ,draw=forestgreen4416044,fill=forestgreen4416044,line width=0.32pt,postaction={pattern=north east lines, pattern color=forestgreen4416044}}
\addlegendentry{WAVES}

\draw[draw=forestgreen4416044,fill=forestgreen4416044,line width=0.32pt,postaction={pattern=north east lines, pattern color=forestgreen4416044}] (axis cs:1.225,0) rectangle (axis cs:1.375,8.80517758737913);
\draw[draw=forestgreen4416044,fill=forestgreen4416044,line width=0.32pt,postaction={pattern=north east lines, pattern color=forestgreen4416044}] (axis cs:2.225,0) rectangle (axis cs:2.375,4.19625356074053);
\draw (axis cs:-0.2,29.737775573088) node[
  scale=0.8,
  anchor=south,
  text=black,
  rotate=90.0
]{\bfseries 19.2};
\draw (axis cs:0.8,80.7487901007673) node[
  scale=0.8,
  anchor=south,
  text=black,
  rotate=90.0
]{\bfseries 70.2};
\draw (axis cs:1.8,25.8060744239661) node[
  scale=0.8,
  anchor=south,
  text=black,
  rotate=90.0
]{\bfseries 15.3};
\draw (axis cs:-0.05,30.8455313754192) node[
  scale=0.8,
  anchor=south,
  text=black,
  rotate=90.0
]{\bfseries 20.3};
\draw (axis cs:0.95,34.94474577694) node[
  scale=0.8,
  anchor=south,
  text=black,
  rotate=90.0
]{\bfseries 24.4};
\draw (axis cs:1.95,29.9678486303485) node[
  scale=0.8,
  anchor=south,
  text=black,
  rotate=90.0
]{\bfseries 19.5};
\draw (axis cs:0.1,32.9287371493322) node[
  scale=0.8,
  anchor=south,
  text=black,
  rotate=90.0
]{\bfseries 22.4};
\draw (axis cs:1.1,30.5604890557012) node[
  scale=0.8,
  anchor=south,
  text=black,
  rotate=90.0
]{\bfseries 20.1};
\draw (axis cs:2.1,48.7146756268473) node[
  scale=0.8,
  anchor=south,
  text=black,
  rotate=90.0
]{\bfseries 38.2};
\draw (axis cs:0.25,29.6775514556552) node[
  scale=0.8,
  anchor=south,
  text=black,
  rotate=90.0
]{\bfseries 19.2};
\draw (axis cs:1.25,16.99824899637788) node[
  scale=0.8,
  anchor=south,
  text=black,
  rotate=90.0
]{\bfseries 6.5};
\draw (axis cs:2.25,33.2331572379831) node[
  scale=0.8,
  anchor=south,
  text=black,
  rotate=90.0
]{\bfseries 22.7};
\draw (axis cs:0.4,11.73963024785066) node[
  scale=0.8,
  anchor=south,
  text=black,
  rotate=90.0
]{\bfseries 1.4};
\draw (axis cs:1.4,19.30517758737913) node[
  scale=0.8,
  anchor=south,
  text=black,
  rotate=90.0
]{\bfseries 8.4};
\draw (axis cs:2.4,14.69625356074053) node[
  scale=0.8,
  anchor=south,
  text=black,
  rotate=90.0
]{\bfseries 4.8};
\end{axis}

\end{tikzpicture}

%% file: Figures/tdoa_methods.tex
\begin{tikzpicture}

\definecolor{darkgoldenrod179950}{RGB}{179,95,0}
\definecolor{darkgreen08242}{RGB}{0,82,42}
\definecolor{darkgrey176}{RGB}{176,176,176}
\definecolor{darkorange23013010}{RGB}{230,130,10}
\definecolor{darkslategrey2674122}{RGB}{26,74,122}
\definecolor{forestgreen4416044}{RGB}{44,160,44}
\definecolor{seagreen2615280}{RGB}{26,152,80}
\definecolor{steelblue33102172}{RGB}{33,102,172}

\begin{axis}[
height =5cm,
width = \linewidth,
legend cell align={left},
legend columns=2,
legend style={fill opacity=0.8, draw opacity=1, text opacity=1, at={(0.4,1.3)}, anchor=north},
tick align=outside,
tick pos=left,
x grid style={darkgrey176},
xmin=-0.5, xmax=2.5,
xtick style={color=black},
xtick={0,1,2},
x tick label style={align=center},
xticklabels={Sim Gel 1, Sim Gel 2, Sim Gel 3}, 
y grid style={darkgrey176},
ylabel={ Estimation Error(\%)},
ymajorgrids,
ymin=0, ymax=105.0,
ytick style={color=black}
]
\draw[draw=forestgreen4416044,fill=forestgreen4416044] (axis cs:-0.33,0) rectangle (axis cs:-0.11,1.41);
\addlegendimage{area legend ,draw=forestgreen4416044, fill=forestgreen4416044}
\addlegendentry{\name}

\draw[draw=forestgreen4416044,fill=forestgreen4416044] (axis cs:0.67,0) rectangle (axis cs:0.89,8.42);
\draw[draw=forestgreen4416044,fill=forestgreen4416044] (axis cs:1.67,0) rectangle (axis cs:1.89,4.84);
\draw (axis cs:-0.22,2.01) node[
  scale=0.8,
  anchor=south,
  text=black,
  rotate=0.0
]{1.41};
\draw (axis cs:0.78,9.19) node[
  scale=0.8,
  anchor=south,
  text=black,
  rotate=0.0
]{8.42};
\draw (axis cs:1.78,5.64) node[
  scale=0.8,
  anchor=south,
  text=black,
  rotate=0.0
]{4.84};
\draw[draw=darkgoldenrod179950,fill=darkorange23013010,postaction={pattern=north east lines, pattern color=black}] (axis cs:-0.11,0) rectangle (axis cs:0.11,38.25);
\addlegendimage{area legend ,draw=darkgoldenrod179950,fill=darkorange23013010,postaction={pattern=north east lines, pattern color=black}}
\addlegendentry{Baseline CC}

\draw[draw=darkgoldenrod179950,fill=darkorange23013010,postaction={pattern=north east lines, pattern color=black}] (axis cs:0.89,0) rectangle (axis cs:1.11,63.45);
\draw[draw=darkgoldenrod179950,fill=darkorange23013010,postaction={pattern=north east lines, pattern color=black}] (axis cs:1.89,0) rectangle (axis cs:2.11,71.26);
\draw (axis cs:0,39.05) node[
  scale=0.8,
  anchor=south,
  text=black,
  rotate=0.0
]{38.25};
\draw (axis cs:1,64.25) node[
  scale=0.8,
  anchor=south,
  text=black,
  rotate=0.0
]{63.45};
\draw (axis cs:2,72.06) node[
  scale=0.8,
  anchor=south,
  text=black,
  rotate=0.0
]{71.26};
\draw[draw=darkgreen08242,fill=steelblue33102172,postaction={pattern=crosshatch, pattern color=black}] (axis cs:0.11,0) rectangle (axis cs:0.33,57.48);
\addlegendimage{area legend ,draw=darkgreen08242,fill=steelblue33102172,postaction={pattern=crosshatch, pattern color=black}}
\addlegendentry{Baseline GCC}

\draw[draw=darkgreen08242,fill=steelblue33102172,postaction={pattern=crosshatch, pattern color=black}] (axis cs:1.11,0) rectangle (axis cs:1.33,65.24);
\draw[draw=darkgreen08242,fill=steelblue33102172,postaction={pattern=crosshatch, pattern color=black}] (axis cs:2.11,0) rectangle (axis cs:2.33,95.37);
\draw (axis cs:0.22,58.28) node[
  scale=0.8,
  anchor=south,
  text=black,
  rotate=0.0
]{57.48};
\draw (axis cs:1.22,66.04) node[
  scale=0.8,
  anchor=south,
  text=black,
  rotate=0.0
]{65.24};
\draw (axis cs:2.22,96.17) node[
  scale=0.8,
  anchor=south,
  text=black,
  rotate=0.0
]{95.37};
\end{axis}

\end{tikzpicture}

%% file: Figures/muscle_plot.tex
\begin{tikzpicture}

\definecolor{darkgoldenrod179950}{RGB}{179,95,0}
\definecolor{darkgreen08242}{RGB}{0,82,42}
\definecolor{darkgrey176}{RGB}{176,176,176}
\definecolor{darkorange23013010}{RGB}{230,130,10}
\definecolor{darkslategrey2674122}{RGB}{26,74,122}
\definecolor{seagreen2615280}{RGB}{26,152,80}
\definecolor{steelblue33102172}{RGB}{33,102,172}

\begin{axis}[
height=4cm,
tick align=outside,
tick pos=left,
width=\linewidth,
x grid style={darkgrey176},
xmin=-0.375, xmax=2.375,
xtick style={color=black},
xtick={0,1,2},
xticklabels={Thigh,Back,Forearm},
y grid style={darkgrey176},
ylabel={Estimation Error (\%)},
ymajorgrids,
ymin=0, ymax=100.0,
ytick style={color=black},
]
\draw[draw=black,fill=steelblue33102172,postaction={pattern=crosshatch, pattern color=black}] (axis cs:-0.25,0) rectangle (axis cs:0.25,13.08);
\draw (axis cs:0,18.0462190261495) node[
  scale=0.8,
  anchor=south,
  text=black,
  rotate=0.0
]{13.1\%};
\draw[draw=black,fill=darkorange23013010,postaction={pattern=north east lines, pattern color=black}] (axis cs:0.75,0) rectangle (axis cs:1.25,5.88);
\draw (axis cs:1,8.67110417285187) node[
  scale=0.8,
  anchor=south,
  text=black,
  rotate=0.0
]{5.9\%};
\draw[draw=black,fill=seagreen2615280,postaction={pattern=none, pattern color=black}] (axis cs:1.75,0) rectangle (axis cs:2.25,7.4);
\draw (axis cs:2,11.8895107112884) node[
  scale=0.8,
  anchor=south,
  text=black,
  rotate=0.0
]{7.4\%};
\path [draw=black, semithick]
(axis cs:0,8.41378097385046)
--(axis cs:0,17.7462190261495);

\addplot [thick, black, mark=-, mark size=6, mark options={solid}, only marks]
table {%
0 8.41378097385046
};
\addplot [thick, black, mark=-, mark size=6, mark options={solid}, only marks]
table {%
0 17.7462190261495
};
\path [draw=black, semithick]
(axis cs:1,3.38889582714813)
--(axis cs:1,8.37110417285187);

\addplot [thick, black, mark=-, mark size=6, mark options={solid}, only marks]
table {%
1 3.38889582714813
};
\addplot [thick, black, mark=-, mark size=6, mark options={solid}, only marks]
table {%
1 8.37110417285187
};
\path [draw=black, thick]
(axis cs:2,3.21048928871163)
--(axis cs:2,11.5895107112884);

\addplot [thick, black, mark=-, mark size=6, mark options={solid}, only marks]
table {%
2 3.21048928871163
};
\addplot [thick, black, mark=-, mark size=6, mark options={solid}, only marks]
table {%
2 11.5895107112884
};
\end{axis}

\end{tikzpicture}

%% file: Figures/same_sub_multi.tex
\begin{tikzpicture}

\definecolor{darkgoldenrod179950}{RGB}{179,95,0}
\definecolor{darkgreen08242}{RGB}{0,82,42}
\definecolor{darkgrey176}{RGB}{176,176,176}
\definecolor{darkorange23013010}{RGB}{230,130,10}
\definecolor{darkslategrey2674122}{RGB}{26,74,122}
\definecolor{seagreen2615280}{RGB}{26,152,80}
\definecolor{steelblue33102172}{RGB}{33,102,172}

\begin{axis}[
height=4cm,
width =\linewidth,
tick align=outside,
tick pos=left,
axis lines=box,
x grid style={darkgrey176},
xmin=-0.375, xmax=2.375,
xtick style={color=black},
xtick={0,1,2},
xticklabels={Day1,Day2,Day3},
y grid style={darkgrey176},
ylabel={Estimation Error(\%)},
ymajorgrids,
ymin=0, ymax=100,
ytick style={color=black}
]
\draw[draw=darkslategrey2674122,fill=steelblue33102172,postaction={pattern=crosshatch, pattern color=black}] (axis cs:-0.25,0) rectangle (axis cs:0.25,12.1851280174506);
\draw[draw=darkgoldenrod179950,fill=darkorange23013010,postaction={pattern=north east lines, pattern color=black}] (axis cs:0.75,0) rectangle (axis cs:1.25,15.7129262663373);
\draw[draw=darkgreen08242,fill=seagreen2615280] (axis cs:1.75,0) rectangle (axis cs:2.25,7.59773710624445);
\draw (axis cs:0,12.3851280174506) node[
  scale=0.8,
  anchor=south,
  text=black,
  rotate=0.0
]{\bfseries 12.19};
\draw (axis cs:1,15.9129262663373) node[
  scale=0.8,
  anchor=south,
  text=black,
  rotate=0.0
]{\bfseries 15.71};
\draw (axis cs:2,7.79773710624445) node[
  scale=0.8,
  anchor=south,
  text=black,
  rotate=0.0
]{\bfseries 7.60};
\end{axis}

\end{tikzpicture}

%% file: Figures/hodogram_comp.tex
\begin{tikzpicture}

\definecolor{darkcyan0119187}{RGB}{0,119,187}
\definecolor{darkcyan0153136}{RGB}{0,153,136}
\definecolor{darkgrey153}{RGB}{153,153,153}
\definecolor{darkgrey176}{RGB}{176,176,176}
\definecolor{gainsboro224}{RGB}{224,224,224}
\definecolor{tomato23811951}{RGB}{238,119,51}

\begin{axis}[
height=4cm,
width=\linewidth,
tick pos=both,
x grid style={darkgrey176},
xmin=-0.413, xmax=1.413,
xtick style={color=black},
xtick={0,1},
xticklabels={High quality,Low quality},
y grid style={gainsboro224},
ylabel={Rejention Rate (\%)},
ymajorgrids,
ymin=0, ymax=115,
ytick style={color=black}
]
\addplot [line width=0.32pt, darkgrey153]
table {%
-0.413 100
1.413 100
};
\draw[draw=black,fill=darkcyan0119187,line width=0.32pt] (axis cs:-0.33,0) rectangle (axis cs:-0.11,62.5);
\draw[draw=black,fill=darkcyan0119187,line width=0.32pt] (axis cs:0.67,0) rectangle (axis cs:0.89,89.2);
\draw (axis cs:-0.22,63) node[
  scale=0.8,
  anchor=south,
  text=black,
  rotate=0.0
]{62.5};
\draw (axis cs:0.78,89.7) node[
  scale=0.8,
  anchor=south,
  text=black,
  rotate=0.0
]{89.2};
\draw[draw=black,fill=tomato23811951,line width=0.32pt,postaction={pattern=north east lines}] (axis cs:-0.11,0) rectangle (axis cs:0.11,69.3);
\draw[draw=black,fill=tomato23811951,line width=0.32pt,postaction={pattern=north east lines}] (axis cs:0.89,0) rectangle (axis cs:1.11,98.4);
\draw (axis cs:0,69.8) node[
  scale=0.8,
  anchor=south,
  text=black,
  rotate=0.0
]{69.3};
\draw (axis cs:1,98.9) node[
  scale=0.8,
  anchor=south,
  text=black,
  rotate=0.0
]{98.4};
\draw[draw=black,fill=darkcyan0153136,line width=0.32pt,postaction={pattern=grid}] (axis cs:0.11,0) rectangle (axis cs:0.33,33.11);
\draw[draw=black,fill=darkcyan0153136,line width=0.32pt,postaction={pattern=grid}] (axis cs:1.11,0) rectangle (axis cs:1.33,86);
\draw (axis cs:0.22,33.61) node[
  scale=0.8,
  anchor=south,
  text=black,
  rotate=0.0
]{33.1};
\draw (axis cs:1.22,86.5) node[
  scale=0.8,
  anchor=south,
  text=black,
  rotate=0.0
]{86.0};
\end{axis}

\end{tikzpicture}

%% file: Related_Work.tex
\section{Related Work}
This section introduces related work from three aspects, which align with the technical contributions and provide comparisons to demonstrate novelty.

\paragraph{Muscle Stiffness Measurement and Estimation.}
Muscle stiffness is an important biomarker for assessing musculoskeletal health, rehabilitation progress, and neuromuscular disorders. Existing approaches estimate muscle stiffness through manual palpation \cite{Davidson2020Jan, Myburgh2008Jun, Rathbone2017Aug}, handheld mechanical probing devices such palpation devices \cite{chuang2012reliability,Myoton026Feb}, and ultrasound shear-wave elastography has become a widely adopted method because shear-wave velocity is strongly correlated with tissue stiffness \cite{Miyamoto2015Apr,Stiver2023Jun,Eby2013Sep}. However, these systems typically require specialized imaging hardware, precise sensor placement, and substantial computational resources, limiting their suitability for continuous and everyday monitoring. Recent estimation approaches also rely on biomechanical models or indirect physiological measurements such as muscle activation signals, which may not directly capture tissue stiffness under varying conditions. In contrast, \name explores a low-frequency active sensing approach that estimates muscle stiffness from shear-wave propagation using a lightweight wearable sensing system designed for continuous deployment.

\paragraph{Wearable Systems for Muscle Health Monitoring.}
Many wearable solutions have been explored for muscle activation and health monitoring.
Electromyography (EMG) is among one of most well explored modality for muscle activity and health. 
EMG senses the electrical changes in muscle that arise from neuron activity and muscle engagement. Change in this signal during muscle activity is studied to estimate movement and muscle fatigue  \cite{Collins1998Apr, Woodward,Newcomer2002Jun,Chang2012Jan,Karthick2018Feb}.
The other sets of works like Mechanomyography (MMG) uses the low frequency vibrations generated by the muscle fibers to track active muscle contraction\cite{mokaya2015myovibe,uwamahoro2021assessment} and fatigue\cite{mokaya2016burnout,tarata2001accelerometer}.
The other set of work use  electric impedance sensing for muscle activity and motion monitoring \cite{zhu2022musclerehab,zhu2021eit}.
However, all these works only measure active muscle contraction induced changes and does not measure the physical property like muscle stiffness.
There are active vibroacoustic based systems\cite{rohal2022finepose,shriram2022sedentary} that brings the muscle stiffness sensing to wearable form factor.
However, these studies only profile relative stiffness changes and do not estimate or predict stiffness.


\paragraph{Physics-Informed and Quality-Aware Sensing Systems.}
Signals from low-cost Internet of Things (IoT) sensing systems are often noisy, impacted by heterogeneous deployment conditions, and measurement artifacts.
As a result, quality-aware sensing and inference strategies have been explored to improve the robustness in sensing applications \cite{zhang2021autoqual, li2022sqee, liu2025earfusion}.
For example, AutoQual quantifies sensor data quality as a data-driven model of upstream sensing system deployment factors that impact downstream task performance, based on expert heuristics \cite{zhang2021autoqual}. 
SQEE quantifies sensing data quality as the uncertainty of downstream machine learning models, with poor sensing conditions typically increasing prediction uncertainty \cite{li2022sqee}.
EarFusion leverages the designed modality-specific Signal Quality Indices to determine the multimodal fusion strategies in real-time \cite{liu2025earfusion}.
Earlier work on acoustic target tracking quantifies sensing data quality by comparing measured propagation delays with those predicted by a physical propagation model and uses the resulting quality estimate to suppress unreliable measurements and prioritize high-quality sensing reports \cite{wang2003acoustic}.

None of these solutions is applicable to the task of shear-wave velocity estimation via TDoA measurements for a few reasons. 
First, in wearable sensing systems, quality-impact factors and target information are often tightly integrated and difficult to separate, as in ambient sensing solutions. 
Second, TDoA estimation is fundamentally a signal-processing problem rather than a machine-learning prediction task. 
As a result, uncertainty estimates derived from downstream predictive models are either unavailable or only become observable after the TDoA has already been estimated.
Last but not least, the target shear-wave signal is often mixed with multiple propagation components and interference sources. 
Therefore, existing quality-aware sensing approaches do not provide this level of propagation-specific assessment.
\name explores new ways to quantify wave-specific quality for deployment guidance and to refine TDoA measurement estimation by designing wave-specific propagation-based metrics.

%% file: Discussion.tex
\section{Discussion}\label{sec:discuss}
This section discusses potential directions and new design spaces that build on this paper.

\paragraph{Opportunistic Sensing via Haptic Interaction.}\label{sec:discussion_interaction}
In this work, we use a bone-conduction transducer to generate shear waves for active sensing. 
Meanwhile, vibration transducers have been widely used for haptic interaction \cite{Huang2022Jan}. 
Therefore, one future direction is to explore opportunistic sensing through interaction, in which the same actuation signals used to provide haptic feedback are repurposed as excitation sources for sensing tissue or surface responses.
This problem is non-trivial because the design for sensing and interaction has different goals that need to be aligned and managed to jointly optimize the effects.
This alignment needs to take into account multiple factors, including sensing quality, user perception/fatigue, and interaction utility, requiring the exploration of new design spaces.

\paragraph{Closed-loop Intervention via Actionable Feedback.}
While this work focuses on measuring muscle stiffness, various forms of closed-loop intervention can be explored in the future. 
As discussed in Section \ref{sec:background}, muscle stiffness is an actionable physiological indicator associated with fatigue, recovery, injury risk, and rehabilitation progress.
Therefore, timely intervention leveraging continuous stiffness measurements would open new doors for muscle health management. 
For example, at a coarse temporal scale (hours to weeks), a long-term stiffness profile can be leveraged to provide personalized recommendations on training, recovery, and rehabilitation plans. At a fine temporal scale (seconds to minutes), haptic interventions (discussed in Section \ref{sec:discussion_interaction}) can provide real-time, muscle-specific feedback to guide exercise execution, stretching, and movement correction. 
Together, these interventions at different scales enable a closed-loop framework that provides comprehensive assistance for muscle health management.


%% file: Conclusion.tex
\section{Conclusion}\label{sec:conclusion}
This paper presents \name, a wearable vibroacoustic muscle-sensing system for muscle stiffness estimation. 
We design \name, a low-cost vibroacoustic active-sensing wearable hardware that uses a bone-conduction transducer and an IMU array to profile shear-wave propagation in the muscle to estimate resting muscle stiffness.
\name uses a physiology-guided deployment and signal processing pipeline to isolate the shear-wave propagation for effective collaborative TDoA estimation.
\name utilizes a physiology- and geometry-guided screening and refinement process to mitigate error. 
We validate \name across tissue-simulant ballistic gels of varied stiffness and test it with multiple users and muscle groups.
\name achieves an average stiffness estimation maximum error rate of 13.1\% across multiple muscle groups.
Our results highlight the potential of wearable vibroacoustic sensing as a practical and scalable approach for long-term muscle health monitoring, opening opportunities for applications in muscular injury prevention and rehabilitation.